\documentstyle[twoside,psfig]{article}

\catcode`\@=11
\long\def\@makefntext#1{
\protect\noindent \hbox to 3.2pt {\hskip-.9pt  
$^{{\eightrm\@thefnmark}}$\hfil}#1\hfill}		

\def\@makefnmark{\hbox to 0pt{$^{\@thefnmark}$\hss}}	
	
\def\ps@myheadings{\let\@mkboth\@gobbletwo
\def\@oddhead{\hbox{}
\rightmark\hfil\eightrm\thepage}   
\def\@oddfoot{}\def\@evenhead{\eightrm\thepage\hfil
\leftmark\hbox{}}\def\@evenfoot{}
\def\sectionmark##1{}\def\subsectionmark##1{}}

\oddsidemargin=\evensidemargin
\newcounter{sectionc}\newcounter{subsectionc}\newcounter{subsubsectionc}
\renewcommand{\section}[1] {\vspace{12pt}\addtocounter{sectionc}{1} 
\setcounter{subsectionc}{0}\setcounter{subsubsectionc}{0}\noindent 
	{\tenbf\thesectionc. #1}\par\vspace{5pt}}
\renewcommand{\subsection}[1] {\vspace{12pt}\addtocounter{subsectionc}{1} 
	\setcounter{subsubsectionc}{0}\noindent 
	{\bf\thesectionc.\thesubsectionc. {\kern1pt \bfit #1}}\par\vspace{5pt}}
\renewcommand{\subsubsection}[1] {\vspace{12pt}\addtocounter{subsubsectionc}{1}
	\noindent{\tenrm\thesectionc.\thesubsectionc.\thesubsubsectionc.
	{\kern1pt \tenit #1}}\par\vspace{5pt}}

\newcounter{appendixc}
\newcounter{subappendixc}[appendixc]
\newcounter{subsubappendixc}[subappendixc]
\renewcommand{\thesubappendixc}{\Alph{appendixc}.\arabic{subappendixc}}
\renewcommand{\thesubsubappendixc}
	{\Alph{appendixc}.\arabic{subappendixc}.\arabic{subsubappendixc}}

\renewcommand{\appendix}[1] {\vspace{12pt}
        \refstepcounter{appendixc}
        \setcounter{figure}{0}
        \setcounter{table}{0}
        \setcounter{lemma}{0}
        \setcounter{theorem}{0}
        \setcounter{corollary}{0}
        \setcounter{definition}{0}
        \setcounter{equation}{0}
        \renewcommand{\thefigure}{\Alph{appendixc}.\arabic{figure}}
        \renewcommand{\thetable}{\Alph{appendixc}.\arabic{table}}
        \renewcommand{\theappendixc}{\Alph{appendixc}}
        \renewcommand{\thelemma}{\Alph{appendixc}.\arabic{lemma}}
        \renewcommand{\thetheorem}{\Alph{appendixc}.\arabic{theorem}}
        \renewcommand{\thedefinition}{\Alph{appendixc}.\arabic{definition}}
        \renewcommand{\thecorollary}{\Alph{appendixc}.\arabic{corollary}}
        \renewcommand{\theequation}{\Alph{appendixc}.\arabic{equation}}
        \noindent{\tenbf Appendix \theappendixc #1}\par\vspace{5pt}}
\newcommand{\subappendix}[1] {\vspace{12pt}
        \refstepcounter{subappendixc}
        \noindent{\bf Appendix \thesubappendixc. {\kern1pt \bfit #1}}
	\par\vspace{5pt}}
\newcommand{\subsubappendix}[1] {\vspace{12pt}
        \refstepcounter{subsubappendixc}
        \noindent{\rm Appendix \thesubsubappendixc. {\kern1pt \tenit #1}}
	\par\vspace{5pt}}

\newcommand{\textlineskip}{\baselineskip=13pt}
\newcommand{\smalllineskip}{\baselineskip=10pt}

\def\eightcirc{
\begin{picture}(0,0)
\put(4.4,1.8){\circle{6.5}}
\end{picture}}
\def\eightcopyright{\eightcirc\kern2.7pt\hbox{\eightrm c}} 

\newcommand{\copyrightheading}[1]
	{\vspace*{-2.5cm}\smalllineskip{\flushleft
	{\footnotesize International Journal of Modern Physics D #1}\\
	{\footnotesize $\eightcopyright$\, World Scientific Publishing
	 Company}\\
	 }}

\newcommand{\publisher}[2]{{\begin{center}\footnotesize\smalllineskip 
	Received #1\\
	Revised #2
	\end{center}
	}}

\def\abstracts#1#2#3{{
	\centering{\begin{minipage}{4.5in}\footnotesize\baselineskip=10pt
	\parindent=0pt #1\par 
	\parindent=15pt #2\par
	\parindent=15pt #3
	\end{minipage}}\par}} 

\newcommand{\bibit}{\nineit}

\renewenvironment{thebibliography}[1]
	{\frenchspacing
	 \ninerm\baselineskip=11pt
	 \begin{list}{\arabic{enumi}.}
	{\usecounter{enumi}\setlength{\parsep}{0pt}
	 \setlength{\leftmargin 12.7pt}{\rightmargin 0pt} 
	 \setlength{\itemsep}{0pt} \settowidth
	{\labelwidth}{#1.}\sloppy}}{\end{list}}

\newcounter{itemlistc}
\newcounter{romanlistc}
\newcounter{alphlistc}
\newcounter{arabiclistc}

\newcommand{\fcaption}[1]{
        \refstepcounter{figure}
        \setbox\@tempboxa = \hbox{\footnotesize Fig.~\thefigure. #1}
        \ifdim \wd\@tempboxa > 5in
           {\begin{center}
        \parbox{5in}{\footnotesize\smalllineskip Fig.~\thefigure. #1}
            \end{center}}
        \else
             {\begin{center}
             {\footnotesize Fig.~\thefigure. #1}
              \end{center}}
        \fi}

\newcommand{\tcaption}[1]{
        \refstepcounter{table}
        \setbox\@tempboxa = \hbox{\footnotesize Table~\thetable. #1}
        \ifdim \wd\@tempboxa > 5in
           {\begin{center}
        \parbox{5in}{\footnotesize\smalllineskip Table~\thetable. #1}
            \end{center}}
        \else
             {\begin{center}
             {\footnotesize Table~\thetable. #1}
              \end{center}}
        \fi}

\def\@citex[#1]#2{\if@filesw\immediate\write\@auxout
	{\string\citation{#2}}\fi
\def\@citea{}\@cite{\@for\@citeb:=#2\do
	{\@citea\def\@citea{,}\@ifundefined
	{b@\@citeb}{{\bf ?}\@warning
	{Citation `\@citeb' on page \thepage \space undefined}}
	{\csname b@\@citeb\endcsname}}}{#1}}

\newif\if@cghi
\def\cite{\@cghitrue\@ifnextchar [{\@tempswatrue
	\@citex}{\@tempswafalse\@citex[]}}
\def\citelow{\@cghifalse\@ifnextchar [{\@tempswatrue
	\@citex}{\@tempswafalse\@citex[]}}
\def\@cite#1#2{{$\null^{#1}$\if@tempswa\typeout
	{IJCGA warning: optional citation argument 
	ignored: `#2'} \fi}}

\def\pmb#1{\setbox0=\hbox{#1}
	\kern-.025em\copy0\kern-\wd0
	\kern.05em\copy0\kern-\wd0
	\kern-.025em\raise.0433em\box0}

\def\fnt#1#2{\footnotetext{\kern-.3em
	{$^{\mbox{\scriptsize #1}}$}{#2}}}

\def\@makefnmark{\hbox to 0pt{$^{\@thefnmark}$\hss}}	
	
\def\ps@myheadings{%
    \let\@oddfoot\@empty\let\@evenfoot\@empty
    \def\@evenhead{\slshape\leftmark\hfil}
    \def\@oddhead{\hfil{\slshape\rightmark}}
    \let\@mkboth\@gobbletwo
    \let\sectionmark\@gobble
    \let\subsectionmark\@gobble
    }
\font\tenrm=cmr10
\font\tenit=cmti10 
\font\tenbf=cmbx10
\font\bfit=cmbxti10 at 10pt
\font\ninerm=cmr9
\font\nineit=cmti9

\font\eightrm=cmr8

\def\qed{\hbox{${\vcenter{\vbox{			
   \hrule height 0.4pt\hbox{\vrule width 0.4pt height 6pt
   \kern5pt\vrule width 0.4pt}\hrule height 0.4pt}}}$}}

\def\be{\begin{eqnarray}}
\def\en{\end{eqnarray}}
\def\bea{\begin{eqnarray}}
\def\ena{\end{eqnarray}}
\def\mM{\mathcal{M}}
\def\sg{\sqrt{-g}}
\def\sgm{\sqrt{-\gamma}}

\begin{document}
\setlength{\textheight}{7.7truein}  

\thispagestyle{empty}

\markboth{\protect{\footnotesize\it Finite-range gravity $\ldots$}}
{\protect{\footnotesize\it Finite-range gravity $\ldots$}}

\normalsize\textlineskip

\setcounter{page}{1}

\copyrightheading{}	

\vspace*{0.88truein}

\centerline{\bf FINITE-RANGE GRAVITY AND ITS ROLE IN GRAVITATIONAL WAVES, }
\vspace*{0.035truein}
\centerline{\bf BLACK HOLES AND COSMOLOGY}
\vspace*{0.37truein}

\centerline{\footnotesize S. V. BABAK}
\baselineskip=12pt
\centerline{\footnotesize\it Department of Physics and Astronomy, Cardiff University,
Cardiff CF24 3YB, UK.}
\vspace*{10pt}
\centerline{\footnotesize  L. P. GRISHCHUK}
\baselineskip=12pt
\centerline{\footnotesize\it Department of Physics and Astronomy, Cardiff University,
Cardiff CF24 3YB, UK}
\baselineskip=10pt
\centerline{\footnotesize\it {\rm and} Sternberg Astronomical Institute, 
Moscow University, Moscow 119899, Russia.}
\vspace*{0.225truein}
\publisher{(received date)}{(revised date)}

\vspace*{0.21truein}
\abstracts{Theoretical considerations of fundamental physics, as well as certain
cosmological observations, persistently point out to permissibility, and
maybe necessity, of macroscopic modifications of the Einstein general 
relativity. The field-theoretical formulation of general relativity helped 
us to identify the phenomenological seeds of such modifications. They take 
place in the form of very specific mass-terms, which appear in addition to 
the field-theoretical analog of the usual Hilbert-Einstein Lagrangian. We 
derive and study exact non-linear equations of the theory, along with its 
linear approximation. We interpret the added terms as masses of $spin-2$ 
and $spin-0$ gravitons. The arising finite-range gravity is a fully consistent 
theory, which smoothly approaches general relativity in the massless limit, 
that is, when both masses tend to zero and the range of gravity tends to 
infinity. We show that all local weak-field predictions of the theory are in 
perfect agreement with the available experimental data. However, some other 
conclusions of the non-linear massive theory are in a striking contrast with 
those of general relativity. We show in detail how the arbitrarily small 
mass-terms eliminate the black hole event horizon and replace a permanent 
power-law expansion of a homogeneous isotropic universe with an oscillatory 
behaviour. One variant of the theory allows the cosmological scale factor to 
exhibit an `accelerated expansion' instead of slowing down to a regular maximum 
of expansion. We show in detail why the traditional, Fierz-Pauli, massive
gravity is in conflict not only with the static-field experiments, but also with 
the available indirect gravitational-wave observations. At the same time, we 
demonstrate the incorrectness of the widely held belief that the non-Fierz-Pauli 
theories possess `negative energies' and `instabilities'.}{}{}

\vspace*{1pt}\textlineskip	


\section{Introduction}
Presently, there seems to be no pressing need in devising theories of
gravitation, alternative to the existing Einstein's general 
relativity. General relativity (GR) is an internally consistent theory,
and it has passed all the performed experimental tests with flying 
colors \cite{Willbook}, \cite{Fock}, \cite{LL}, \cite{MTW}, \cite{W}. 
And yet, there are some clouds on the horizon. 
On the theoretical side, the M/string theory considerations persistently
point out to possible macroscopic modifications of GR, particularly in the 
form of various ``mass-terms". On the observational side, there exists
some discomfort in understanding the large-scale structure and evolution of 
the Universe, including some indications to the possibility of its 
present ``accelerated expansion". So far, theorists enjoy 
playing with the cosmological $\Lambda$-term and various highly 
speculative forms of matter, but the credibility of these models 
can soon be exhausted. The old question arises again, whether there 
do exist well-motivated consistent alternative theories of macroscopic
gravity, with non-trivial observational consequences.

At the first sight, general relativity is an isolated theory with
no immediate neighbours. In particular, it seems that GR cannot be 
modified without raising the order of differential field equations. Indeed, 
in the geometrical formulation of GR, which operates with the curved space-time
metric tensor $g_{\mu \nu}$, there is no structure that can be added
to the usual Hilbert-Einstein Lagrangian. The only possibility is the
$\Lambda$-term: ${\sqrt {-g}} \Lambda$, but this structure can be included
in the definition of GR, and we know all the phenomenological
consequences of the $\Lambda$-term, and in any case the $\Lambda$-term 
is not a ``mass-term". The situation changes drastically when one looks at GR 
from the field-theoretical perspective. The old remark of Feynman \cite{F} on
the intrinsic value of equivalent formulations of a fundamental theory 
proves to be very profound. One gets the possibility to analyse the problems 
which otherwise could not be even properly formulated. We believe that
general relativity does indeed contain the seeds of its own modification, 
and the field-theoretical formulation of GR helped us to identify these seeds.
The modification of GR, which looks almost unavoidable from the viewpoint 
of the field-theoretical approach, leads to the appearance of very specific
mass-terms. The resulting theory is a fully consistent finite-range 
gravitational theory. General relativity is a smooth limit of this theory when 
the range of gravity tends to infinity. The theory is in perfect agreement 
with all local weak-field experiments, such as experiments in the Solar system, 
and satisfies the requirement, formulated long ago \cite{BD}, of ``physical 
continuity". However, some other consequences of the theory are truly 
striking. It is surprising to see that some of the crucial conclusions of 
GR are so much vulnerable to pretty innocent modifications of GR. For instance,
the existence of a black hole event horizon, and a permanent power-law
expansion of the matter-dominated Universe, get invalidated by the 
arbitrarily small mass-terms. We introduce and explain 
this finite-range gravitational theory in the present paper.    

The fundamental quantity in the field-theoretical GR is a symmetric 
second-rank tensor field $h^{\mu \nu}(x^{\alpha})$. The gravitational field
$h^{\mu \nu}(x^{\alpha})$ is defined in a flat space-time with the line-element 
\be
{\rm d} \sigma^2 = \gamma_{\mu \nu} {\rm d}x^{\mu} {\rm d} x^{\nu}.
\label{sigma}
\en
The curvature tensor constructed from $\gamma_{\mu \nu}(x^{\alpha})$ is 
identically zero:
\be
\label{R}
\breve R_{\alpha\beta\mu\nu}(\gamma_{\rho\sigma}) = 0. 
\en
In flat space-time, one is always free to choose Lorentzian coordinates, 
in which case Eq. (\ref{sigma}) takes on the Minkowski form
\be
\label{Mi}
{\rm d} \sigma^2 = \eta_{\mu \nu} {\rm d}x^{\mu} {\rm d} x^{\nu} =
c^2{\rm d}t^2 - {\rm d}x^2 - {\rm d}y^2 - {\rm d}z^2.
\en
The flat space-time is not a choice of some artificial ``prior geometry", 
but is a reflection of experimental facts. As far
as the present-day physics knows, the intervals of space and durations of 
time, in absence of all fields including gravity, satisfy the relationships
of the Minkowski 4-dimensional interval (\ref{Mi}). If there existed any
observational evidence to something different, we would have started 
from a different metric.

The Lagrangian of the field-theoretical GR depends on the 
gravitational field variables $h^{\mu \nu}(x^{\alpha})$ and their first 
derivatives. (We present more details in Sec.~2) The variational 
principle gives rise to the dynamical field equations, which are fully 
equivalent to the Einstein equations. 
The transition to the geometrical formulation of GR proceeds through the
introduction of the tensor $g^{\mu \nu}(x^{\alpha})$ and the inverse tensor 
$g_{\mu \nu}(x^{\alpha})$: $g^{\mu \rho}g_{\nu \rho} = \delta^{\mu}_{\nu}$.
The quantities $g^{\mu \nu}$ are calculable from the gravitational field 
variables $h^{\mu \nu}$ and the metric tensor $\gamma^{\mu \nu}$ 
according to the rule  
\bea
\label{g}
\sqrt{-g}g^{\mu\nu}= \sqrt{-\gamma}(\gamma^{\mu\nu} + h^{\mu\nu})~, 
\ena
where $g=det|g_{\mu\nu}|$, $\gamma=det|\gamma_{\mu\nu}|$, and 
$\gamma^{\mu \rho}\gamma_{\nu \rho}=\delta^{\mu}_{\nu}$.
The tensor density $\sqrt{-g}g^{\mu\nu}$ participates in the matter Lagrangian,
realizing the universal coupling of gravity to all other fields, but apart of 
that, it is simply a short-hand notation for the quantity in the 
right-hand-side (r.h.s.) of Eq.~(\ref{g}). In the geometrical formulation 
of GR, tensor $g_{\mu \nu}(x^{\alpha})$ is interpreted as the metric tensor of 
a curved space-time:  
\be
\label{ds}
{\rm d}s^2 = g_{\mu \nu} {\rm d}x^{\mu} {\rm d} x^{\nu}.
\en
In terms of $g_{\mu \nu}$, the field equations acquire the familiar form of the
geometrical Einstein's equations. From the viewpoint of the field-theoretical 
formulation, the tensor $g_{\mu \nu}(x^{\alpha})$ is the effective metric 
tensor; it defines the intervals of space and time measured in the presence 
of the universal gravitational field $h^{\mu \nu}(x^{\alpha})$. 
The field-theoretical approach to GR has a long and fruitful history.
For a sample of references, see \cite {Pap}, \cite{Gupta}, \cite{Kr}, 
\cite{Thirring}, \cite{Ros}, \cite{weinb}, \cite{Deser}, \cite{Feyn}, 
\cite{GPP}, \cite {GZ}, including a history review \cite{lastreview}, 
and many papers cited therein.

It was shown \cite{BG} that the gravitational Lagrangian of the 
field-theoretical GR must include, in addition to the field-theoretical 
analog of the Hilbert-Einstein term, the extra term
\be
\label{ad}
\sqrt{-\gamma}\left[-\frac{1}{4}\breve R_{\alpha\rho\beta\sigma}
(h^{\alpha\beta}h^{\rho\sigma} - h^{\alpha\sigma} h^{\rho\beta})\right].
\en
This term does not affect the field equations, but is needed for the
variational derivation of the gravitational energy-momentum 
tensor $t^{\mu \nu}$. The variational (metrical) energy-momentum tensor 
is the response of a physical system to variations of the metric tensor
$\gamma_{\mu \nu}$, caused by arbitrary coordinate transformations. 
Obviously, such variations of the metric tensor should 
obey the constraint (\ref{R}). The variational procedure incorporates 
the constraint (\ref{R}) by adding to the Lagrangian an extra term: 
$\Lambda^{\alpha \beta\rho\sigma} \breve R_{\alpha\rho\beta\sigma}$,
where $\Lambda^{\alpha \beta\rho\sigma}$ are undetermined 
Lagrange multipliers. The constraint (\ref{R}) has to be enforced at the 
end of the variational derivation of the field equations and the 
energy-momentum tensor. It has been proven \cite{BG} that the Lagrange 
multipliers must have the unique form 
\[
\Lambda^{\mu \nu \alpha \beta} = - \frac{1}{4}(h^{\alpha\beta} h^{\mu\nu} -
h^{\alpha \nu} h^{\beta\mu}), 
\]
in order for the derived energy-momentum tensor $t^{\mu \nu}$  
to satisfy all the necessary mathematical and physical requirements, 
including the absence of second derivatives of the field variables 
in the $t^{\mu \nu}$. 

As was explained above, the quantity $\breve R_{\alpha\rho\beta\sigma}$ 
in Eq.~(\ref{ad}) is the curvature tensor of a flat space-time. If it were
something other than that, the theory would not be GR. 
However, it is natural to assume that the Lagrangian may also include
an additional term similar to (\ref{ad}), but where  
the quantity $\breve R_{\alpha\rho\beta\sigma}$ is the curvature tensor of
an abstract space-time with a constant non-zero curvature. 
Space-times of constant curvature are as symmetric as flat space-time, 
but contain a parameter $K$ with dimensionality of $[length]^{-2}$:  
\be
\label{cc}
\breve R_{\alpha\beta\mu\nu}=K(\gamma_{\alpha\mu}\gamma_{\beta\nu}-
\gamma_{\alpha\nu}\gamma_{\beta\mu}).
\en
If one uses (\ref{cc}) in (\ref{ad}), the generated additional term in the 
Lagrangian is  
\be
\label{fp}
\sqrt{-\gamma}\frac{K}{2}(h^{\alpha \beta}h_{\alpha\beta} - h^2).
\en
Clearly, the new theory is not GR, but what this theory is ? 
Quite surprisingly, one recognizes in (\ref{fp}) the 
Fierz-Pauli \cite{FP} mass-term. 
Having discovered that the structure (\ref{ad}) generates mass-terms,
we have asked about the most general form of such terms. It is easy
to show that there exist only two independent quadratic combinations: 
$h^{\alpha \beta}h_{\alpha\beta}$ and $h^2$. Therefore, we arrive at a
2-parameter family of theories with the additional mass-terms in the
gravitational Lagrangian: 
\be
\label{two}
\sqrt{-\gamma}\left[ k_1h^{\rho\sigma}h_{\rho\sigma} + k_2 h^2\right],
\en 
where $k_1$ and $k_2$ have dimensionality of $[length]^{-2}$. 
Fierz and Pauli, as well as many other authors after them, were considering 
the (internally contradictory) ``linear gravity", whereas in our case 
the tensor $h^{\mu\nu}$ is 
the full-fledged non-linear gravitational field. The 2-parameter 
class of theories with the additional mass-terms (\ref{two}) is 
what we shall study in the present paper. We consider the mass-terms as
phenomenological, even though their deep origin can be quantum-mechanical 
or multi-dimensional. 

The structure of the paper and its conclusions are as follows.

In Sec.2 we derive exact non-linear equations, as well as gravitational 
energy-momentum tensor, for the gravitational field in absence
of any matter sources. Since almost all calculations in gravitational
physics are performed in geometrical language, and we will need some of
the results, we introduce the notion of a quasi-geometrical description
of the finite-range gravity.\footnote{Geometry in physics, like communism 
in politics, is not very dangerous, if introduced in well-measured doses.} 
Specifically, we retain the usual presentation of the Einstein part of 
the equations in terms of $g_{\mu \nu}$, but in the massive part, which 
originates from (\ref{two}) and cannot be written in 
terms of $g_{\mu \nu}$ only, we trade $h_{\mu \nu}$ for $g_{\mu \nu}$ and
$\gamma_{\mu \nu}$, according to the rule (\ref{g}). The important point is 
the symmetries of the theory. Equations of the field-theoretical GR enjoy two 
different symmetries. The first one (general covariance, or diffeomorphism) 
is the freedom to use arbitrary coordinates and the associated 
transformations of, both, the metric tensor $\gamma^{\mu \nu}$ and the field
tensor $h^{\mu \nu}$. The second symmetry is the freedom to use the 
(true) gauge transformations, which do not touch coordinates and the
metric tensor, but transform the field variables only \cite{GPP}. It is 
this second symmetry that gets violated by the mass-terms, while the first 
symmetry survives.  

In Sec.3 we formulate exact equations for the gravitational field in
the presence of matter sources. Again, we are often using the quasi-geometrical 
description. This means, in particular, that in the matter part of the 
field equations we retain the geometrical energy-momentum tensor 
$T_{\mu \nu}$, i.e. the matter energy-momentum tensor defined 
as the variational derivative of the matter Lagrangian with respect to 
$g^{\mu \nu}$, as opposed to the field-theoretical 
energy-momentum tensor $\tau_{\mu\nu}$, defined as the variational 
derivative of the matter Lagrangian with respect to $\gamma^{\mu\nu}$. 
The content of Sec.3 will be needed in Section 7 and, partially, 
in Section 5.  

In Sec.4 we discuss the linearised approximation of the theory and give 
physical interpretation to the parameters $k_1$ and $k_2$. In accord with
the analysis of Ogievetsky and Polubarinov \cite{OP}, 
and Van Dam and Veltman \cite{VDV}, these
parameters give rise to the two fundamental masses: the mass $m_2$ 
of the $spin-2$ graviton, and the mass $m_0$ of the $spin-0$ graviton. 
Strictly speaking, the corresponding wave-equations contain two fundamental
lengths, rather than two fundamental masses. Concretely, the equations
contain two parameters, $\alpha^2$ and $\beta^2$, with dimensionalities 
of $[length]^{-2}$:
\be
\label{ab}
\alpha^2 = 4k_1, ~~~~~ \beta^2 = -2k_1 \frac{k_1+4k_2}{k_1+k_2}, 
\en
but $\alpha$ and $\beta$ can be thought of as inverse Compton wavelengths 
of the two gravitons with the masses 
\be
\label{mm}
m_2 = \frac{\alpha \hbar}{c}, ~~~~~m_0 = \frac{\beta \hbar}{c}. 
\en
The interpretation of the free parameters in terms of masses implies that
$\alpha^2$ and $\beta^2$ are strictly positive quantities. However, the 
Lagrangian itself does not require this restriction, and we will exploit 
this freedom in the cosmological Section 7. 

One very special choice of the parameters $k_1$ and $k_2$ is $k_2 = -k_1$. 
This choice of the parameters brings the
Lagrangian (\ref{two}) to the Fierz-Pauli form (\ref{fp}). It is this
case that has led to a lively debate 
on the unacceptability of a ``massive graviton".
Although the Lagrangian (\ref{fp}) itself does smoothly vanish in the
limit $k_1 \rightarrow 0$, the corresponding solutions and local 
weak-field physical predictions 
(for instance, the deflection angle of light propagating in the 
gravitational field of the Sun) do not approach those of GR. 
In other words, this particular massive theory 
disagrees with the original massless theory even in the limit of
vanishingly small mass $m_2$ and, hence, in the limit of 
arbitrarily long Compton wavelength $1/\alpha$. The finite, and
independent of the mass $m_2$, difference in local predictions became 
known as the Van Dam-Veltman-Zakharov discontinuity \cite{VDV}, 
\cite{Zakh}, \cite {FadSl}, \cite{Iwas}, \cite{Vain}, \cite{Visser}, 
\cite{KogMP}, \cite{Por}, \cite{DDGV}, \cite{CarG}, \cite{DesT}, 
\cite{Gruz}. This puzzling conclusion about discontinuity is 
described \cite{Velt} as something that seems counter-intuitive to certain 
physicists. We have to confess that the usual presentation of this conclusion 
seems counter-intuitive to us, as well. We believe that the issue should be 
looked upon from a different angle. When taking the massless limit of 
a massive theory, one should do what the logic requires to do, namely, 
to send both masses to zero. Then, both Compton wavelengths 
tend to infinity, and one recovers, as expected, the local weak-field 
predictions of GR. If, instead, one takes $k_2 = -k_1$ (whatever
the motivations behind this choice might be), the mass $m_0$ becomes 
infinitely large (see Eqs. (\ref{ab}), (\ref{mm})) and the corresponding 
Compton wavelength $1/\beta$ is being sent to zero. Any local experiment 
is now supposed to be performed at scales much larger than one of the 
characteristic lengths, $1/\beta$. In this situation, the deviations
from GR should be expected on the grounds of physical intuition. There is 
no wonder that the subsequent limit $1/\alpha \rightarrow \infty$ does 
not cure these deviations. This situation may look like a counter-intuitive 
discontinuity. 

To explore the difference
in local predictions, there is no need to propagate light in the Solar
system. It is sufficient to consider the geodesic deviation equation for
free test bodies separated by small distances.   
We do this study below in the paper. In particular, the geodesic deviation 
equation illustrates the difference between GR and finite-range gravity in 
the domain of gravitational-wave predictions.

In Sec.5 and Appendices B, C, we study weak gravitational waves. Certain 
modifications of GR are well anticipated. In the field-theoretical GR, 
the $spin-0$ gravitational waves (represented by the trace 
$h=h^{\mu \nu} \eta_{\mu \nu}$) exist as gauge solutions. They contribute 
neither to the gravitational energy-momentum tensor $t^{\mu \nu}$, nor to 
the deformation pattern of a ring of test particles in the geodesic deviation
equation. The same is true for the $helicity-0$ polarization state
(represented by the spatial trace $h^{ij} \eta_{ij}$) of the $spin-2$ graviton.
In the finite-range gravity, as one could expect, both these degrees of 
freedom become essential. They provide additional contributions to the 
energy-momentum flux carried by the gravitational wave, and the extra 
components of motion of the test particles. However, gravitational wave 
solutions, their energy-momentum characteristics, and observational predictions 
of GR are fully recovered in the massless limit $\alpha \rightarrow 0$, 
$\beta \rightarrow 0$ of the theory. 
We show that the Fierz-Pauli case is very peculiar and unacceptable. Even in
the limit of $\alpha \rightarrow 0$, there remains a nonvanishing ``common mode" 
motion of test particles in the plane of the wave front. The extra component
of motion is accounted for by the corresponding 
additional flux of energy from the source; typically, of the same order of
magnitude as the GR flux. This analysis, together with 
the Solar system arguments, leads to the important conclusion. Whatever 
the sophisticated ``brane world" motivations of the M/string theory may be, 
if they lead to the phenomenological mass-term of the Fierz-Pauli type, 
the corresponding variant of the theory should be rejected
as being in conflict with the static-field experiments and with the already 
available indirect gravitational-wave observations of binary pulsars.
We do not think that this conclusion can be invalidated by
any ``non-perturbative effects". At the same time, by doing concrete 
calculations, we dispel the deeply-rooted myth that the non-Fierz-Pauli 
theories should suffer from ``negative energies" and ``instabilities". 

The fully non-linear finite-range gravity is considered in the next two 
Sections. In Sec.6 
we analyse static spherically-symmetric solutions. We summarise 
the weak-field approximation in Appendix~\ref{bhb}. 
There, we demonstrate that the GR solutions and physical predictions are 
recovered in the massless limit $\alpha \rightarrow 0$, $\beta \rightarrow 0$, 
and, with the help of the geodesic deviation equation, we confirm the 
observational unacceptability of the Fierz-Pauli coupling. 
The main thrust of Sec.6 is the non-linear (would-be black hole) 
solutions. The case of arbitrary relationship
between  $\alpha$ and $\beta$ is difficult to analyse in full generality. 
The equations are somewhat simpler when the masses are assumed to be equal, 
i.e. $\alpha = \beta$. We call this choice the
Ogievetsky-Polubarinov (OP) case. We analyse this case in great detail, and
present more general considerations whenever possible, demonstrating that 
the qualitative conclusions remain valid for $\alpha \neq \beta$. 
A single dimensionless parameter in the OP case is $\alpha M$, where $M$ is 
the Schwarzschild mass (using $G =1$ and $c=1$), which is supposed to be a 
very small number.  We start with intermediate scales, that is, with 
Schwarzschild distances $R$ which are much larger than $2M$, but much 
smaller than $1/\alpha$. Combining 
analytical and numerical techniques, we demonstrate that the solution of 
the massive theory is practically indistinguishable from that of GR
for all $R$ sufficiently larger than $2M$, but, obviously, smaller
than $1/\alpha$. As expected, for $R$ larger than $1/\alpha$, the solution
takes on the form of the Yukawa-type potentials; this is why the theory is 
called finite-range gravity. However, the massive solution strongly 
deviates from that of GR not only at very large distances, but also in the 
vicinity of $R=2M$. This is a consequence of the non-linear character
of the field equations. The hypersurface $R=2M$ is the location of the 
(globally defined) event horizon of the Schwarzschild black hole in GR. 
We carefully explore the vicinity of $R=2M$, as well as $0 \le R < 2M$, 
the region that would have been the interior of the 
Schwarzschild black hole. We show that the smaller the parameter $\alpha M$, 
the closer to $R=2M$ one can descend (from large $R$) along the essentially 
Schwarzschild solution. We show that the deviations from GR 
near $R=2M$ are so radical that the event horizon does not form, and the
solution smoothly continues to the region $R < 2M$. The further 
continuation of the solution terminates at $R=0$, where the curvature 
singularity develops. Since the $\alpha M$ can be extremely small, 
the redshift of the photon emitted at $R=2M$ can be extremely large, 
but it remains finite. In contrast to GR, the infinite redshift is reached 
at the singularity $R=0$, and not at $R=2M$.
The conclusion of this study is quite dramatic. In the astrophysical sense, 
the resulting solution still looks like a black hole; in the region
of space just outside the $R=2M$, the gravitational field 
is practically indistinguishable from the Schwarzschild solution. However, 
all conclusions that rely specifically on the existence of the black hole event 
horizon, are likely to be abandoned. It is very remarkable and surprising that 
the phenomenon of black hole should be so unstable with respect to the 
inclusion of the tiny mass-terms (\ref{two}), whose Compton wavelengths can
exceed, say, the present-day Hubble radius.

Section 7 is devoted to cosmological solutions for a homogeneous isotropic
universe. Matter sources are taken in the simplest form of perfect fluids 
with fixed equations of state. First, we show that if the mass of
the $spin-0$ graviton is zero, i.e. $\beta^2 = 0$, the cosmological
solutions of the massive theory are exactly the same as those of GR,
independently of the mass of the $spin-2$ graviton, that is, independently
of the value of $\alpha^2$. This result could be expected due to 
the highest spatial symmetry of the problem under consideration; the
$spin-2$ degrees of freedom have no chance to reveal themselves.
Then, we proceed to cases with $\beta^2 \neq 0$. Since we prefer to deal 
with technically simple equations, we consider a particular case 
$4\beta^2 = \alpha^2$. This case is studied in full details, 
but we also show that the qualitative
results are general and are valid for $4\beta^2 \neq \alpha^2$. Combining 
analytical approximations and numerical calculations, we demonstrate 
that the massive solution has a long interval of evolution where it
is practically indistinguishable from the Friedmann solution of GR.
However, the deviations from GR are dramatic
at very early times and very late times. The unlimited expansion is
being replaced by a regular maximum of the scale factor, whereas the
singularity is being replaced by a regular minimum of the scale factor. 
The smaller $\beta$, the higher maximum and the deeper minimum. In other
words, astonishingly, the arbitrarily small mass-terms (\ref{two}) give 
rise to the oscillatory behaviour of the cosmological scale factor.  
 
Following the logic of interpretation of the theory in terms of masses,
we assume in the most of the paper that the signs of $\alpha^2$ and 
$\beta^2$ are positive. However, as mentioned above,
the general structure of the Lagrangian (\ref{ad}) does not imply this.
It is interesting to observe that if we allow $\alpha^2$ and $\beta^2$ to
be negative (which would probably require to think of the
massive gravitons in terms of ``tachyons"), the late time evolution of the 
scale factor exhibits an 
``accelerated expansion", instead of slowing down towards the maximum. 
This behaviour of the scale factor is similar to the one governed 
by a positive cosmological $\Lambda$-term. The physical significance of 
this result is presently unclear, but the problem deserves further study.
In any case, cosmological modifications proposed here are justified
better, than in many inconsistent ``ad hoc" models that appeared 
in the literature. 

We briefly summarise our results in the concluding Sec.8 and
relegate some technical details of the paper to Appendices A, B, C, D.


\section{Source-free gravitational field}\label{sec11a}

The gravitational contribution $S^g$ to the total action is
$$
S^g= \frac 1{c} \int L^g \;d^4x. 
$$
The gravitational Lagrangian density $L^g$ consists of two parts - the GR 
part and the massive part:
\bea
L^g = L^{g}_{GR} + L^{g}_{mass}. \label{genform}
\ena

As was explained in Introduction, the GR part itself consists of two
terms \cite{BG}, which are i) the field-theoretical analog of the 
Hilbert-Einstein Lagrangian and ii) the term incorporating the 
constraint (\ref{R}):
\bea
L^g_{GR} =  -\frac{\sqrt{-\gamma}}{2\kappa}\left\{\frac{1}{2}
{\Omega^{-1}_{\rho\sigma\alpha\beta}}^{\omega\tau}
{h^{\rho\sigma}}_{;\tau} {h^{\alpha\beta}}_{;\omega} - \frac{1}{4}
(h^{\rho \sigma} h^{\alpha \beta} - h^{\alpha \sigma} h^{\beta \rho})
\breve R_{\alpha\rho\beta\sigma} \right\}, 
\ena
where $\kappa= 8 \pi G/c^4$. The field-theoretical analog of the 
Hilbert-Einstein term has the 
form similar to the kinetic energy of classical mechanics; the Lagrangian is 
manifestly quadratic in the generalised velocities  ${h^{\mu\nu}}_{;\tau}$. 
[We remind the reader that the raising and lowering of indeces of the 
field $h^{\mu \nu}$, and its covariant differentiation denoted by a
semicolon ``;", are performed with the help of the metric tensor 
$\gamma_{\mu \nu}$ and its Christoffel symbols $C^{\alpha}_{\rho \sigma}$.] 
The generalised coordinates $h^{\mu\nu}$ are present only in the tensor 
${\Omega^{-1}_{\rho\sigma\alpha\beta}}^{\omega\tau}$. For the reference,
we reproduce here the compact expression of this tensor, but we refer
to \cite{BG} for details,
\bea
{\Omega^{-1}_{\mu\nu\rho\sigma}}^{\tau\omega}= \frac 1{4} \frac{\sqrt{-\gamma}}
{\sqrt{-g}}[(\delta^{\tau}_{\mu}\delta^{\pi}_{\nu} +
\delta^{\tau}_{\nu}\delta^{\pi}_{\mu})(\delta^{\omega}_{\rho}
\delta^{\lambda}_{\sigma} +
\delta^{\omega}_{\sigma}\delta^{\lambda}_{\rho})g_{\pi\lambda}
- \nonumber \\
g^{\tau\omega}(g_{\mu\rho}g_{\nu\sigma} + g_{\nu\rho}g_{\mu\sigma}
- g_{\mu\nu}g_{\rho\sigma})]~. \label{Omega-1}
\ena

As was explained in Introduction, the massive part of the Lagrangian is
given by
\bea
L^g_{mass} =  -\frac{\sqrt{-\gamma}}{2\kappa}\left\{
k_1h^{\rho\sigma}h_{\rho\sigma} +k_2 h^2\right\},  \label{Lmg}
\ena
where, obviously, $h \equiv h^{\alpha \beta}\gamma_{\alpha \beta}$. 

Having defined the gravitational Lagrangian, we are in the position
to derive the source-free field equations: 
\bea
\label{feq}
\frac{\delta L^g}{\delta h^{\alpha\beta}}= \frac{\partial L^g}
{\partial h^{\alpha\beta}} - \left(\frac{\partial L^g}{\partial 
{h^{\alpha\beta}}_{;\sigma}}\right)_{;\sigma} = 0~.
\ena
Certainly, the GR part alone generates the equations completely 
equivalent to the Einstein equations:
\bea 
\label{evac}
\frac{1}{2} \left[(\gamma^{\mu \nu} +h^{\mu\nu})(\gamma^{\alpha\beta}+
h^{\alpha\beta}) - (\gamma^{\mu \alpha} +h^{\mu\alpha})(\gamma^{\nu\beta}+
h^{\nu\beta})\right]_{; \alpha ; \beta} = \kappa t^{\mu\nu},
\ena  
where $t^{\mu \nu}$ is the gravitational energy-momentum tensor satisfying 
all the necessary requirements (see formula (65) in \cite{BG}). 
\footnote{It appears that the attitude towards the field-theoretical 
GR is approaching the last phase of the quite usual response, when the  
dialog begins with the objection ``this is impossible and cannot be true", 
goes through ``this is interesting but has not been proven", and finishes 
with the reassuring ``this is correct and wonderful, and I have done this 
long ago".}~ To write equations (\ref{evac}) in the 
geometrical language, one composes combination of equations 
(\ref{feq}) by multiplying them with the factor 
$\delta_{\mu}^{\alpha}\delta^{\beta}_{\nu} -\frac1{2} g^{\alpha\beta}
g_{\mu\nu}$ and uses relationship (\ref{g}) (for details, see
\cite{BG}). As a result, one arrives at the geometrical Einstein's equations    
\bea
G_{\mu\nu} =0, \label{1psgf}
\ena
where
$G_{\mu\nu}$ is the Einstein tensor
\bea
G_{\mu\nu} \equiv 
R_{\mu\nu} - \frac1{2}g_{\mu\nu}R 
\label{1Gmn}
\ena
and $R_{\mu\nu}$ is the Ricci tensor constructed from $g_{\mu\nu}$ in the
usual manner. 

The massive part $L^g_{mass}$ makes its own contribution to the field
equations,
\bea
-\frac{2\kappa}{\sqrt{-\gamma}} \frac{\delta L^{g}_{mass}}{\delta h^{\mu\nu}}=
2k_1 h_{\mu\nu} + 2k_2 \gamma_{\mu\nu}h ~, \label{8cf}
\ena
and to the gravitational energy-momentum tensor: 
\bea
\label{massemt}
\kappa t^{\mu \nu}_{mass} &=& (k_1 +2k_2) h^{\mu\nu}h -
2 (k_1 + k_2 h)h^{\mu}_{\alpha} h^{\nu \alpha}- 
2k_1 h_{\alpha \beta} h^{\nu\alpha}h^{\mu\beta} +  \nonumber \\
& &\frac1{2}(\gamma^{\mu \nu}+2 h^{\mu \nu}) (k_1h^{\alpha\beta}
h_{\alpha\beta} + k_2 h^2).
\ena 
The field equations (\ref{1psgf}) get modified. Taking into account 
(\ref{8cf}) and repeating the steps described above for the GR case, one 
arrives at the source-free equations of the finite-range theory:
\bea
G_{\mu\nu} + M_{\mu\nu}=0, \label{psgf}
\ena
where
\bea
M_{\mu\nu} \equiv  
\left(\delta^{\alpha}_{\mu}\delta^{\beta}_{\nu} -
\frac1{2} g^{\alpha\beta}g_{\mu\nu}\right)
\left( 2k_1h_{\alpha\beta} +2k_2 \gamma_{\alpha\beta}h \right). 
\label{1M}
\ena
Replacing, with the help of (\ref{g}), $h^{\mu \nu}$ in favour of $g^{\mu\nu}$ 
and $\gamma^{\mu\nu}$, one obtains the quasi-geometric form of $M_{\mu \nu}$:
\bea
M_{\mu\nu} &=& 2 \gamma_{\rho\alpha}\gamma_{\sigma\beta}
\left(\delta^{\alpha}_{\mu}\delta^{\beta}_{\nu} -
\frac1{2} g^{\alpha\beta}g_{\mu\nu}\right)
\left[ k_1 \left(\frac{\sqrt{-g}}{\sqrt{-\gamma}}g^{\rho\sigma}- 
\gamma^{\rho\sigma} \right) + \right. \nonumber \\  
& & \left. k_2 \gamma^{\rho\sigma}\left(\frac{\sqrt{-g}}
{\sqrt{-\gamma}}g^{\tau\psi} \gamma_{\tau\psi} -4 \right) \right]. \label{M}
\ena
Thus, in the source-free case, we have to solve the quasi-geometric 
equations (\ref{psgf}), instead of the GR equations (\ref{1psgf}). 
In the purely field-theoretic formulation, we would have to solve the 
modified Eq. (\ref{evac}), where the l.h.s. of Eq. (\ref{evac}) contains 
the additional term ${\mM}^{\mu \nu}$ (formula (\ref{mM}) below), 
while the r.h.s. 
of Eq. (\ref{evac}) contains the additional term $\kappa t^{\mu\nu}_{mass}$
(formula (\ref{massemt}) above).

The choice of coordinates and, hence, the form of the metric tensor 
$\gamma_{\mu \nu}$, is entirely in our hands. In what follows, 
we will be using the Lorentzian coordinates (\ref{Mi}) or, 
where convenient, spatially-spherical coordinates,
\be
\label{sph}
{\rm d}\sigma^2=c^2{\rm d}t^2-{\rm d}r^2-r^2({\rm d}\theta^2+
sin^2\theta {\rm d}\phi^2).
\en

One difference between Eqs.~(\ref{psgf}) and Eqs.~(\ref{1psgf}) is apparent.
Let us denote by a stroke ``$|$'' the covariant derivative defined with the 
help of the effective metric tensor $g_{\mu\nu}$ and its Christoffel symbols 
$\Gamma^{\alpha}_{\rho \sigma}$. Then, the Bianchi identities read 
\bea
g^{\mu \alpha}G_{\alpha\nu |\mu}\equiv G^{\mu}_{\ \nu |\mu} \equiv 0. 
\label{Bianki}
\ena
In other words, this particular combination of equations (\ref{1psgf})
is satisfied identically. This is not so in the case of equations 
(\ref{psgf}). Applying the same differentiation, one arrives at the non-trivial 
consequences of equations (\ref{psgf}):
\bea
g^{\mu\alpha}M_{\alpha\nu |\mu} \equiv M^{\mu}_{\ \nu |\mu}= 0.
\label{MBian}
\ena
Although Eqs. (\ref{MBian}) are merely the consequences of the 
full set of Eqs. (\ref{psgf}), and therefore contain no new information, 
it proves convenient, as will be shown below, to use them instead 
of some members of the original set of Eqs. (\ref{psgf}). 

It is interesting to note that Eqs.~(\ref{MBian}) can also be written 
in the totally
equivalent form, which employs the field variables $h^{\mu \nu}$ and the
metric tensor $\gamma^{\mu \nu}$: 
\bea
{\mM^{\mu\nu}}_{;\nu}=0, \label{fbian}
\ena
where
\bea
{\mM}^{\mu\nu}&\equiv& 2k_1 h^{\mu\nu} - (k_1 + 2k_2)\gamma^{\mu\nu}h +
2k_1h^{\nu\beta}h^{\mu}_{\ \beta} + 2k_2 h^{\mu\nu}h -  \nonumber \\
& &\frac1{2}\gamma^{\mu \nu} (k_1h^{\alpha\beta}
h_{\alpha\beta} + k_2 h^2). \label{mM}
\ena
We will prove the equivalence of (\ref{MBian}) and (\ref{fbian}) 
in Appendix A. The representation (\ref{fbian}) 
will be especially helpful in the cosmological Sec.~7. 

\section{Gravitational field with matter sources}
\label{sec11b}

The total action in the presence of matter sources is 
$$
S=\frac 1{c} \int (L^g + L^m)\; d^4x, 
$$
where $L^m$ is the Lagrangian density for matter fields. $L^m$ includes 
also the interaction of matter fields with the gravitational 
field. One or several matter 
fields are denoted by $\phi_A$, where $A$ is some general index.
We assume the universal coupling of all matter fields to the 
gravitational field. Specifically, we assume that $L^m$
depends on the gravitational field variables $h^{\mu\nu}$ in a 
particular manner, namely, through the combination 
$\sqrt{-g}g^{\mu\nu}$: 
\bea
L^m=L^m\left[\sqrt{-\gamma} (\gamma^{\mu\nu} + h^{\mu\nu}); 
(\sqrt{-\gamma} (\gamma^{\mu\nu} + h^{\mu\nu}))_{,\alpha}; 
\phi_A; \phi_{A,\alpha}\right]. \label{mcoupl}
\ena
The adopted coupling of matter to gravity is exactly the same
as in GR. Therefore the matter field equations, 
\bea
\frac{\delta L^m}{\delta \phi_A}=0,  \label{meqmm}
\ena
are also exactly the same as in GR.

We can now derive the gravitational field equations with matter sources, 
\bea
-\frac{2\kappa}{\sqrt{-\gamma}} \frac{\delta L^{tot}}{\delta h^{\mu\nu}} =
- \frac{2\kappa}{\sqrt{-\gamma}}\frac{\delta L^{g}}{\delta h^{\mu\nu}}
- \frac{2\kappa}{\sqrt{-\gamma}}\frac{\delta L^{m}}{\delta h^{\mu\nu}}=0.
\label{feqs}
\ena
Everything what comes out of $L^g$ is already known. The contribution 
of $L^m$ to the gravitational field equations can be worked out by taking 
the advantage of the specific form of Eq.~(\ref{mcoupl}). Indeed, one can 
write  
\bea
\frac2{\sqrt{-\gamma}} \frac{\delta L^m}{\delta h^{\mu\nu}} = 
2 \frac{\delta L^m}{\delta (\sqrt{-g}g^{\rho\sigma})} = 
T_{\mu\nu} -\frac1{2}
g_{\mu\nu}g^{\alpha\beta} T_{\alpha\beta}. \label{m78}
\ena 
where $T_{\mu\nu}$ is the (geometrical) energy-momentum tensor of the matter.
It is defined as the variational derivative of $L^m$ with respect to 
$g^{\mu\nu}$: 
\[
T_{\mu\nu} = \frac2{\sqrt{-g}} \frac{\delta L^m}{\delta g^{\mu\nu}}.
\]
Multiplying the field equations (\ref{feqs}) by the factor 
$\delta_{\mu}^{\alpha}\delta^{\beta}_{\nu} -
\frac1{2} g^{\alpha\beta} g_{\mu\nu}$ and rearranging the terms, we
arrive at the gravitational field equations of the finite-range gravity 
with matter sources:
\bea
 G_{\mu \nu} + M_{\mu\nu} = \kappa T_{\mu\nu}.\label{feqps}
\ena

Let us now discuss the consequences of these equations related to 
the Bianchi identities $G^{\mu}_{\ \nu |\mu}\equiv 0$. It is
important to remember that the conservation equations
\be
\label{Tconser}
T^{\mu}_{\ \nu |\mu}=0
\en
are satisfied as soon as the matter equations
of motion (\ref{meqmm}) are satisfied. In other words, eqs.~(\ref{Tconser}) 
are satisfied independently of the gravitational
field equations. Therefore, taking the $|$-covariant divergence of
Eq.~(\ref{feqps}) and assuming that Eq.~(\ref{meqmm}) are fulfilled, 
we obtain equations (\ref{MBian}) and (\ref{fbian}), i.e. exactly the 
same equations as in the source-free case.

\section{Linearised theory of the source-free gravitational field}
\label{sec12}

The proper physical interpretation of the free parameters $k_1$ and $k_2$
is revealed from the linearised field equations. The
linearisation means that the quantities $h^{\mu\nu}$ are regarded small, 
and only terms linear in $h^{\mu\nu}$ are retained in the field equations.
To find the linear version of Eq.~(\ref{psgf}) one can use the linear version
of Eq.~(\ref{g}):
\be
\label{gL}
g^{\mu \nu} \approx \gamma^{\mu\nu}+(h^{\mu\nu}-\frac{1}{2} \gamma^{\mu\nu}h).  
\en
It is also possible to start from the modified Eq.~(\ref{evac}) remembering 
that the total $t^{\mu \nu}$ is not less than quadratic in $h^{\mu\nu}$. By 
either route, one arrives at the linearised version of equations for the 
finite-range gravity:  
\bea
\frac{1}{2} \left[{h^{\mu\nu ;\alpha}}_{;\alpha} +
\gamma^{\mu\nu}{h^{\alpha\beta}}_{;\alpha ;\beta} -
{h^{\nu\alpha ;\mu}}_{;\alpha}- {h^{\mu\alpha ;\nu}}_{;\alpha}\right] + \nonumber \\
\left[2k_1h^{\mu\nu}-\left(k_1 + 2k_2\right) \gamma^{\mu\nu} h \right]=0.
\label{Lpsgf}
\ena
The useful consequence of Eq.~(\ref{Lpsgf}) is derived by taking the
$;$-covariant divergence of Eq.~(\ref{Lpsgf}). Since the $;$-covariant 
divergence of the GR part (first square bracket) is identically zero, we 
obtain
\bea
\left[2k_1 h^{\mu\nu}-\left(k_1 + 2k_2\right)\gamma^{\mu\nu} h
\right]_{;\nu}=0. \label{Leq2}
\ena
Clearly, Eq.~(\ref{Leq2}) is the linearised version of the exact 
Eq.~(\ref{fbian}). 

Apparently, the first study of the two-parameter set of 
equations (\ref{Lpsgf}) has been done by Fierz and Pauli \cite{FP} who 
assumed that $k_2 = - k_1$. We will consider here the general case of 
arbitrary parameters. Following Van Dam and Veltman \cite{VDV}, we will 
start from transforming Eq.~(\ref{Lpsgf}) to the more suggestive set 
of equations. 

Taking the covariant divergence of (\ref{Leq2}) one derives 
\bea
2k_1{h^{\mu\nu}}_{;\mu ;\nu}= \left(k_1 + 2 k_2\right)\Box h, \label{L2.5}
\ena
where the symbol $\Box$ denotes the d'Alembert operator in arbitrary
(in general, curvilinear) coordinates: 
$\Box Z= \gamma^{\alpha\beta}Z_{;\alpha ;\beta}$.
Taking the trace of Eq.~(\ref{Lpsgf}) one derives 
\bea
\label{Lh1} 
\Box h + 2 {h^{\alpha\beta}}_{;\alpha ;\beta} - 4 (k_1+4k_2)h=0. 
\ena 
Combining Eq.~(\ref{L2.5}) and Eq.~(\ref{Lh1}) in
order to exclude ${h^{\alpha\beta}}_{;\alpha ;\beta}$,
one obtains the equation for the trace $h$:
\bea
\Box h - 2k_1 \frac{k_1+4k_2}{k_1+k_2}h =0. \label{Leq3}
\ena

Obviously, when writing down the equation (\ref{Leq3}), we assume that 
$k_2\ne -k_1$. Otherwise, i.e. in the Fierz-Pauli case $k_1 +k_2 =0$, the 
full set of equations (\ref{Lpsgf}) is equivalent to
\bea
h=0 \label{k-1.0},\;\;
{h^{\mu\nu}}_{;\nu}=0,\;\;
\Box h^{\mu\nu} + 4k_1 h^{\mu\nu} =0 . \label{k=-1}
\ena
We shall discuss the observational consequences of the Fierz-Pauli theory
in Sections 5 and 6. Meanwhile, we shall return to the general 
case $k_2 \ne -k_1$.

It is convenient to introduce the quantity $H^{\mu\nu}$ according to the  
relationship
\bea
H^{\mu\nu}= h^{\mu\nu} - \frac{k_1+k_2}{3k_1} \gamma^{\mu\nu} h -
\frac{k_1+k_2}{6k_1^2} h^{;\mu ;\nu} +
\frac{k_1+k_2}{12 k_1^{2}} \gamma^{\mu\nu} \Box h. 
\label{dech}
\ena
The trace and the covariant divergence of $H^{\mu\nu}$ vanish 
due to the equations (\ref{Leq2}) and (\ref{Leq3}),  
\bea
\gamma_{\mu\nu}H^{\mu\nu}=0, \;\;\;
{H^{\mu\nu}}_{;\nu}=0. \label{TT}
\ena
In terms of $H^{\mu\nu}$, and taking into account (\ref{Leq2}) 
and (\ref{Leq3}), Eqs.~(\ref{Lpsgf}) transform to 
\bea
\Box H^{\mu\nu} + 4k_1 H^{\mu\nu} =0.  \label{Leq4}
\ena

We now introduce $\alpha^2$ and $\beta^2$ according to the 
definitions (\ref{ab}). Then, the full set of equations (\ref{Lpsgf})
is equivalent to 
\bea
2k_1{h^{\mu\nu}}_{;\nu}-\left(k_1 +2k_2\right)h^{;\mu}=0, \label{dDL}\\
\Box h +\beta^2 h =0,  \label{lsp0} \\
\Box H^{\mu\nu} + \alpha^2 H^{\mu\nu} =0. \label{lsp2}
\ena
The wave-like form of the resulting equations (\ref{lsp2}) and (\ref{lsp0})
justifies the interpretation of the parameters $k_1$ and $k_2$,
\[
k_1 = \frac{\alpha^2}{4}, ~~~~~ k_2= -\frac{\alpha^2(\alpha^2+2 \beta^2)}
{8(2 \alpha^2+ \beta^2)},
\]
in terms of masses (\ref{mm}). The constraints (\ref{TT}) allow one to associate
the tensor field $H^{\mu\nu}$ with the $spin-2$ graviton, whereas the scalar 
quantity $h$ can be associated with the $spin-0$ graviton.


\section{Weak gravitational waves}
\label{gw}

We shall start with a brief summary of gravitational waves in GR, and we shall do this 
in the framework most appropriate for the further comparison with the case of massive 
theory. In this Section, it is convenient to work in Lorentzian coordinates (\ref{Mi}). 
This means that the metric tensor $\gamma_{\mu\nu}$ simplifies to $\eta_{\mu\nu}$,
and the $;$-covariant derivative simplifies to the ordinary derivative, denoted 
by a comma.

\subsection{Weak gravitational waves in general relativity.}\label{gwa}

The linearised source-free equations (\ref{Lpsgf}) take the form 
\be
\label{Lgw}
{h^{\mu\nu ,\alpha}}_{,\alpha} + \eta^{\mu\nu}{h^{\alpha\beta}}_{,\alpha ,\beta} - 
{h^{\nu\alpha ,\mu}}_{,\alpha}- {h^{\mu\alpha ,\nu}}_{,\alpha} = 0.
\en
A general plane-wave solution to Eq.~(\ref{Lgw}) is given by 
\be
\label{plw}
h^{\mu\nu}= a^{\mu\nu} e^{i  k_{\alpha} x^{\alpha}} +
(c^{\mu}q^{\nu}+c^{\nu}q^{\mu} - \eta^{\mu\nu}c^{\alpha}q_{\alpha}) 
e^{i  q_{\alpha} x^{\alpha}} + c.c., 
\en
where $c.c$ denotes the complex conjugate part; $k_{\alpha}k^{\alpha} = 0$; 
$q_{\alpha}q^{\alpha}$ can be of any sign or zero; 4 quantities $c^{\mu}$ are arbitrary; 
10 components of the matrix $a^{\mu\nu}$ are constrained by 4 conditions: 
\be
\label{ak}
a^{\mu\nu}k_{\nu} = 0.
\en
Conditions (\ref{ak}) allow one to express $a^{00}$ and $a^{0i}$ in terms of 6 
independent components of the spatial matrix $a^{ij}$:
\be
\label{a0mu}
a^{00} =\frac{1}{k_0^2} a^{ij}k_ik_j,~~~~~a^{0i}= -\frac{1}{k_0} a^{ij}k_j.
\en
The matrix $a^{ij}$ itself can be written in the general form
\be
\label{atild}
a^{ij}=\tilde {a}^{ij} +b^ik^j+b^jk^i +b \eta^{ij},
\en  
where 4 quantities $b, b^i$ are arbitrary, while the matrix $\tilde{a}^{ij}$ has 
only 2 independent components (sometimes called TT-components) as it 
satisfies 4 extra conditions:   
\be
\label{TTa}
\tilde{a}^{ij}k_j = 0, ~~~~~~ \tilde{a}^{ij}\eta_{ij} = 0. 
\en

The gravitational energy-momentum tensor $t^{\mu\nu}$, in its lowest (quadratic) 
approximation, 
depends only on the TT-components of the matrix $a^{ij}$ (see Appendix C). 
Specifically, 
\be
\label{gwenm}
t^{\mu\nu}= \frac{1}{4 \kappa} k^\mu k^\nu \left[2 \tilde{a}^{ij} 
\tilde{a}^{*}_{ij} \right],
\en
where we have dropped the purely oscillatory terms.
For a plane wave propagating in $z$-direction, 
i.e. for $k_{\alpha} = (k_0,~0,~0,~k_3)$ and $k_0^2-k_3^2 = 0$, the energy-momentum 
tensor (\ref{gwenm}) depends only on 
$$
\tilde{a}^{11}=\frac{1}{2}(a^{11}-a^{22})=-\tilde{a}^{22}~~~~
{\rm and}~~~~\tilde{a}^{12}= a^{12}.
$$  

Numerical values of $a^{\mu\nu}$ are determined by the source of 
gravitational waves. In the presence of matter sources, the wave-equations 
are  
\be
\label{Lgwm}
{h^{\mu\nu ,\alpha}}_{,\alpha} + \eta^{\mu\nu}{h^{\alpha\beta}}_{,\alpha ,\beta} - 
{h^{\nu\alpha ,\mu}}_{,\alpha}- {h^{\mu\alpha ,\nu}}_{,\alpha} = 2\kappa T^{\mu\nu}.
\en
It is assumed that the field $h^{\mu\nu}(x^{\alpha})$ is Fourier-expanded:
\be
\label{fexp}
h^{\mu\nu}(x^{\alpha})=\frac 1{(2\pi)^4}\int a^{\mu\nu}(k_{\alpha}) 
e^{ik_{\alpha}x^{\alpha}} d^4k 
\ena
Then, the amplitudes $a^{\mu\nu}(k_{\alpha})$ are determined by Fourier
components of the retarded solution to Eq.~(\ref{Lgwm}). Assuming that the 
distance $R_0$ to the source is large, one obtains
\be
\label{amplGR} 
a^{\mu\nu}(k_{\alpha})= \frac{1}{L} \hat{T}^{\mu\nu}(k_{\alpha}),
\en
where 
\be
\hat{T}^{\mu\nu}(k_{\alpha})=\int\left( \int T^{\mu\nu}(t_r,r_0) d^3 r_0\right)
e^{-ik_{\alpha}x^{\alpha}} d^4x, ~~~~~ 
\frac{1}{L} = \frac{2 \kappa}{4 \pi R_0}, 
\nonumber 
\en
and $t_r$ is retarded time. Therefore, we obtain 
\be
\label{atild2}
\tilde{a}^{11}= - \tilde{a}^{22}=  
\frac{1}{2L} [\hat{T}^{11}(k_{\alpha})-
\hat{T}^{22}(k_{\alpha})],~~~ 
\tilde{a}^{12}= \frac{1}{L} \hat{T}^{12}(k_{\alpha}).
\en

The action of a gravitational wave on free test particles can be discussed 
in terms of the geodesic deviation equation:
\bea
\frac{D^2 \xi^{\alpha}}{ds^2}= R^{\alpha}_{\ \mu\nu\sigma}u^{\mu}u^{\nu}
\xi^{\sigma}, \nonumber
\ena
where $\xi^{\alpha}$ is the separation between world-lines of two nearby freely 
falling particles. Assuming that the reference particle in the origin of the
coordinate system is at rest, i.e. $u^{\alpha}=(1,~0,~0,~0)$, 
the geodesic deviation equation reduces to 
\bea
\label{lgdein} 
\frac{d^2\xi^{i}}{c^2dt^2}=R^{i}_{\ 0j0}\xi^{j}. 
\ena
The Riemann tensor is given by
\bea
R_{\alpha\mu\beta\nu}&=& \left[ \frac 1{2} (h_{\alpha\nu,\mu,\beta}+
h_{\mu\beta,\alpha,\nu}- h_{\alpha\beta,\mu,\nu}-h_{\mu\nu,\alpha,\beta}) -
\right.\nonumber \\
& & \left. \frac 1{4} \left( \eta_{\alpha\nu}h_{,\mu\beta}+\eta_{\mu\beta}
h_{,\alpha,\nu}-\eta_{\alpha\beta}h_{,\mu,\nu}-\eta_{\mu\nu}
h_{,\alpha,\beta}\right)\right] \label{linriem}
\ena
Calculating (\ref{linriem}) with the help of (\ref{plw}) one finds that the 
terms with $c^{\mu}$ cancel out identically. Moreover, using (\ref{a0mu}) and 
(\ref{atild}) one finds that the components $R^{i}_{\ 0j0}$, participating
in Eq.~(\ref{lgdein}), depend only on $\tilde{a}^{ij}$:
\be
\label{riemTT}
R_{i0j0} = \frac{1}{2} k_0^2 \tilde{a}_{ij} e^{i k_{\alpha}x^{\alpha}}.
\en
Therefore, for a wave propagating in $z$-direction, equations (\ref{lgdein})
read:
\bea
\frac{d^2\xi^1}{c^2dt^2}&=& -\frac{1}{2}k_0^2\left[\frac{1}{2}
(a^{11}-a^{22})\xi^1+ a^{12} \xi^2\right]e^{i  k_0 x^0},
\nonumber \\
\frac{d^2\xi^2}{c^2dt^2}&=& -\frac{1}{2}k_0^2\left[a^{12} \xi^1 -\frac{1}{2}
(a^{11}-a^{22})\xi^2\right]e^{i  k_0 x^0},
\nonumber \\
\frac{d^2\xi^3}{c^2dt^2}&=& 0
\label{mrelmotws}.
\ena
Recalling equations (\ref{atild2}), one relates the deformation pattern of a
set of test particles to the energy-momentum tensor $T^{\mu\nu}$ of the 
gravitational-wave source:
\bea 
\label{defT}
\frac{d^2\xi^1}{c^2dt^2}&=& -\frac{1}{2} k_0^2 \frac{1}{L} 
\left[\frac{1}{2} \left(\hat{T}^{11}-
\hat{T}^{22}\right)\xi^1 + \hat{T}^{12} \xi^2\right] e^{i  k_0 x^0},
\nonumber \\
\frac{d^2\xi^2}{c^2dt^2}&=& -\frac{1}{2} k_0^2 \frac{1}{L} 
\left[\hat{T}^{12} \xi^1 - \frac{1}{2} \left(\hat{T}^{11}- 
\hat{T}^{22}\right)\xi^2\right] e^{i  k_0 x^0},
\nonumber \\
\frac{d^2\xi^3}{c^2dt^2}&=& 0.
\ena

\subsection{Gravitational waves in the finite-range gravity.}\label{gwb}

Source-free equations are now Eqs. (\ref{lsp0}), (\ref{lsp2}) and (\ref{dDL}).
We shall consider the Fierz-Pauli case, i.e. Eq.~(\ref{k=-1}), separately. 
Plane-wave solutions  to (\ref{lsp0}) and (\ref{lsp2}) are given by
\bea
h&=&C e^{i  q_{\alpha} x^{\alpha}} + c.c.~, 
\label{scwave}\\
H^{\mu\nu}&=& a^{\mu\nu} e^{i  k_{\alpha} x^{\alpha}} + c.c.~, 
\label{tenwave}
\ena
where the wave-vectors $q_{\alpha}$ and $k_{\alpha}$ satisfy the conditions 
\bea
q_{\alpha} q^{\alpha} = \beta^2,~~~~ k_{\alpha} k^{\alpha} = \alpha^2. \nonumber
\ena
As a consequence of (\ref{TTa}), 10 members of the matrix $a^{\mu\nu}$ are
restricted by 5 constraints:
\be
\label{5con} 
a^{\mu\nu} \eta_{\mu\nu} =0, \;\;\;\;\;\;\; a^{\mu\nu}k_{\nu} =0.
\en

Having found $h$ and $H^{\mu\nu}$ we can write down the original quantities
$h^{\mu\nu}$, using for this purpose the relationship (\ref{dech}). In doing 
that, it is convenient to introduce a new notation:  
\[
A = -C \frac{\alpha^2}{2(2 \alpha^2 +\beta^2)}.
\]
Then,
\bea
h^{\mu\nu}= a^{\mu\nu}e^{i  k_{\alpha} x^{\alpha}}-
\eta^{\mu\nu} A e^{i  q_{\alpha} x^{\alpha}}
+ 2 \frac{q^{\mu}q^{\nu}}{\alpha^2}A e^{i  q_{\alpha} x^{\alpha}}- 
\eta^{\mu\nu}\frac{\beta^2}{\alpha^2} A e^{i  q_{\alpha} x^{\alpha}} +c.c.~.
\label{lad1}
\ena
With this $h^{\mu\nu}$, Eq.~(\ref{dDL}) is satisfied automatically.
 
One can note that the term with $\eta^{\mu\nu}$ in Eq.~(\ref{lad1}) (and indeed
the original term with $\gamma^{\mu\nu}$ in Eq.~(\ref{dech})) is split 
into two parts. This has been done on purpose. One can see that the last two terms 
in Eq.~(\ref{lad1}) have the structure of the $c^{\mu}$-terms in Eq.~(\ref{plw})
(with $c^{\mu} \propto q^{\mu}$).
We know that these terms do not contribute to the observational effects of
the geodesic deviation equation, because of cancellation of these terms in the 
Riemann tensor (\ref{linriem}). Also, these terms do not contribute 
to the gravitational energy-momentum tensor (see Appendix C). This fact 
simplifies calculations, but it also demonstrates that the appearance 
of $\alpha^2$ in the denominator of the field variables does not necessarily
represent any danger by itself, even if $\alpha^2$ is eventually sent to zero. 
 
The gravitational wave amplitudes $a^{\mu\nu}$ and $A$ of the finite-range
gravity are determined by the source emitting gravitational waves (for 
detailed calculation see Appendix~\ref{gwc}).

\subsection{Observable manifestations of gravitational waves in the 
finite-range gravity.}\label{gwd}

The geodesic deviation equation (\ref{lgdein}) is valid regardless of whether 
the participating field $h^{\mu\nu}$ is a solution to equations of GR or to 
equations of the finite-range gravity. However, solutions $h^{\mu\nu}$ are 
different in these two theories and, therefore, explicit expressions for 
$R_{i0j0}$ are also different. Using  (\ref{lad1}) and (\ref{5con}), we find 
in the massive theory:
\bea
\label{Rmass}
R_{i0j0}&=&\frac 1{2}\left(k_0^2 a_{ij} -k_ik_j a^{lm}\eta_{lm} + 
k_i a_j^l k_l +k_ja_i^lk_l \right) e^{i  k_{\alpha} x^{\alpha}} + \nonumber \\
& & \frac{1}{2}A\left(q_0^2 \eta_{ij} + 
q_{i}q_{j}\right)e^{i  q_{\alpha} x^{\alpha}}.
\ena
Therefore, Eq.~(\ref{lgdein}) takes the form
\bea
\frac{d^2\xi^{i}}{c^2dt^2}&=&\frac{1}{2}\left(
k_0^2 a^i_j \xi^j -k^i a^{lm} \eta_{lm} k_j \xi^j + k^ia_j^l k_l \xi^j +
a^{il}k_lk_j \xi^j \right) e^{i  k_{0} x^{0}} +  \nonumber \\
& &\frac{1}{2}A\left(q_0^2 \xi^i  + q^{i}q_j \xi^j \right) e^{i  q_0 x^0}.
\label{masgeq}
\ena

As before, we consider a wave propagating in $z$-direction. This means that
$k_0^2 - k_3^2 = \alpha^2$, $q_0^2 - q_3^2 = \beta^2$, and a consequence of
Eq.~(\ref{5con}) reads $a^{11} + a^{22} + \frac{\alpha^2}{k_0^2} a^{33} =0$. 
For concreteness, we will be interested in response of the test particles
to waves of a given fixed frequency, so we put $q_0 = k_0$ in Eq.~(\ref{masgeq}).
Then, Eqs.~(\ref{masgeq}) take on a simpler form: 
\bea
\frac{d^2\xi^1}{c^2dt^2}&=& -\frac{1}{2} k_0^2\left[(a^{11} -A)\xi^1 +a^{12}\xi^2 +
\frac{\alpha^2}{k_0^2}a^{13}\xi^3\right]e^{i  k_0 x^0}, \nonumber \\
\frac{d^2\xi^2}{c^2dt^2}&=& -\frac{1}{2} k_0^2\left[a^{12}\xi^1 +(a^{22} -
A)\xi^2 + \frac{\alpha^2}{k_0^2}a^{23}\xi^3\right]e^{i  k_0 x^0}, \nonumber \\
\frac{d^2\xi^3}{c^2dt^2}&=& -\frac{1}{2} \alpha^2 \left[a^{13}\xi^1 +
a^{23}\xi^2 + \frac{\alpha^2}{k_0^2}a^{33} \xi^3\right]e^{i  k_0 x^0} 
+\frac{1}{2} \beta^2 A \xi^3 e^{i  k_0 x^0}.
\label{mrelmotprel}
\ena
Clearly, the small terms proportional to $\alpha^2$ and $\beta^2$ provide
the extra components of motion as compared with the GR 
behaviour (\ref{mrelmotws}).

The next step is to use in Eq.~(\ref{mrelmotprel}) the calculated amplitudes 
(\ref{amnA}), and to explore the massless limit. In doing that, 
we will be taking into account 
a consequence of Eq.~(\ref{Tcons1}) which reads: 
\[
\hat{T} \equiv \hat{T}^{00}-
\hat{T}^{11}- \hat{T}^{22}-\hat{T}^{33}=-\hat{T}^{11}-\hat{T}^{22}-
\frac{\alpha^2}{k_0^2} \hat{T}^{33}
\]
A little calculation shows that
\[
a^{11} - A = \frac{1}{L} \left[ \frac{1}{2}(\hat{T}^{11}-
\hat{T}^{22}) -\frac{1}{2} \frac{\alpha^2}{k_0^2} \hat{T}^{33}\right],
\]
\[
a^{22} - A = \frac{1}{L} \left[- \frac{1}{2}(\hat{T}^{11}-
\hat{T}^{22}) -\frac{1}{2} \frac{\alpha^2}{k_0^2} \hat{T}^{33}\right],
\]
\[
\frac{\alpha^2}{k_0^2} a^{33} \approx 
\frac{1}{L} \left[\frac{1}{3}\hat{T}\right]. 
\]
Therefore, Eqs.~(\ref{mrelmotprel}) reduce to 
\bea
\frac{d^2\xi^1}{c^2dt^2}&=&- \frac{1}{2} k_0^2\frac{1}{L} 
\left[\frac1{2}(\hat{T}^{11}-\hat{T}^{22})\xi^1 +\hat{T}^{12} \xi^2\right] 
e^{ik_0x^0} +O(\alpha^2, \beta^2),\nonumber\\
\frac{d^2\xi^2}{c^2dt^2}&=&- \frac{1}{2} k_0^2\frac{1}{L} 
\left[ \hat{T}^{12} \xi^1 -\frac1{2}(\hat{T}^{11}-\hat{T}^{22})\xi^2 \right] 
e^{ik_0x^0} +O(\alpha^2, \beta^2), \nonumber\\
\frac{d^2\xi^3}{c^2dt^2}&=& -\frac{1}{L} \left[\frac{\alpha^2}{2} 
\left(\hat{T}^{13}\xi^1+ \hat{T}^{23} \xi^2 \frac1{3}\hat{T}\xi^3 +O(\alpha^2)\right)
+ \right.\nonumber \\
& &\left.\frac{\beta^2}{2} \left(\frac1{6}\hat{T}\xi^3+O(\beta^2)\right)\right] e^{ik_0x^0}. 
\label{masgdfin}
\ena
In the massless limit, when both $\alpha^2$ and $\beta^2$ tend to zero,
Eqs.~(\ref{masgdfin}) approach Eqs.~(\ref{defT}), and, hence, all the  
observational manifestations of gravitational waves in the finite-range
gravity tend to those of GR. In full accord with physical intuition, 
the gravitational energy-momentum tensor, as a function of the 
g.w. source, also tends to that of GR, without any ``negative energies",
etc. (see Appendix C).  

\subsection{The Fierz-Pauli case.}\label{gwe}
 
It was shown above that the smooth transition to GR is achieved when
both parameters $\alpha^2$ and $\beta^2$ are sent to zero. The Fierz-Pauli
coupling violates this requirement, as it postulates that $k_1+k_2 =0$. 
From the viewpoint of the 2-parameter finite-range gravity this choice
of $k_1$ and $k_2$ corresponds to the limit of $\beta^2 = \infty$. In these
circumstances, strong deviations from GR should be expected on the grounds 
of physical intuition, independently of the value of the remaining free parameter 
$\alpha^2$. Although the deviations are expected, and even in the limit
$\alpha^2 \rightarrow 0$, it is interesting to study the Fierz-Pauli 
theory on its own, regardless of its place in the 2-parameter family.   

The starting point of the discussion is equations (\ref{Lpsgws}), in 
which $k_2$ is taken to be equal to $-k_1$. In particular, the source-free 
equations reduce to the set of equations (\ref{k=-1}). The quantities $H^{\mu\nu}$
coincide with $h^{\mu\nu}$, as is seen from  
Eq.~(\ref{dech}). The wave-equations in the presence of matter sources can 
be derived anew, but in fact they can also be recovered from the existing 
equations (\ref{dDL}), (\ref{mlhws}), (\ref{mlhttws}), 
if one takes the limit $k_1+k_2 = 0$ ($\beta^2 \rightarrow \infty$). 
In particular, Eq.~(\ref{mlhws}) now reads:
\be
h =\frac{1}{3 \alpha^2}2 \kappa T. \label{mlhwsfp}  
\en
The field degree of freedom represented by $h$ ($spin-0$ graviton) has lost 
the ability to be radiated away. Moreover, $h$ vanishes everywhere outside the 
matter source, and $h$ can be non-zero only within the region occupied 
by matter with $T \ne 0$. As for equations (\ref{mlhttws}), they are exactly 
the same as before, but $H^{\mu\nu} \equiv h^{\mu\nu}$. 

Far away from the radiating source, the gravitational wave field can still be
written in the form of Eq.~(\ref{lad1}), with $a^{\mu\nu}$ satisfying the
constraints (\ref{5con}), but the amplitude $C$ in this equation is strictly 
zero. The retarded solution to Eq.~(\ref{mlhttws}) produces 
the same amplitudes $a^{\mu\nu}$ as in formula (\ref{amnA}), 
but $A \equiv 0$ and $A \beta^2 \equiv 0$.     
The necessary change to the geodesic deviation equation (\ref{masgeq}) 
consists in dropping out the term with $A$. For a wave propagating in 
$z$-direction, Eqs. (\ref{mrelmotprel}) retain their form, but with 
$A=0$ and $A \beta^2 =0$. Since
\[
\hat{T}^{11} +\frac{1}{3} \hat{T} = \frac{1}{2} \left( \hat{T}^{11} - 
\hat{T}^{22} \right) -\frac{1}{6} \hat{T}- \frac{\alpha^2}{2 k_0^2}\hat{T}^{33},
\]
and
\[
\hat{T}^{22} +\frac{1}{3} \hat{T} = -\frac{1}{2} \left( \hat{T}^{11} - 
\hat{T}^{22} \right) -\frac{1}{6} \hat{T}- \frac{\alpha^2}{2 k_0^2}\hat{T}^{33},
\]
equations (\ref{mrelmotprel}), in the limit $\alpha^2 \rightarrow 0$, 
take the form 
\bea 
\label{defTfp} 
\frac{d^2\xi^1}{c^2dt^2}&=& -\frac{1}{2} k_0^2 \frac{1}{L} 
\left[\frac{1}{2} \left(\hat{T}^{11}- \hat{T}^{22}\right)\xi^1 -
\frac{1}{6} \hat{T} \xi^1 + \hat{T}^{12} \xi^2\right] e^{i  k_0 x^0},
\nonumber \\
\frac{d^2\xi^2}{c^2dt^2}&=& -\frac{1}{2} k_0^2 \frac{1}{L} 
\left[\hat{T}^{12} \xi^1 -\frac{1}{2} \left(\hat{T}^{11}- 
\hat{T}^{22}\right)\xi^2 - \frac{1}{6} \hat{T} \xi^2\right] e^{i  k_0 x^0},
\nonumber \\
\frac{d^2\xi^3}{c^2dt^2}&=& 0.
\ena

Equations (\ref{defTfp}) of the Fierz-Pauli theory should be compared with 
the equations (\ref{defT}) of GR. We see that the observational manifestations 
of gravitational waves differ from those in GR even in the limit 
of $\alpha^2 \rightarrow 0$. [Certain observational restrictions on
gravitational waves propagating with the speed different from $c$ have been   
discussed previously \cite{will}, \cite{finn}.] The deformation pattern of test 
particles acquires the additional ``common mode" motion with the amplitude 
proportional to $\hat{T}/6$. This centrally-symmetric motion can be associated
with the survived $helicity-0$ polarisation state of the $spin-2$ graviton.
For typical astrophysical sources, this extra component of motion is not smaller 
than the GR components. This means that the gravitational wave source should be 
emitting, at least, a factor of 2 different amount of energy, as compared with GR. 
The future gravitational wave observations will be capable of putting direct 
experimental limits on the presence of the ``common mode" component. However, 
the existing observations of binary pulsars are already sufficient to reject 
this particular modification of GR. Indeed, it is known that the gravitational 
wave flux from the binary pulsar PSR 1913+16 cannot deviate from the GR prediction 
by a 1 percent \cite{taylor}, let alone to be a factor of 2 different. The 
important lesson to be learnt from this study is the manner in which 
gravitational-wave considerations constrain the possible massive theories. 
The decisive factor is the potentially large difference in the radiation process 
itself, and not the tiny discrepancies, altogether vanishing in the limit 
$\alpha^2 \rightarrow 0$, in the propagation speeds of gravitational waves. 
 

\section{Black holes} \label{BH} 

The main result of this Section is the astonishing replacement of the
Schwarzschild solution by a solution without an event horizon. Below, we 
derive and explain this non-linear solution of the massive gravity. 
However, we begin with the linear approximation to the 
problem of static spherically-symmetric gravitational field, both, in GR 
and in the massive gravity. We summarise the linear approximation in
Appendix D. There, we show that the conclusions of both theories are
practically identical at the intermediate distances from the central source.
We later use this linearised approximation as the starting point for the
numerical and analytical non-linear treatment.

Since the gravitational constant $G$ enters the equations only in the product 
with the mass $M$ of the central source, we write $M$ instead of $GM$, 
effectively putting $G=1$. We also put $c=1$. The products $\alpha M$ and
$\beta M$ are dimensionless.

\subsection{The general non-linear equations}
\label{bhc}

A static spherically-symmetric gravitational field depends only on 
$r =  \sqrt{x^2 +y^2 +z^2}$, and therefore it is convenient to use spherical 
coordinates (\ref{sph}) and the metric tensor 
\be
\label{sphm}
\gamma_{00}=1,\; \;\gamma_{11}=-1,\; \;  \gamma_{22}=-r^2,\; 
\gamma_{33}=-r^2 \sin^2\theta. 
\en
The non-zero components of the gravitational field $h^{\mu\nu}$ can be written as
$$
h^{00}=A(r), \;\;\; h^{11}=-B(r), \;\;\; h^{22}=-D(r), \;\;\; 
h^{33}=- \frac{D(r)}{\sin^2\theta},
$$
where three functions, $A(r), ~B(r), ~D(r)$, should be found from the 
field equations. Since we try to use as many results as possible from the text-book 
calculations, mostly performed in the geometrical language, we introduce  
three new functions, $f(r), ~f_1(r), ~R(r)$, according to the relationships 
$$
A= \left(\frac{R}{r}\right)^2 \sqrt{\frac{f_1}{f}} - 1,\;\;\;
B= \left(\frac{R}{r}\right)^2 \sqrt{\frac{f}{f_1}} - 1,\;\;\;
D= \frac1{r^2}\left(\sqrt{ff_1}-1\right).
$$
The rationale behind this notation is the following one. The tensor $g^{\mu\nu}$, 
calculable from Eq.~(\ref{g}), and the inverse tensor $g_{\mu\nu}$, calculable
from $g^{\mu\nu}$, will now be described by simple and familiar expressions:
$$
g^{00}=\frac1{f}, \; \; \; g^{11}=-\frac1{f_1}, \; \;\;
g^{22}=-\frac1{R^2}, \; \;\; g^{33}=-\frac1{R^2\sin^{2}\theta}
$$
and
\be
\label{gmnr}
g_{00}=f, \; \; \; g_{11}=-f_1, \; \;\; g_{22}=-R^2, \; \;\; 
g_{33}=-R^2\sin^{2}\theta. 
\en
Indeed, one can check these relationships by using the definition of $g_{\mu\nu}$, 
which follows from Eq.~(\ref{g}):
\bea
\label{gspher}
g_{00}&=& (1-r^2h^{22})\sqrt{\frac{1-h^{11}}{1+h^{00}}} = f(r), \;\;\;\;
g_{11}= -(1-r^2h^{22})\sqrt{\frac{1+h^{00}}{1-h^{11}}} = -f_1(r),\nonumber\\
g_{22}&=&\frac 1{\sin^2\theta}g_{33}= -r^2\sqrt{(1+h^{00})(1-h^{11})} =-R^2(r).
\ena

Taking into account the notations (\ref{gmnr}), one can calculate the
non-zero components of the Einstein tensor $G^{\mu}_{\ \nu}$:  
\bea
G^{0}_{\ 0}&=& \frac1{f_1} \left[ -2\frac{\ddot R}{R} -
\left(\frac{\dot R}{R}\right)^2 + \frac{\dot R}{R}\frac{\dot f_{1}}{f_1}+
\frac{f_1}{R^2}\right], \nonumber \\ 
G^{1}_{\ 1}&=& \frac1{f_1} \left[-\left(\frac{\dot R}{R}\right)^2 -
\frac{\dot R}{R}\frac{\dot f}{f} + \frac{f_1}{R^2}\right], \nonumber \\
G^{2}_{\ 2} & = & G^{3}_{\ 3} = \frac1{2f_1} \left[
- \frac{\ddot f}{f} - 2\frac{\ddot R}{R} +
\frac1{2} \left(\frac{\dot f}{f}\right)^2 + \frac1{2}
\frac{\dot f}{f}\frac{\dot f_{1}}{f_1} -
\frac{\dot R}{R}\frac{\dot f}{f} +
\frac{\dot R}{R}\frac{\dot f_{1}}{f_1} \right], \nonumber
\ena
where an over-dot denotes the derivative with respect to $r$.

In the massive theory, we will be using the quasi-geometric 
equations (\ref{psgf}). For static spherically-symmetric fields,
we have only three independent equations: 
\bea
G^{0}_{\ 0}+M^{0}_{\ 0}=0,\label{BH1s}\\
G^{1}_{\ 1}+M^{1}_{\ 1}=0,\label{BH2s}\\
G^{2}_{\ 2}+M^{2}_{\ 2}=0.\label{BH3s}
\ena
The consequence of these equations, Eq.~(\ref{MBian}), can be written 
in the form of Eq.~(\ref{fbian}), which amounts to a single equation 
\be 
{\mM}^{rr}_{\ \ ;r}=0. \label{BHsys}
\en
Before proceeding to the massive theory, it is instructive to review the 
derivation of the Schwarzschild solution in GR.   

In GR, one puts $M^{\mu}_{\ \nu}= 0$ and solves the massless field equations 
$G^{\mu}_{\ \nu}=0$. Then, equation $G^{2}_{\ 2}=0$ is not independent. Due to 
Bianchi identities, this equation can be obtained as the combination of equations 
$G^{0}_{\ 0}=0$ and $G^{1}_{\ 1}=0$. (From this point of view, Eq.~(\ref{BHsys})
of the massive theory is the ``extra" equation, non-existent in GR.) One is 
left with two independent equations for three unknown functions of $r$: 
$f,~f_1,~R$. From equations $G^{0}_{\ 0}=0$ and $G^{1}_{\ 1}=0$
one finds $f$ and $f_1$ in terms of $R$:
\bea
f = a\left(1 - \frac{R_g}{R}\right), \;\;
f_1 = {\dot R}^2 \left(1 - \frac{R_g}{R}\right)^{-1},\label{schw}
\ena
where $a$ and $R_g$ are constants of integration, while $R$ remains an arbitrary 
function of $r$. The effective line-element takes the form
\bea
\label{Schw}
{\rm d}s^2&=&  a\left(1 - \frac{R_g}{R}\right){\rm d}t^2 - \frac{\dot R^2}
{\left(1 - \frac{R_g}{R}\right)}{\rm d}r^2 - R^2 {\rm d}\Omega^2 \nonumber \\
&=& \left(1 - \frac{2M}{R}\right){\rm d}t^2 -
\frac1{\left(1 - \frac{2M}{R}\right)}{\rm d}R^2 - R^2{\rm d}\Omega^2,
\ena
which is the familiar Schwarzschild solution. The function $R(r)$ has been 
announced an independent coordinate variable $R$, whereas the constants $a=1$ 
and $R_g=2M$ have been found from comparison of (\ref{Schw}) with 
the Newtonian gravity at $R\rightarrow \infty$.

In contrast to GR, in the massive theory, there is no functions of $r$ left 
arbitrary. All three functions of $r$: $f,~f_1,~R$, are determined by 
three independent equations: 
(\ref{BH1s}), (\ref{BH2s}), (\ref{BH3s}). The mass contributions  
$M^{\mu}_{\ \nu}$ are supposed to be calculated in terms of the 
functions $f,~f_1,~R$ and the metric tensor (\ref{sphm}). In order to 
facilitate the comparison of the finite-range solution with the Schwarzschild 
solution, it is convenient to
re-define the field variables and the metric tensor. First, we invert the
function $R=R(r)$ to $r=r(R)$ and denote $r(R) \equiv X(R)$. Second,
we introduce $F(R)$ according to the definition
\bea
F\equiv \frac{(\dot R)^2}{f_1}= \frac 1{{X'}^2 f_1},\nonumber
\ena
where a prime denotes the derivative with respect to $R$:  $X'=dX/dR$.
Then, the tensor $g_{\mu\nu}$ reads 
\be
\label{gnew}
g_{00}=f, \;\;\;\;\; g_{11}=-\frac 1{{X'}^2 F}, \;\;\;\;\;
g_{22}=-R^2, \;\;\;\;\;   g_{33}=-R^2 sin^2\theta~, 
\en
so that the effective line-element 
$$
{\rm d}s^2=f(R){\rm d}t^2 - \frac1{F(R)}{\rm d}R^2 -R^2 {\rm d}\Omega^2 
$$
takes the general structure of the Schwarzschild line-element (\ref{Schw}).  
In GR, the functions $f(R), ~F(R)$ are given by
\bea
f(R) =F(R) = 1-\frac{2M}{R},
\label{mlsolx}
\ena
whereas they are expected to be given by some other formulas in the finite-range 
gravity. As for the metric components (\ref{sphm}), they transform into functions 
of $R$: 
\be
\label{gammanew}
{\rm d}\sigma^2= {\rm d}t^2 -{X'}^2{\rm d}R^2- X^2{\rm d}\Omega^2.
\en
The field equations, including the mass contributions, will now be built from
the quantities entering Eqs. (\ref{gnew}), (\ref{gammanew}). From the 
field-theoretical viewpoint, we have simply performed a coordinate 
transformation $r=r(R)$ of the radial coordinate, and have applied 
this coordinate transformation to the metric tensor 
and to the gravitational field components. Obviously, the field equations 
derived from Eqs. (\ref{sphm}), (\ref{gmnr}) in terms of $f(r),~ f_{1}(r), 
~R(r)$, and the field equations derived from Eqs. (\ref{gammanew}),
(\ref{gnew}) in terms of $f(R), ~F(R), ~X(R)$, are exactly the same 
equations, if one takes into account the corresponding change of notations. 

We will now write down explicitly the exact non-linear equations. 
In doing that, we use the notation $\zeta$ for the mass ratio:
\be
\label{zeta}
\zeta = \frac{\beta^2}{\alpha^2} = \frac{{m_0}^2}{{m_2}^2} .
\en
The first two equations (\ref{BH1s}), (\ref{BH2s}) take the form 
\bea
-F\left[\frac 1{R^2}+ \frac{F'}{F}\frac1{R}\right] +\frac 1{R^2} =
-M^{0}_{\ 0}, \label{bhsx1}\\
-F\left[\frac 1{R^2}+ \frac{f'}{f}\frac1{R}\right] +\frac 1{R^2}
= -M^{1}_{\  1}, \label{bhsx2}
\ena
where
\bea
M^0_{\ 0} &=& \frac{\alpha^2}{2(\zeta+2)}\sqrt{X'^2\frac{F}{f}}\left[
\frac 3{4}\left(\frac{R}{X}\right)^2 \left(\frac 1{X'^2Ff} - X'^2Ff\right)
- \right. \nonumber \\
& & \left.(1-\zeta)\left(\frac{X}{R}\right)^2 \frac{f}{X'^2F} +(2\zeta +1)f\right]
- \nonumber \\
& &\frac{3\beta^2}{2(\zeta+2)}\left[\frac 1{2}\left(X'^2F-\frac 1{f}\right) +
\left(\frac{X}{R}\right)^2\right],\\
M^1_{\ 1} &=& \frac{\alpha^2}{2(\zeta+2)}\sqrt{X'^2\frac{F}{f}}\left[
-\frac 3{4}\left(\frac{R}{X}\right)^2 \left(\frac 1{X'^2Ff} - X'^2Ff\right)
- \right.\nonumber \\
& & \left.(1-\zeta)\left(\frac{X}{R}\right)^2 \frac{f}{X'^2F} +
 (2\zeta +1)\frac 1{X'^2F}\right] + \nonumber \\
& &\frac{3\beta^2}{2(\zeta+2)}\left[\frac 1{2}\left(X'^2F-\frac 1{f}\right) -
\left(\frac{X}{R}\right)^2\right]. \nonumber
\ena
Obviously, in the massless GR, that is, for $M^0_{\ 0} =M^1_{\ 1} =0$, the exact
solution to these equations is the familiar formula (\ref{mlsolx}).

We now turn to the third equation (\ref{BH3s}). It proves more
illuminating to use Eq.~(\ref{BHsys}) instead of Eq.~(\ref{BH3s}).
The reason being that Eq.~(\ref{BHsys}) gives directly the ``extra"
equation, which is absent in GR. Explicitly, Eq.~(\ref{BHsys}) has the form 
\be
\frac{\alpha^2}{2\zeta+4}\frac 1{X'}\left(\frac{R}{X}\right)^2
\left(-2\frac{X''}{X'}c_1 +\frac{X'}{X}c_2 +
\frac{f'}{f}c_0 -\frac{F'}{F}c_1 +
\frac{c_R}{R}\right)=0, \label{bhsx3}
\en
where
\bea
c_R &=&3\left(\frac{R}{X}\right)^2\left[-3X'^2Ff + \frac 1{X'^2Ff}\right] +
2(2\zeta+1)\left[\left(\frac{R}{X}\right)^2 -\frac 1{X'^2F} + f\right] - \nonumber \\
& &12\zeta\sqrt{X'^2Ff}, \nonumber\\
c_0&=&-\frac 3{4}\left(\frac{R}{X}\right)^2\left[3X'^2Ff + \frac 1{X'^2Ff}\right] +
(1-\zeta)\left(\frac{X}{R}\right)^2 \frac {f}{X'^2F} + \nonumber \\
& &(2\zeta+1)f - 3\zeta\sqrt{X'^2Ff}, \nonumber\\
c_1&=&\frac 3{4}\left(\frac{R}{X}\right)^2\left[3X'^2Ff + \frac 1{X'^2Ff}\right] +
(1-\zeta)\left(\frac{X}{R}\right)^2 \frac {f}{X'^2F} - \nonumber \\
& & (2\zeta+1)\frac 1{X'^2F} + 3\zeta\sqrt{X'^2Ff}, \nonumber\\
c_2 &=& 3\left(\frac{R}{X}\right)^2\left[X'^2Ff - \frac 1{X'^2Ff}\right] +
4(1-\zeta)\left(\frac{X}{R}\right)^2 \frac {f}{X'^2F} + \nonumber \\
& &12\zeta\left(\frac{X}{R}\right)^2\sqrt{\frac{f}{X'^2F}}. \nonumber
\ena
Since we assume that $\alpha \neq 0$, the common factor in Eq.~(\ref{bhsx3}) can
be ignored, so that Eq.~(\ref{bhsx3}) simplifies to
\bea
-2\frac{X''}{X'}c_1 +\frac{X'}{X}c_2 + \frac{f'}{f}c_0 -\frac{F'}{F}c_1 +
\frac{c_R}{R}=0 \label{bhsx3'}.
\ena
To double-check our analytical calculations, we have verified that a 
direct consequence of equations (\ref{BH1s}), (\ref{BH2s}),  
(\ref{BH3s}), is indeed Eq.~(\ref{bhsx3'}), as it should be.    
Thus, our final goal is to find three functions $f(R),~F(R),~X(R)$ from
three equations (\ref{bhsx1}), (\ref{bhsx2}), (\ref{bhsx3'}). 

The parameters $\alpha$ and $\beta$ enter Eq.~(\ref{bhsx3'}) only through the 
ratio (\ref{zeta}). One convenient choice of $\zeta$ is $\zeta = 1$, i.e.
$\beta^2 = \alpha^2$ and, equivalently, ${m_0}^2 = {m_2}^2$. This choice of 
parameters has attracted some interest in the literature, because, in this
case, Eqs. (\ref{system2.0}), which constitute the linear version of 
the ``extra" equations (\ref{fbian}), take the form of 
${h^{\mu\nu}}_{,\nu} =0$. In terms of $g^{\mu\nu}$, these 
last equations read $(\sqrt{-g} g^{\mu\nu})_{\;,\nu} =0$. In GR, these
equations define the set of harmonic coordinate systems, so successfully 
used by Fock \cite{Fock}. Our attention to this choice of $\zeta$ is guided 
simply by a technical simplification of exact equations that we want to 
solve. It is clear from the structure of equations, and some specific 
analytical and numerical evaluations performed for the cases $\zeta \ne 1$,
that the choice of $\zeta =1$ does not incur any loss of generality to our 
conclusions. The most of our analytical and numerical calculations will deal 
with the case $\zeta =1$. (The linear massive theory with $\zeta = 1$, and 
some non-linear theories with ``subsidiary" conditions, have been considered 
in a number of papers \cite{OP}, \cite{Freund}, \cite{Visser}, \cite{Logunov}.) 
When performing numerical calculations, we have used 
the D02CBF--NAG Fortran Library Routine which integrates ordinary
differential equations from $R_{in}$ to $R_{end}$ using a variable-step Adams
method.

\subsection{Weak-field approximation in the case $\zeta=1$}
\label{bhd}

It is convenient to begin with the intermediate distances from the central
source, where the behaviour of the sought-for solution is known from the linear 
theory (see Appendix~\ref{bhb}). When $\alpha = \beta$, the function $\Psi$, 
Eq.~(\ref{Psi}), vanishes and Eqs. (\ref{sfh00}), (\ref{sfh11}), (\ref{sfh22}) 
simplify to 
$$
h^{00}= \frac{4M}{r}+O(M\alpha), \;\;\; h^{11}=0, \;\;\;
h^{22}=\sin^2 \theta h^{33}= 0.
$$
Using Eqs. (\ref{gspher}), we can find, first, the functions $f(r),~f_1(r),~R(r)$:
\bea
f\approx 1-\frac{2M}{r}, \;\; f_1\approx 1+\frac{2M}{r}, \;\;
R \approx r +M, \nonumber
\ena
and, then, the functions $f(R),~F(R),~X(R)$:  
\bea
F\approx f\approx 1-\frac{2M}{R} ,\;\;\; X\approx R - M\label.\label{z1indat}
\ena
Thus, in the intermediate region, i.e. for $R$ satisfying the inequalities 
\be
\label{interz}
1\ll\frac{R}{M}\ll\frac 1{\alpha M},  
\en
the behaviour of our non-linear solution is given by Eq.~(\ref{z1indat}). 
Certainly, the exact solution of GR, Eq.~(\ref{mlsolx}), subject to the 
transformation to harmonic coordinates $X= R-M$, is also described by 
formulas (\ref{z1indat}), but with the symbols 
of approximate equality being replaced by the symbols of equality.
In harmonic coordinates, the Schwarzschild solution takes the form:
\be
\label{Schwh}
{\rm d}s^2= \frac{X-M}{X+M} {\rm d}t^2 - \frac{X+M}{X-M} {\rm d}X^2 - 
(X+M)^2{\rm d}\Omega^2. \nonumber 
\en
The approximate solution (\ref{z1indat}) helps us to formulate the initial 
conditions for numerical integration of equations (\ref{bhsx1}), 
(\ref{bhsx2}), (\ref{bhsx3'}). As mentioned above, we reduce 
these equations to the case $\zeta =1$. For purely technical reasons 
of computational error and resolution, we begin with the unrealistically
large value of the dimensionless parameter
$\alpha M$: $\alpha M = \sqrt{2}\times 10^{-6}$. Later on we will discuss 
the variations 
of this parameter. The starting point of integration is $R_{in}=5\times 10^3M$, 
so that the inequality (\ref{interz}) is satisfied there pretty well. The 
initial 
values of the participating functions at $R_{in}$ are given by 
\bea
F=f=1-\frac{2M}{R_{in}},\;\;\;
X=R_{in}-M,\;\;\;
X'=1.\label{indata}
\ena
The initial value of $X'$ needs to be specified as well, since the equation 
for $X(R)$ (\ref{bhsx3'}) is a second-order differential equation.

\begin{figure}[tbh]
\vspace*{13pt}
\centerline{\psfig{file=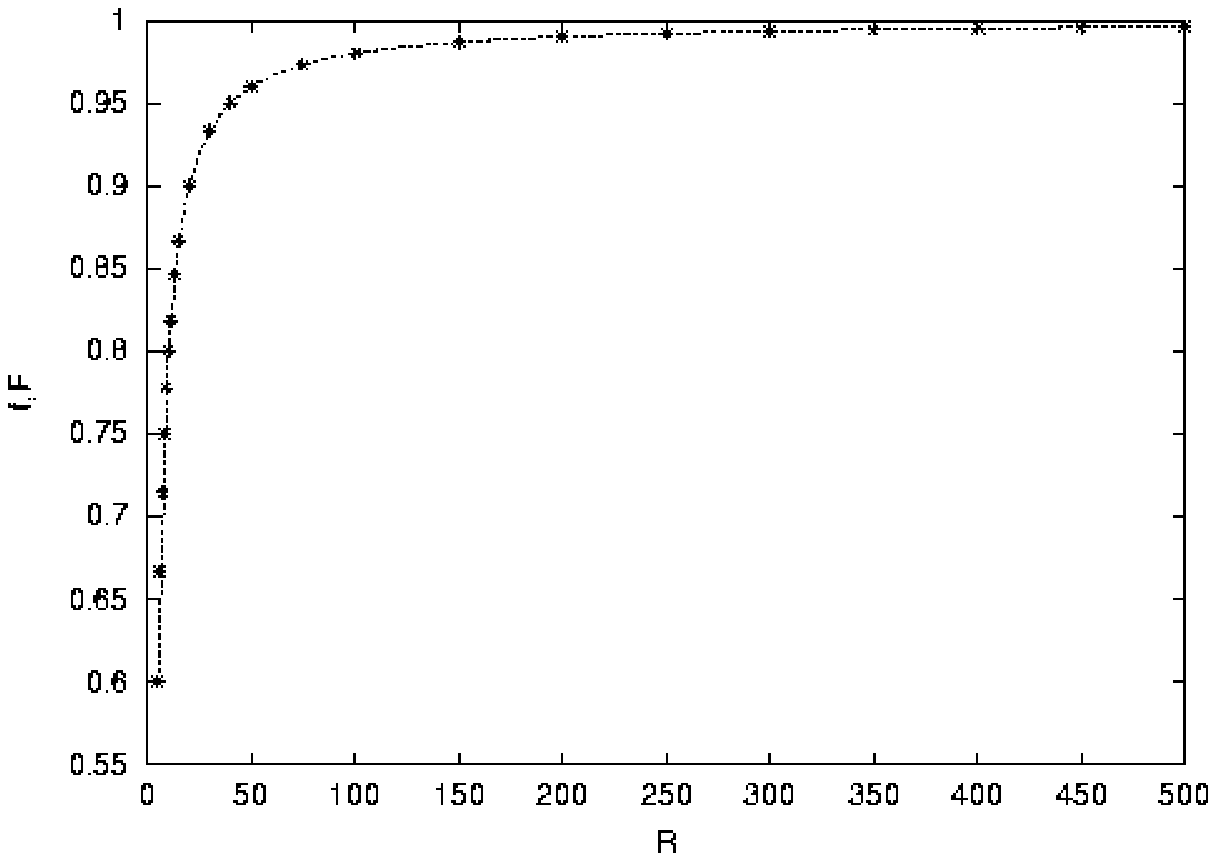}}
\vspace*{13pt}
\fcaption{Numerical solutions in the region $5M\le R\le 500M$.
The dashed line is the Schwarzschild solution. The values of $f$ and $F$ in 
the massive gravity are shown, respectively, by $+$ and and $\times$ marks, 
which almost superimpose on each other.} 
\label{z1r15.5e3foFml}
\end{figure}

As expected, at intermediate distances, the solution of the finite-range 
gravity is practically indistinguishable from the Schwarzschild solution.
In Fig.~\ref{z1r15.5e3foFml}, we show the values of $f$ and $F$ numerically 
calculated at discrete values of $R$ ($R$ is given in units of $M$). For 
comparison, the dashed line shows the Schwarzschild functions $f= F = 1-2M/R$.

\begin{figure}[tbh]
\vspace*{13pt}
\centerline{\psfig{file=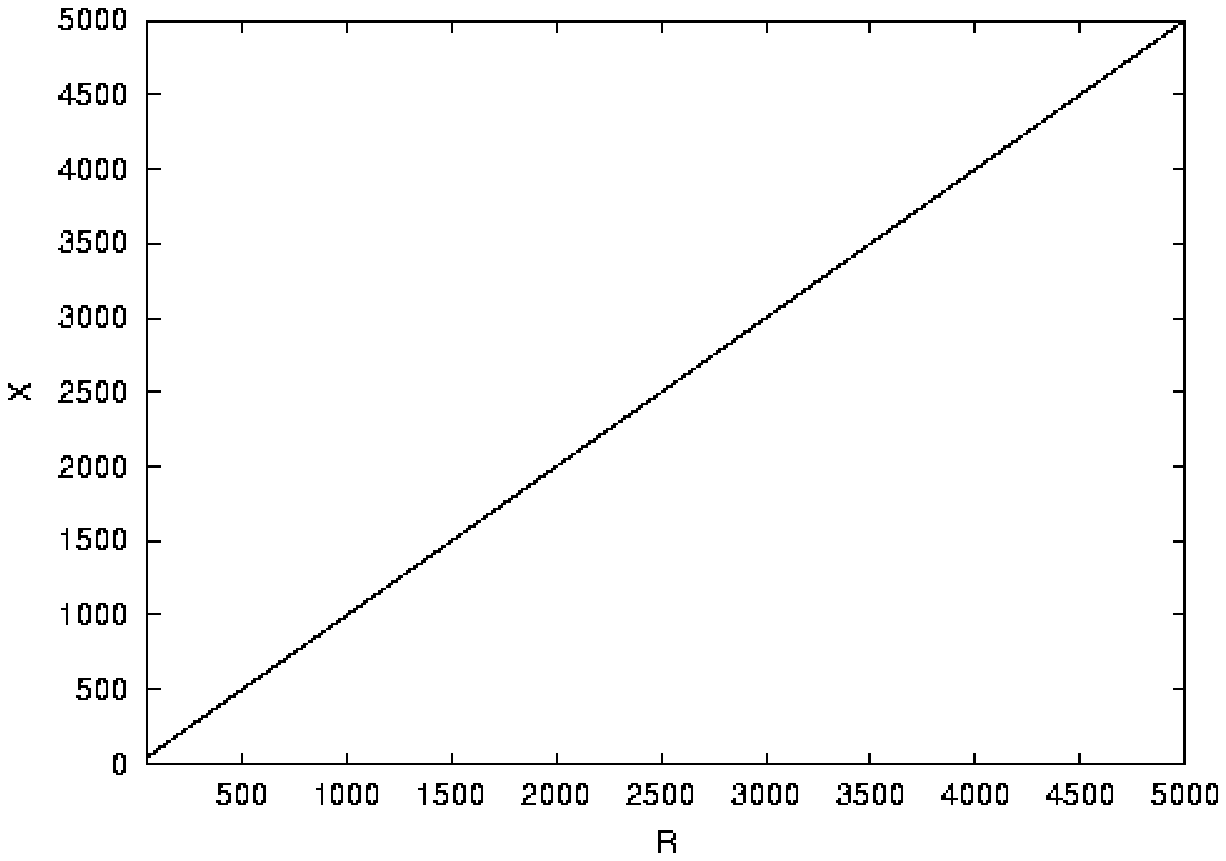}}
\vspace*{13pt}
\fcaption{Numerical solution for the function $X(R)$
in the region $50M\le R\le 5\times 10^3M$. }
\label{z1r15.5e3X}
\end{figure}

The equally good agreement takes place between the numerically calculated 
function $X(R)$ of the massive gravity and the function $X= R-M$, 
which is a solution of the harmonic-coordinate conditions of GR. One can 
see in Fig.~\ref{z1r15.5e3X} that the function $X(R)$ (in units of $M$) 
is indistinguishable from a straight line $X=R-M$ for all covered values
of $R$. 

Certainly, the displayed numerical graphs are in agreement with analytical 
calculations. In the intermediate zone, equation (\ref{bhsx3'}) is well 
approximated by its linear part:  
\bea
-2\frac{X''}{X'}+4\frac{X}{X'}\frac{1}{R^2F}-\frac{f'}{f}-
\frac{F'}{F}-\frac{4}{R}=0.\label{bhlinreg}
\ena
[It is not easy to recognise equation (\ref{bhlinreg}) as a linear part of 
Eq.~(\ref{bhsx3'}), but a straightforward way to verify 
Eq.~(\ref{bhlinreg}) is to use the fact that this equation 
is Eq.~(\ref{dDL}): ${h^{\mu\nu}}_{;\nu}=0$.]
In the intermediate zone, equations (\ref{bhsx1}), (\ref{bhsx2})
can also be approximated by simpler equations, as the terms $M^{0}_{\ 0}$, 
$M^{1}_{\  1}$ can be neglected. This leads to the approximate solution of
these equations: $f = F = 1-2M/R$. Using these expressions for $f$ and $F$
in Eq.~(\ref{bhlinreg}), one obtains a second-order differential equation
for $X(R)$:
\[
X^{''}+ \frac{2 X^{'}(R-M)}{R(R-2M)} -\frac{2X}{R(R-2M)} =0.
\]
The general solution to this equation is given by 
\bea 
X(R) =a_1(R-M)+a_2\left[\frac{R-M}{2}\ln |1-\frac{2M}{R}| +M \right], 
\label{Xgr} 
\ena
where $a_1$ and $a_2$ are arbitrary constants. Solution (\ref{Xgr}) was 
first derived by Fock \cite{Fock} in his study of harmonic coordinates for the 
Schwarzschild metric. With this approximate solution for $f, ~F, ~X$, 
one can verify that the neglected terms in equations (\ref{bhsx1}), (\ref{bhsx2}), 
(\ref{bhsx3'}) are indeed smaller than the retained ones.

\begin{figure}[tbh]
\vspace*{13pt}
\centerline{\psfig{file=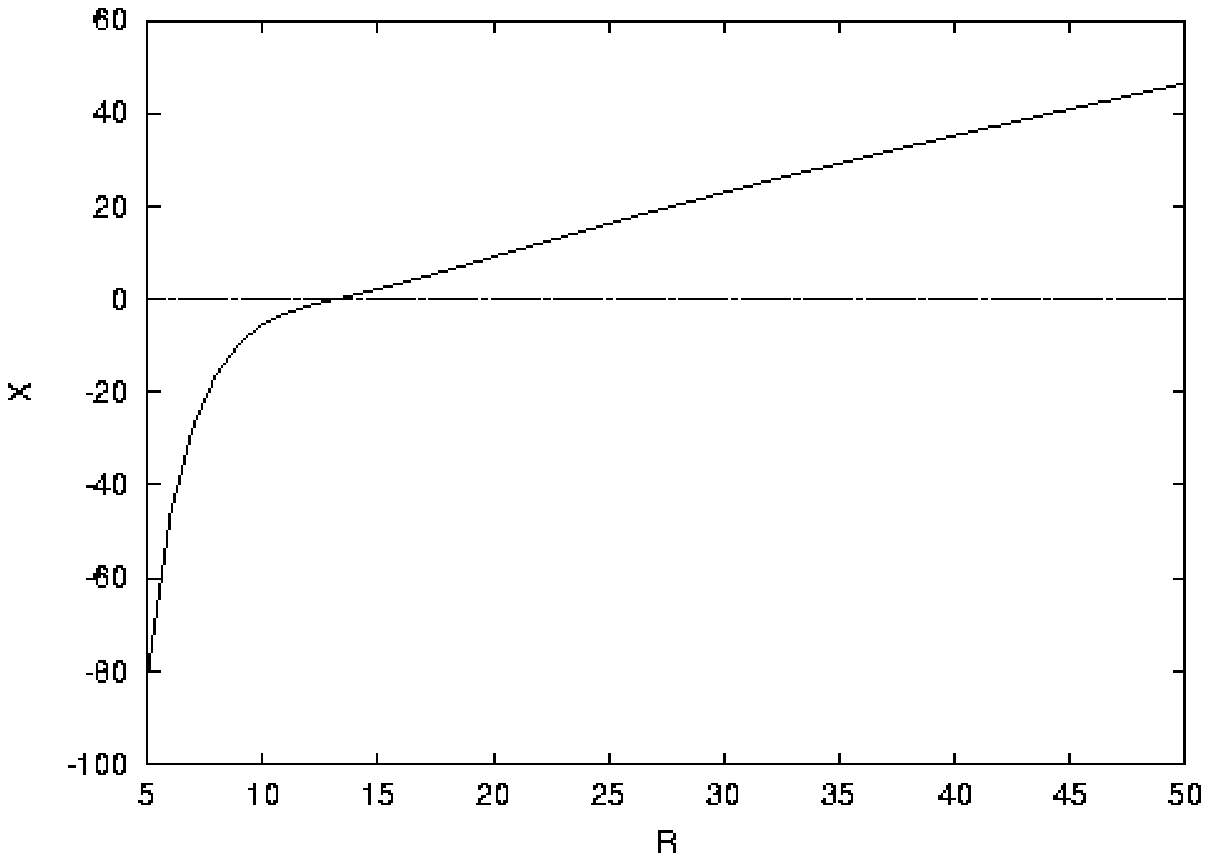}}
\vspace*{13pt}
\fcaption{Continuation of the numerical solution for $X(R)$ to the 
region $5M\le R\le 50M$.}
\label{z1r5.50X}
\end{figure}

In general relativity, it is sufficient to use only one branch of the
solution (\ref{Xgr}), choosing $a_1=1, ~a_2=0$ \cite{Fock}. Our initial
conditions (\ref{indata}) do also imply $a_1=1, ~a_2=0$ in Eq.~(\ref{Xgr}). 
This is why
our numerical solution of exact equations (\ref{bhsx1}), (\ref{bhsx2}), 
(\ref{bhsx3'}) is practically indistinguishable, everywhere in the intermediate
zone, from the Fock's exact solution: $f = F = 1-2M/R, ~X=R-M$.
However, in the massive gravity, formula (\ref{Xgr}) is only an approximate
solution. The second branch of this formula, with the logarithmically divergent 
term, suggests that the function $X(R)$ may start deviating from the straight
line at some sufficiently small $R$. This is indeed the case. In 
Fig.~\ref{z1r5.50X} we show the continuation of the numerical graph for
$X(R)$ from the region covered in Fig.~\ref{z1r15.5e3X} to the region 
$5M\le R \le 50M$.

It is seen from this graph that the function $X(R)$ crosses zero at 
some point near $R=13M$, and then sharply drops down to large negative
values. For comparison, one should recall that the GR function $X = R-M$ would have
crossed zero only at $R=M$. The sharp decrease of $X(R)$ continues at smaller $R$.
This behaviour of $X(R)$ feeds back to the equations  
(\ref{bhsx1}), (\ref{bhsx2}) and drastically changes the functions $f(R)$ and
$F(R)$. Clearly, this strong deviation from GR takes place at values of $R$ 
approaching $2M$, i.e. in the region where the intermediate zone approximation 
ceases to be valid.

\subsection{No black holes in massive gravity}
\label{bhe}

\begin{figure}[tbh]
\vspace*{13pt}
\centerline{\psfig{file=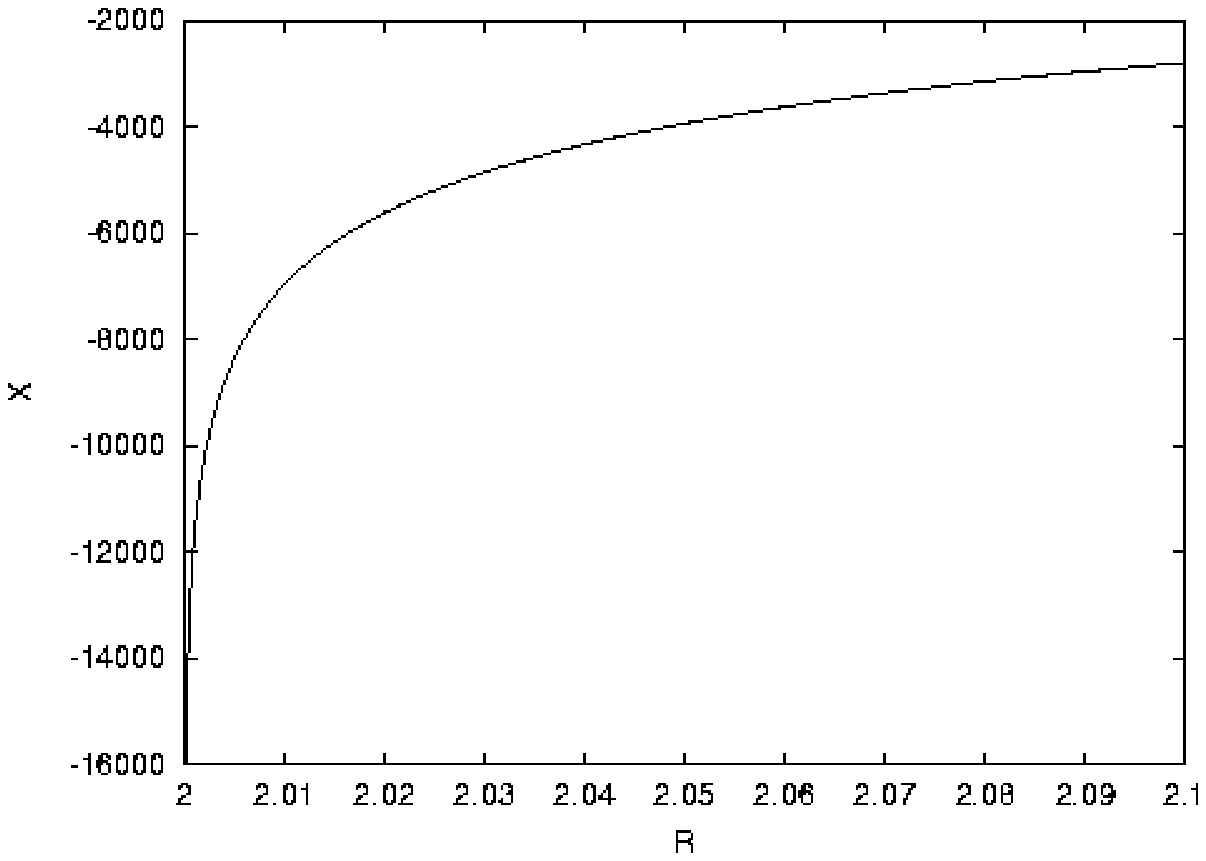}}
\vspace*{13pt}
\fcaption{Numerical function X(R) in the region $2.0001M\le R\le 2.1M$.}
\label{z1r2.5Xan}
\end{figure}

\begin{figure}[tbh]
\vspace*{13pt}
\centerline{\psfig{file=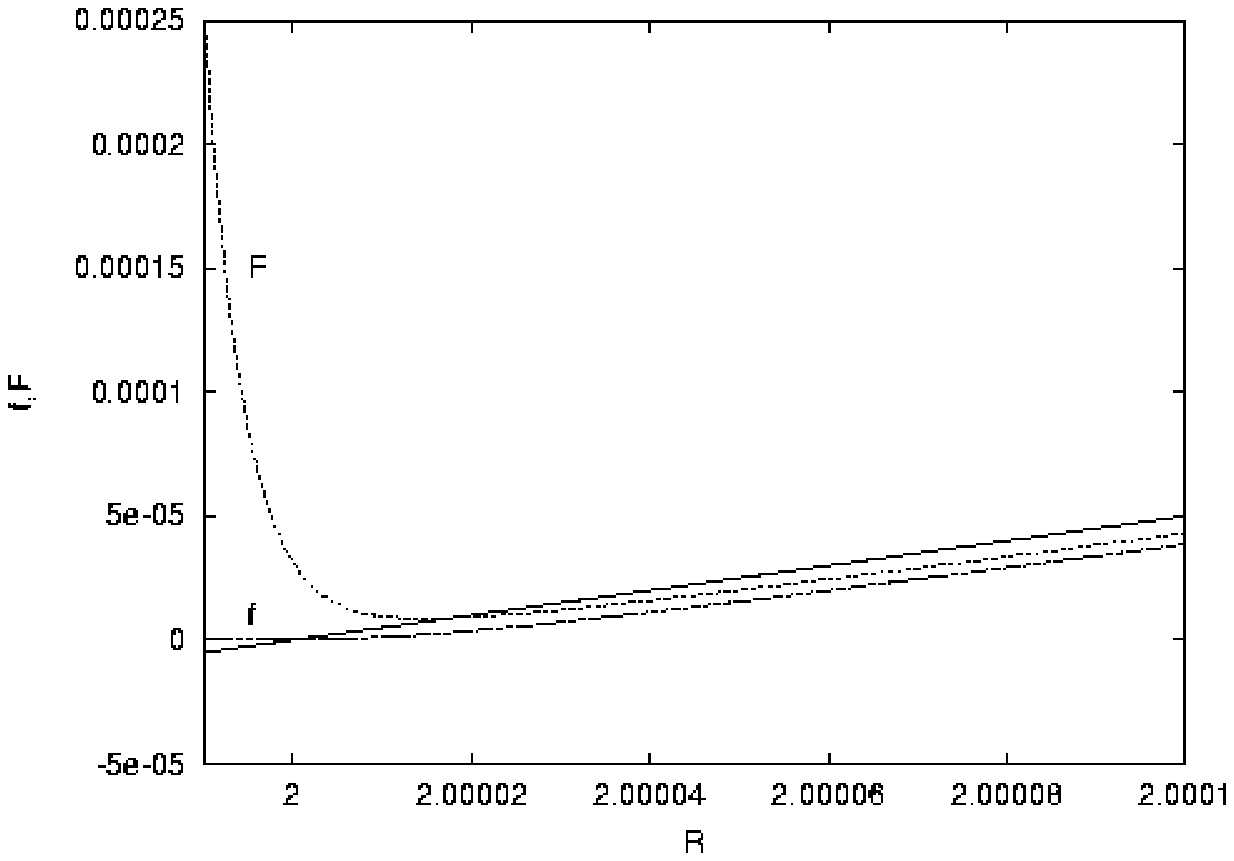}}
\vspace*{13pt}
\fcaption{The solid line is the Schwarzschild solution $f = F= 1-2M/R$. The
dashed line is the numerical solution for $f(R)$, and the dotted line is the
numerical solution for $F(R)$.}
\label{z1arounhor}
\end{figure}

In Fig.~\ref{z1r2.5Xan} we show the continuation of $X(R)$ to the interval
$2.0001M\le R\le 2.1M$.
Given the initial conditions (\ref{indata}), the GR function 
$X =R -M$ would be positive and very close to  $M$ in the interval 
of $R$ covered by Fig.~\ref{z1r2.5Xan}. But in the massive theory, $X(R)$ 
is negative and continues to sharply decline below the 
level of $-1.4 \times 10^{4} M$. Since equations (\ref{bhsx1}), 
(\ref{bhsx2}), (\ref{bhsx3'}) are coupled differential equations, it 
is natural to expect that a strong deviation from
GR of one of the functions will be accompanied by strong deviations of other 
functions. Indeed, in Fig.~\ref{z1arounhor} we show the continuation of 
numerical graphs for $f(R)$ and $F(R)$ to still smaller $R$, including the 
point $R=2M$.

\begin{figure}[tbh]
\vspace*{13pt}
\centerline{\psfig{file=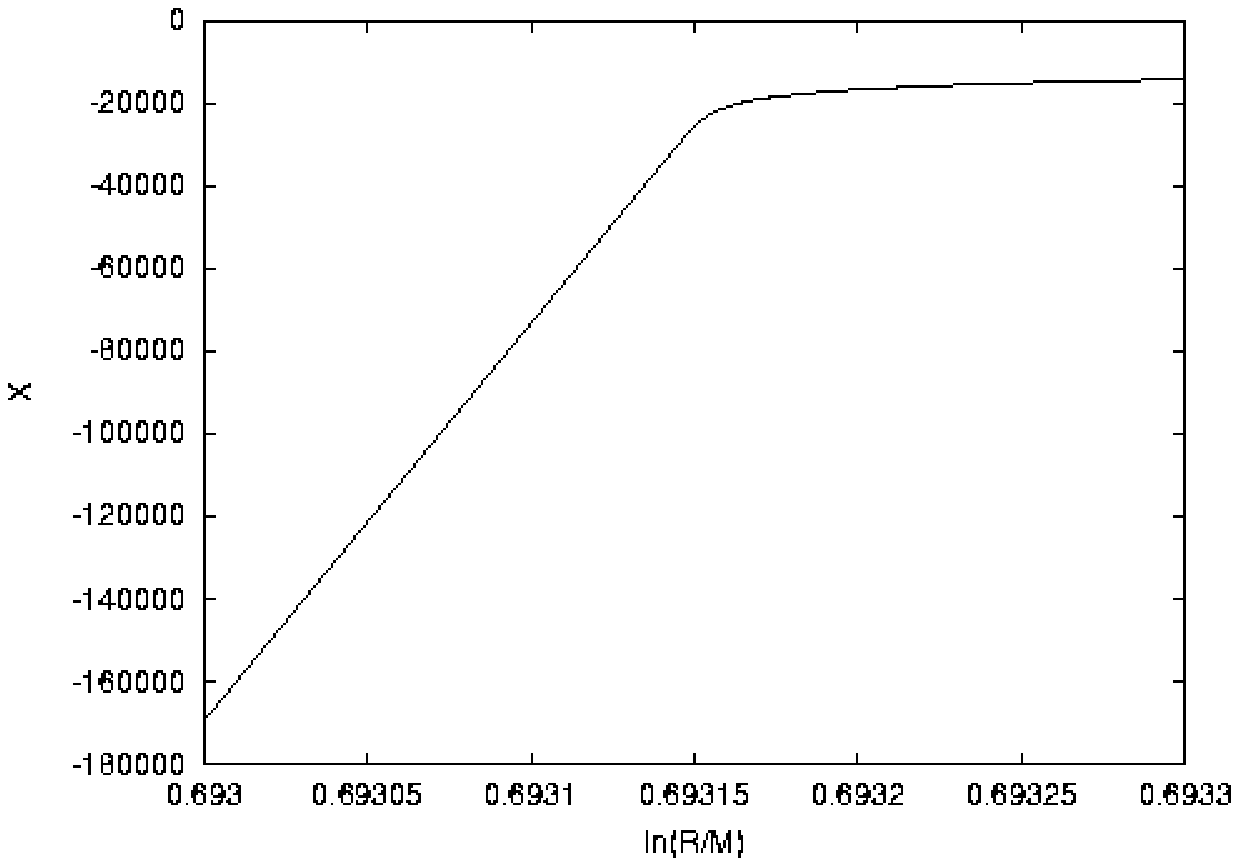}}
\vspace*{13pt}
\fcaption{The graph shows $X(R)$ versus $\ln(R/M)$   
in a very narrow region around $2M$. The point $R=2M$ corresponds to 
$ln(R/M)=0.693147$. }
\label{Xat2M}
\end{figure}

It is seen from this graph that on the way to the point $R=2M$ the function 
$F(R)$ reaches a minimum, and then starts increasing again. The function
$f(R)$ does not cross zero, and, presumably, approaches zero asymptotically,
i.e. for $R \rightarrow 0$. The continuation of $X(R)$ to the vicinity of 
$R=2M$ is shown in Fig.~\ref{Xat2M}.
It is difficult to analyse analytically the immediate 
vicinity of the point $R=2M$, but the asymptotic analytical description is 
possible for much smaller $R$, i. e. $R \ll 2M$. This description can then be 
extrapolated to $R=2M$, in pretty good agreement with the numerical analysis. 
We will now give this analytical description and will compare it with numerical 
calculations.
It is likely, and we will justify this later, that the function $X(R)$
has the general form 
\bea
X(R) = a M\; ln\left(\frac{R}{2M}\right) -b M
\label{Xanal}
\ena
at $R\ll 2M$, where $a$ and $b$ are some constants. This behaviour
is suggested by formula (\ref{Xgr}) in the limit $R \ll 2M$. If so, one can 
use Eq.~(\ref{Xanal}) for evaluation of the leading terms in $M^{0}_{\ 0}$ and 
$M^{1}_{\  1}$ in the limit of small $R$. This evaluation shows that 
\[
M^{0}_{\ 0} \approx -\frac{1}{4} \alpha^2 {X^{\prime}}^2 F, ~~~~~
M^{1}_{\  1} \approx \frac{1}{4} \alpha^2 {X^{\prime}}^2 F. 
\]
Then, equations (\ref{bhsx1}), (\ref{bhsx2}) take the 
form:

\bea
-F\left[\frac 1{R^2}+ \frac{F'}{F}\frac1{R}\right] +\frac 1{R^2} =
\nu \frac{F}{R^2}, \nonumber\\
-F\left[\frac 1{R^2}+ \frac{f'}{f}\frac1{R}\right] +\frac 1{R^2}
= -\nu \frac{F}{R^2} , \nonumber
\ena
where
\be
\label{nu}
\nu = \frac{1}{4} a^2 (\alpha M)^2.
\en
One can now find the exact solution to these approximate equations. It is 
given by
\bea
\label{FfsmallR1}
F(R) &=& C_F \left(\frac{R}{2M}\right)^{-1-\nu} + \frac{1}{1+\nu},  \\
f(R) &=& C_f \left(\frac{R}{2M}\right)^{-1 + \nu} + \frac{C_f}{(1+\nu) C_F} 
\left(\frac{R}{2M}\right)^{2 \nu}, \label{FfsmallR2}
\ena
where $C_F$ and $C_f$ are arbitrary constants. This solution allows one to identify
the leading terms in the equation (\ref{bhsx3'}) for $X$. Specifically,
the main contributions to Eq.~(\ref{bhsx3'}) are provided by the last terms in the
expressions for $c_R, ~c_0, ~c_1$. The term with $c_2$ is subdominant. The leading
terms in Eq.~(\ref{bhsx3'}) combine to produce the approximate equation   
\[
X^{\prime \prime} + \frac{1}{R} X^{\prime} =0. 
\]
We see that expression (\ref{Xanal}) is indeed a general solution to this equation.
Having found the functions $F(R), ~f(R),~X(R)$, one can now check that the 
neglected terms in all three equations (\ref{bhsx1}), (\ref{bhsx2}), (\ref{bhsx3'})  
are indeed small in comparison with the retained ones. The approximate solution
(\ref{FfsmallR1}), (\ref{FfsmallR2}), (\ref{Xanal}) is asymptotically exact in
 the limit $R \rightarrow 0$, i.e. in the vicinity of singularity, which we 
will discuss later. The arbitrary constants $C_F,~C_f,~a,~b$ can only be found 
from comparison of this analytical solution with numerical calculations. 

\begin{figure}[tbh]
\vspace*{13pt}
\centerline{\psfig{file=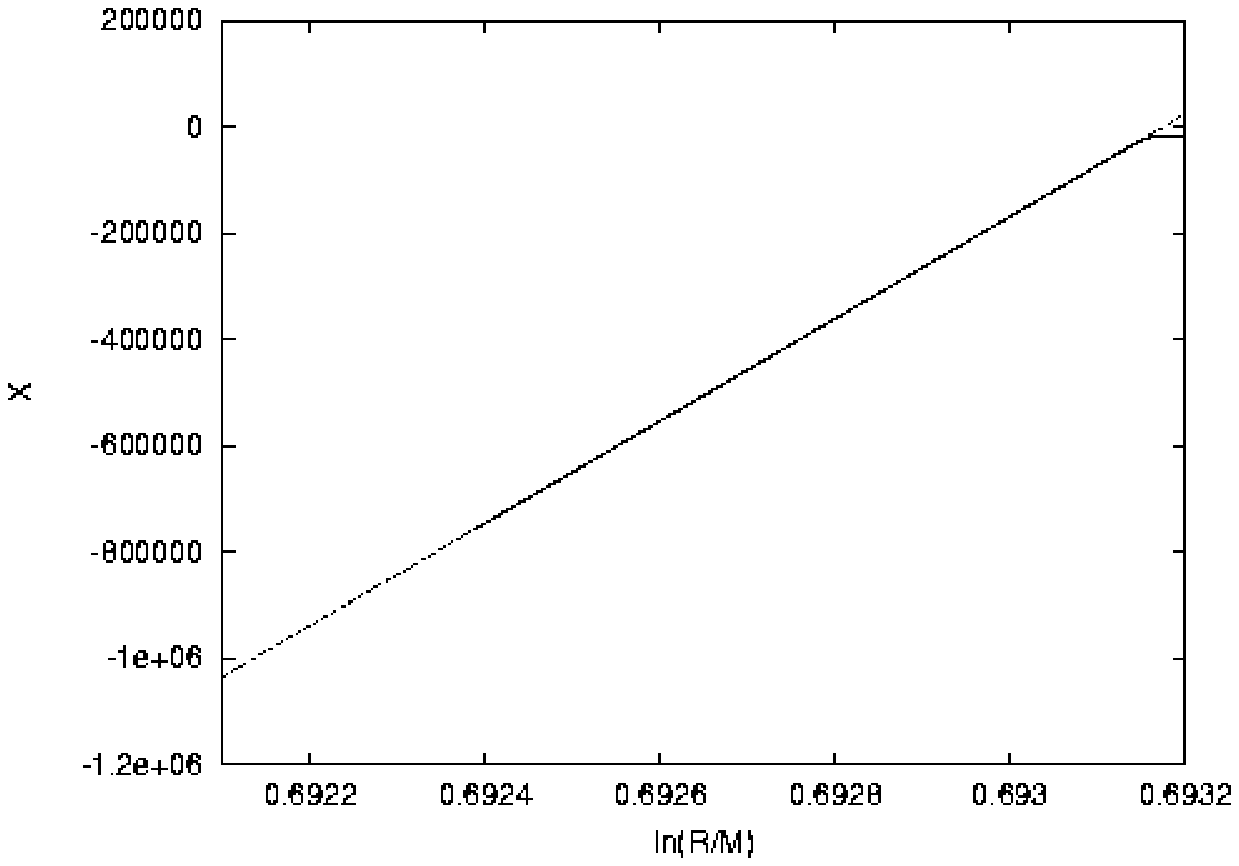}}
\vspace*{13pt}
\fcaption{The graph shows $X(R)$ versus $\ln(R/M)$ at values of $R$ somewhat smaller 
than $2M$ and including $R=2M$. The solid line is the numerical solution, while 
the dashed line is its analytical approximation (\ref{Xanal}) with $a =9.62265\times 10^8$,
$b=2.8028\times 10^4 $.}
\label{z1zagorXan}
\end{figure}

\begin{figure}[tbh]
\vspace*{13pt}
\centerline{\psfig{file=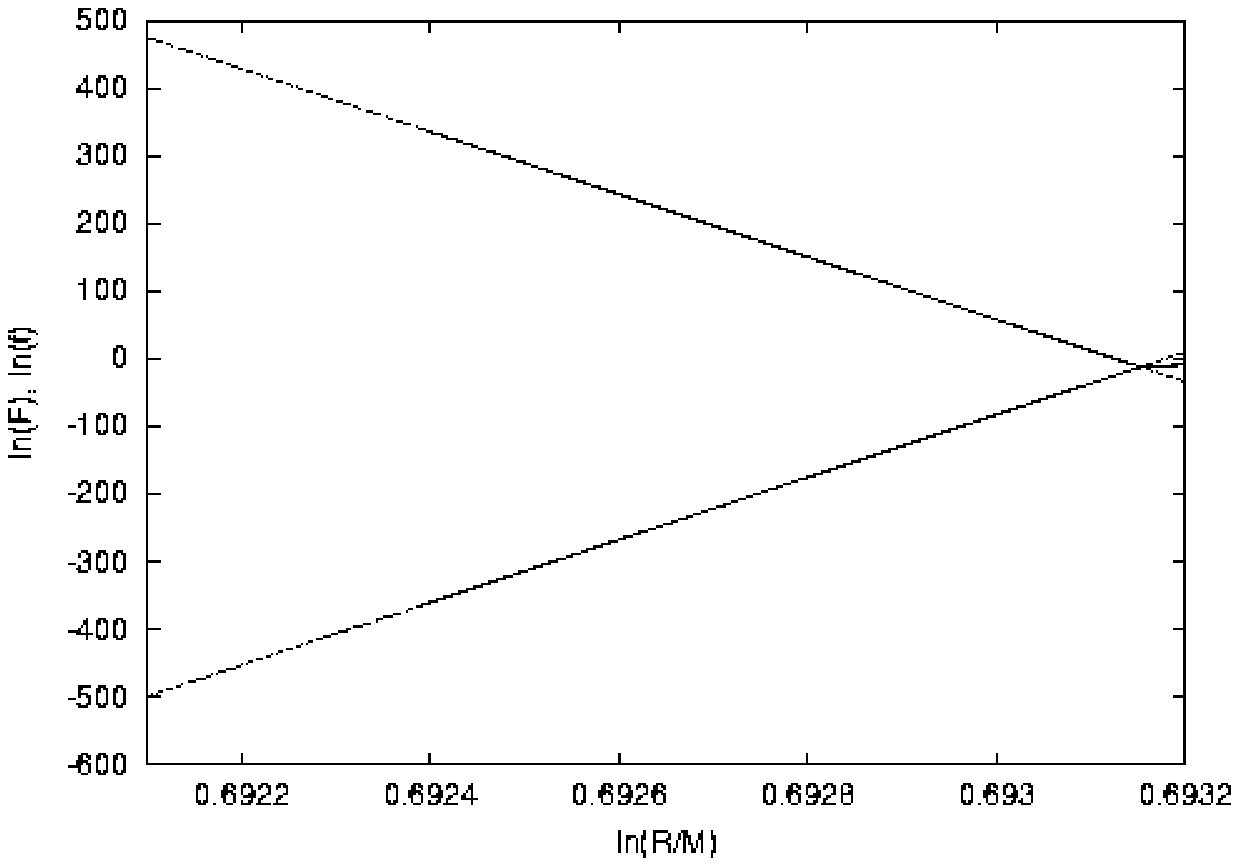}}
\vspace*{13pt}
\fcaption{The functions $\ln F(R)$ and $\ln f(R)$ are plotted versus $\ln(R/M)$.
The upper solid line is the numerical solution for $F(R)$, while 
the dotted line is its analytical approximation (\ref{FfsmallR1}) 
with $C_F = 9.8$.
The lower solid line is the numerical solution for $f(R)$, while the dashed 
line is its analytical approximation (\ref{FfsmallR2}) with $C_f = 1.06$.}
\label{z1zagorfoFan2}
\end{figure}

In Figs.~\ref{z1zagorXan},~\ref{z1zagorfoFan2}, we show the continuation of the
numerically calculated functions $X(R), F(R), f(R)$ to the values of $R$ somewhat
smaller than $2M$. Since the functions change very rapidly, we switch the display 
from $F(R), f(R)$ to their logarithms. These graphs can be approximated by
the analytical formulas (\ref{FfsmallR1}), (\ref{FfsmallR2}), (\ref{Xanal}), 
which allow us to evaluate the constants $C_F,~C_f,~a,~b$ at the covered 
interval of $R$.

The found constant $a$ determines the parameter $\nu$, Eq.~(\ref{nu}). Since
$\alpha M = \sqrt{2} \times 10^{-6}$, the numerical value of $\nu$ is 
$\nu = 4.62977\times 10^5$, 
so that one can neglect 1 in comparison with $\nu$ in Eq.~(\ref{FfsmallR1}), 
(\ref{FfsmallR2}).   
Clearly, the deviations of the functions $f(R), ~F(R)$ from their
behaviour in GR are caused by the single dimensionless parameter: 
$\alpha M$, which was chosen to be $\alpha M = \sqrt{2} \times 10^{-6}$. 
In particular, the 
numerical values of $f(R), ~F(R), ~ X(R), ~X^{'}(R)$ at $R=2M$ are roughly
expressible as various simple powers of the number $\alpha M = \sqrt{2} 
\times 10^{-6}$. 
We have varied this parameter and have checked numerically that the general 
behaviour of solutions remains the same, but significant deviations 
of $f(R), ~F(R)$ from their GR behaviour begin closer and closer to $R =2M$ 
for smaller and smaller $\alpha M$. So, the field configuration resembles a 
black hole, in astrophysical sense, better and better, when 
$\alpha M$ decreases.  We have also checked 
that the qualitative behaviour of numerical solutions remains the same for 
some other choices of $\zeta$: $\zeta =2$ and $\zeta =3$.   
The most interesting conclusion of this investigation is that the functions
$f(R), ~F(R)$ remain regular and positive all the way down to $R=0$.
For static spherically-symmetric metrics (\ref{gmnr}), the location of the
(globally defined) event horizon is determined \cite{Wald} by the condition  
\[
f(R)=0. 
\]
In the massive gravity, the function $f(R)$ reaches zero only in the center 
$R=0$, where, as we will show shortly, the physical singularity occurs. The fact
that the function $f(R)$ does not vanish at any $R>0$ shows that the static
spherically-symmetric solutions of the massive gravity do not possess a regular
event horizon. It is truly surprising that it requires so little to get rid 
of the black hole event horizon - just the inclusion of arbitrarily small 
mass-terms (\ref{two}) in the highly non-linear gravitational equations. 
\footnote{Many would say that if a black hole (in its strict, 
mathematical sense) is the answer to a physical question, it must have been a 
very strange question. For example, the idea of unlimited collapse to a black hole
seemed (apparently) so disgusting to Landau that in order to avoid this conclusion
he was prepared to sacrifice the laws of quantum statistics \cite{Land}.
But it seems surprising to us that the disappearance of the 
event horizon may be caused by arbitrarily small mass-terms.} 

The Riemann invariant $I=g_{\mu\nu}g^{\alpha\beta}g^{\rho\sigma}g^{\phi\lambda}
R^{\mu}_{\alpha\rho\phi}R^{\nu}_{\beta\sigma\lambda}$ diverges at $R=0$ both in
GR and in the finite-range gravity. In GR, one uses the Schwarzschild solution
(\ref{mlsolx}) to calculate $I$: $I = 48 M^2 / R^6$. In the finite-range 
gravity, one uses the asymptotic formulas (\ref{FfsmallR1}), (\ref{FfsmallR2})
 in the limit of $R \rightarrow 0$.
The result of this calculation gives
\[
I \approx \frac{16 M^2}{R^6} (2 \nu^2 - 2\nu +3) 
C_F^2 \left(\frac{R}{2M} \right)^{-2 \nu}. 
\]
The singularity of the massive gravity at $R=0$ is a reflection of 
the assumed point-like nature of the source of the field and the source-free
form of equations everywhere outside of the source. One can expect that for 
realistic extended sources, the singularity will be replaced by a very compact 
distribution of matter.


\section{Cosmology} \label{sec13a}

The cosmological solutions are based on smoothly distributed matter,
so we shall work with the full set of (quasi-geometric) equations 
with matter sources (\ref{feqps}):
\bea
 G^{\mu}_{\ \nu} + M^{\mu}_{\ \nu} = \kappa T^{\mu}_{\ \nu}.
\label{pseudosys}
\ena
The conservation equations (\ref{Tconser}) are satisfied independently of
(\ref{pseudosys}), and therefore the consequences of Eq.~(\ref{pseudosys}) 
are given by Eq.~(\ref{MBian}) or, equivalently, by Eq.~(\ref{fbian}). 

We will be using the Lorentzian coordinates (\ref{Mi}), and we will be 
interested in simplest homogeneous isotropic cosmological
models. This means that the gravitational field components $h^{\mu\nu}$ depend 
only on time $t$ and have a diagonal form:
$$
h^{00}=A(t), \;\;\; h^{11}=h^{22}=h^{33}=-B(t).
$$
Since we want to use as many text-book calculations as possible, we introduce new
field variables $a(t)$ and $b(t)$ according to the definitions
$$
A= \frac{a^3}{b}-1, \;\;\;\; B= ab-1
$$
Then, the tensor $g^{\mu\nu}$, calculable from Eq.~(\ref{g}), has the 
following non-zero components 
$$
g^{00}=\frac1{b^2},\;\;\;\; g^{11}=g^{22}=g^{33}=-\frac1{a^2},
$$
and the inverse tensor is
$$
g_{00}= b^2, \;\;\;\; g_{11}=g_{22}=g_{33}=-a^2.
$$
The effective line-element acquires a familiar form
\be
\label{eff}
{\rm d} s^2 = b^{2}(t)c^2{\rm d}t^2 - a^{2}(t)({\rm d}x^2 + {\rm d}y^2 + {\rm d}z^2).
\en
The Einstein tensor $G^{\mu}_{\ \nu}$ calculated from the effective metric
(\ref{eff}) has the following non-zero components: 
\bea
G^{0}_{\  0}&=& \frac3{b^2}\left( \frac{a'}{a}\right)^2, \\ 
G^{1}_{\ 1}=G^{2}_{\ 2}=G^{3}_{\ 3}&=&
\frac1{b^2}\left[ 2\left(\frac{a'}{a}\right)^{'} +
3 \left( \frac{a'}{a}\right)^2 - 2 \frac{a'}{a}\frac{b'}{b}\right], 
\ena
where a prime denotes the derivative with respect to $ct$: $\prime = {\rm d}/ c{\rm d}t$.

As for the matter sources, we adopt a perfect fluid model with the (geometrical) 
energy-momentum tensor
\bea
T^{\mu\nu} =(\varepsilon + p)u^{\mu}u^{\nu} -p g^{\mu\nu}. \nonumber
\ena
Since $u^{i}=0$ and $u^0=1$, the non-zero components of $T^{\mu}_{\ \nu}$ are 
$T^{0}_{\ 0} =\varepsilon(t)$ and 
$T^{1}_{\ 1}= T^{2}_{\ 2}= T^{3}_{\ 3}= - p(t)$. The conservation equations
(\ref{Tconser}) reduce to a single equation  
\bea
 \varepsilon' + 3\frac{a'}{a}(p + \varepsilon)=0. \label{ctdot}
\ena
As the final simplification, we assume that the fluid is described by the 
equation of state $p(t)=q \varepsilon(t)$, where $q$ is a constant. (For our
purposes it will be sufficient to consider $-1 < q <1$.) Then equation 
(\ref{ctdot}) can be integrated to produce 
\bea
\label{1stint}  
\varepsilon(t) = \frac{\varepsilon_0}{a^{3(q+1)}}, 
\ena
where $\varepsilon_0$ is an arbitrary constant with the dimensionality
of energy density. More realistic models of matter
assume piece-wise equations of state, whereby the constant $q$ is 
different at different intervals of cosmological evolution.

\subsection{Homogeneous isotropic solutions in GR} 

It is instructive to start from the simplest Friedmann solutions of GR. The two 
independent Einstein equations are
\bea
\frac3{b^2}\left( \frac{a'}{a}\right)^2 = \kappa \varepsilon, \label{GRe} \\ 
\frac1{b^2}\left[ 2\left(\frac{a'}{a}\right)^{'} +
3 \left( \frac{a'}{a}\right)^2 - 2 \frac{a'}{a}\frac{b'}{b}\right]=
-\kappa q \varepsilon. \nonumber 
\ena
The second equation is satisfied identically, if the first equation and 
Eq.~(\ref{1stint}) are satisfied. Using the relationship (\ref{1stint}) in 
Eq.~(\ref{GRe}), and introducing the independent variable $\tau$ according to 
the definition 
\[
{\rm d} \tau = b(t) {\rm d}t, 
\]
equation (\ref{GRe}) can be integrated to yield 
\be
\label{atau}
a(\tau) = \left(\frac{\tau}{\tau_1}\right)^{\frac{2}{3(q+1)}},
\en
where
\[
c\tau_1 = \frac{2}{\sqrt{3} (q+1)} l_0 ~~~~{\rm and}~~~~
l_0 = \frac{1}{\sqrt{\kappa \varepsilon_0}}.
\]
Returning back to Eq.~(\ref{1stint}) with $a(\tau)$ from Eq.~(\ref{atau}), 
one finds
\be
\label{epsi}
\kappa \varepsilon = \frac{4}{3(q+1)^2 c^2 \tau^2}.
\en
Thus, in GR, the function $b(t)$ remains arbitrary. The effective 
line-element (\ref{eff}) can be written in the form 
\be
\label{2eff}
{\rm d} s^2 = c^2{\rm d} \tau^2 - a^{2}(\tau)({\rm d}x^2 + {\rm d}y^2 + {\rm d}z^2),
\en
where $a(\tau)$ is called the scale factor. The independent variable 
$\tau$ is the absolute time elapsed since the singularity at $\tau =0$. 
The values of measurable quantities, i.e. the matter energy
density $\varepsilon(\tau)$, Eq.~(\ref{epsi}), and the Hubble radius 
\[
l_{H}(\tau) =c/H(\tau) = \frac{3 (q+1)}{2} c\tau, 
\]
are completely determined by the value of the absolute time $\tau$. The 
constant $\tau_1$ (or, for this matter, the constant $l_0 /c$) has the
dimensionality of $[time]$. The constant $\tau_1$ marks the moment of 
time $\tau$ when the scale factor $a(\tau)$, Eq.~(\ref{atau}), 
reaches $a =1$, and the energy density $\varepsilon(\tau)$ reaches 
$\varepsilon_0$, but the numerical value of $a(\tau)$ has no physical significance. 
At any chosen moment of time $\tau$, by adjusting the constant $\tau_1$, one 
can make $a > 1$ or $a < 1$, while solutions with differing constants $\tau_1$ 
are observationally indistinguishable. As we shall see below, this situation 
changes in the massive gravity.  

\subsection{Exact cosmological equations in the finite-range gravity} 

The massive contributions $M^0_{\ 0}$ and $M^1_{\ 1}=M^2_{\ 2}=M^3_{\ 3}$
are directly calculable from their definitions (\ref{M}). With the massive
terms taken into account, the two independent field equations 
(\ref{pseudosys}) read:
\bea
3\left( \frac{\dot a}{a}\right)^2 + \frac3{8} \frac{\alpha^2}{\zeta +2} 
\left[ \frac{a^3}{b^3} - \frac{b}{a} + 2\zeta\left( \frac{1}{b^2} +
2 \frac{b}{a} -\frac{3}{a^2}\right) \right] = \kappa \varepsilon, 
\label{c00'}\\
2\left(\frac{\dot a}{a}\right)^{.} + 3 \left( \frac{\dot a}{a}\right)^2- \nonumber \\
\frac1{8} \frac{\alpha^2}{\zeta +2}\left[ 3\frac{a^3}{b^3} + \frac{b}{a} -  
4\frac{a}{b} + 2\zeta \left(\frac{3}{b^2} -2 \frac{b}{a} -4\frac{a}{b} +
\frac{3}{a^2} \right) \right] = - \kappa q\varepsilon,  \label{c11'}
\ena
where an over-dot denotes the derivative with respect to $c\tau$: 
$\; \cdot = {\rm d}/c{\rm d} \tau = {\rm d}/b(t) c {\rm d} t$, and $\zeta$ is
defined in Eq.~(\ref{zeta}). Since the matter
energy density $\varepsilon$ is determined by Eq.~(\ref{1stint}), the two 
unknown functions $a(\tau), ~b(\tau)$ are fully determined by the two equations 
(\ref{c00'}), (\ref{c11'}). In the finite-range cosmology, in contrast to GR, 
the function $b$ is not arbitrary.     

A direct consequence of Eqs. (\ref{c00'}), (\ref{c11'}) has the form of
${\dot M}^0_{\ 0} +3 ({\dot a}/a)[M^0_{\ 0} -M^1_{\ 1}] =0$. This is the 
single nonvanishing equation from the set of equations (\ref{MBian}). The 
left-hand-side of this equation can be transformed to a total time-derivative. 
This fact can be seen more easily from the equivalent form of this equation, 
stemming from  Eq.~(\ref{fbian}): 
\bea
{{\mM}^{00}}_{;0}=  \frac3{16}\frac{\alpha^2}{\zeta +2}
\frac{d}{dt}\left[ 3\frac{a^6}{b^2} - (1-4\zeta)a^2b^2 -2 (2\zeta +1)a^4
+ 8\zeta \frac{a^3}{b} - 8\zeta \right] =0.  \label{cmdot}
\ena
Equation (\ref{cmdot}) says that the combination of terms in the square 
brackets must be a constant. The value of this constant 
is determined by the observation that the zero gravitational field, i.e. 
$h^{\mu\nu} =0$ and, hence, $b= 1/a, ~a^2 = \pm 1$, should also
be a solution of this equation. On this ground, one finds that the integration 
constant should be equal to zero. As a result, Eq.~(\ref{cmdot}) 
yields to the following algebraic relationship between $a$ and $b$:
\be
\label{banda}
3 a^6 -a^2 b^4 -2a^4b^2 +4 \zeta(a^2b^4 - a^4b^2 +2a^3b- 2b^2) =0.
\en
In principle, this equation allows one to express $b$ in terms of $a$
for arbitrary $\zeta$. Then, the only differential equation to be solved
is one of the two equations (\ref{c00'}), (\ref{c11'}); say, the first one. 
Although this strategy solves the cosmological problem in principle, it 
is not easy to implement it analytically, for arbitrary $\zeta$. This is 
why we shall concentrate on particular simplifying choices of the 
parameter $\zeta$.

The first interesting case is $\zeta =0$. According to the definition
(\ref{zeta}), this case corresponds to $\beta^2 =0$ and, hence, to the zero 
mass of the $spin-0$ graviton. If $\zeta =0$, Eq.~(\ref{banda}) requires 
$b = \pm a$. Then, both, $M^0_{\ 0}$ and $M^1_{\ 1}$ vanish identically, and
equations (\ref{c00'}), (\ref{c11'}) retain their GR form. Thus, 
in the case of $\zeta =0$, the finite-range cosmology is exactly the 
same as the GR cosmology, independently of the mass of the $spin-2$ 
graviton. Therefore, deviations from the GR cosmology can arise only if 
the parameter $\beta$ is non-zero. 

Introducing
\[ 
y= \frac{a}{b}, 
\]
one can rewrite Eq.~(\ref{banda}) in the form
\be
\label{banda2}
y^4 +\frac{8\zeta}{3a^2} y^3 - \frac{2}{3a^4}\left(a^4 +2 \zeta a^4 +
4\zeta\right) y^2 + \frac{1}{3} \left(4\zeta - 1 \right) =0.
\en
This equation helps one to identify one more special case: $\zeta = 1/4$.
In this case, the 4th-order algebraic equation (\ref{banda2}) reduces to 
the 2nd-order equation, with the solution 
\bea
y \equiv \frac{a}{b} = \frac{-1 +\sqrt{7+9a^4}}{3a^2}. \label{a/b1/4}
\ena
(We have eliminated one of solutions by demanding $y\ge 0$.) 
We shall study this case analytically and numerically in considerable 
details. However, we will also be presenting, whenever possible, more 
general relationships, valid for $\zeta \ne 1/4$. In terms of the 
function $y$, equation (\ref{c00'}) can be written as
\bea
3\left( \frac{\dot a}{a}\right)^2+ 
\frac3{8} \frac{\alpha^2}{\zeta +2} \left[ y^3
- (1-4\zeta )\frac{1}{y} + \frac{2\zeta}{a^2}\left( y^2 -3 \right)
\right] =  \frac{\kappa \varepsilon_0}{a^{3(q+1)}}~, \label{c00f'} 
\ena

\subsection{The early-time and the late-time evolution in the finite-range 
cosmology}

\begin{figure}[tbh]
\vspace*{13pt}
\centerline{\psfig{file=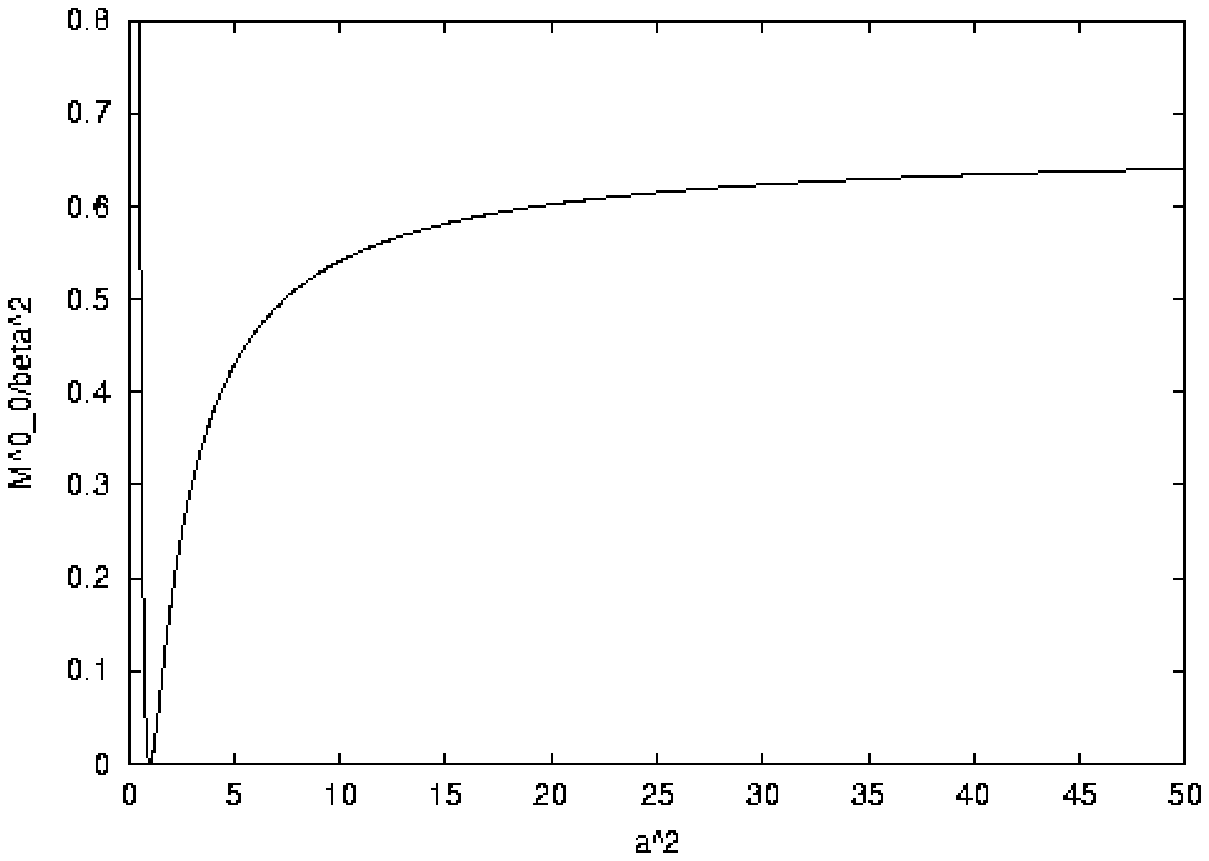}}
\vspace*{13pt}
\fcaption{${M^0}_0/\beta^2$ as a function of $a^2$ for $\zeta =1/4$.}
\label{M00}
\end{figure}

Equation (\ref{a/b1/4}) demonstrates that, in the massive gravity, the numerical 
value of $a(\tau)$ becomes important. The asymptotic formulas for $y(a^2)$ and,
hence, for $M^0_{\ 0} (a^2)$, depend on whether $a^2 \gg 1$ or $a^2 \ll 1$. 
In the former limit, the approximate expression of Eq.~(\ref{a/b1/4}) is
\[
y \approx 1, ~~~~~~a^2 \gg 1.
\]
This approximate solution is valid for any $\zeta$. Then, the approximate
expression for $M^0_{\ 0}$ is 
\[
M^0_{\ 0} \approx \frac{3}{2(\zeta +2)} \beta^2, ~~~~~~a^2 \gg 1.
\]
In the latter limit, the approximate expression of Eq.~(\ref{a/b1/4}) is 
\[
y \approx \frac{1}{3a^2} \left( -1 + \sqrt{7} \right), ~~~~~~a^2 \ll 1.
\]
The generalization of this solution to $\zeta \ne 1/4$ is
\[
y \approx \frac{4 \zeta}{3a^2} \left( -1 + \sqrt{1 +\frac{3}{2\zeta}}\right),~~~~~
a^2 \ll 1,
\]
and we ignore other solutions to Eq.~(\ref{banda2}) in this limit. Then, the 
approximate expression for $M^0_{\ 0}$ ($ \zeta = 1/4$) in the limit of 
small $a^2$ is
\[
M^0_{\ 0} \approx \frac{2(7\sqrt{7} -10)}{81} \frac {\beta^2}{a^6}, ~~~~~~
a^2 \ll 1.
\]
The full behaviour of $M^0_{\ 0}$ ($\zeta =1/4$) as a function of $a^2$ is described 
by a smooth curve that descends as $M^0_{\ 0} \propto a^{-6}$ from $+$infinity 
at the origin, then comes to the minimum, equal to zero, 
at $a^2 = 1$, and rises again to reach asymptotically, 
for $a^2 \rightarrow \infty$, the constant level $M^0_{\ 0} = 2 \beta^2 /3$.
This behaviour is illustrated in a numerical plot of Fig.~\ref{M00}.

We shall first show that there is a long interval of evolution where the
finite-range cosmology is practically indistinguishable from the GR 
cosmology. In addition to the already defined scales $l_H(\tau)=a/\dot{a}$ and 
$l_0=1/\sqrt{\kappa\varepsilon_0}$,
we introduce the finite-range scale $l_{\beta} = 1/\beta$. For simplicity, we will
be confined to the case $\zeta =1/4$, so that the finite-range scale associated
with $\alpha$ is simply related to the introduced one: $l_{\alpha} = 1/\alpha= 
l_{\beta}/2$. Let us consider the interval of evolution when $a^2 \gg 1$, 
but the massive term can be neglected. If $a^2 \gg 1$, the second (massive) 
term in Eq.~(\ref{c00f'}) is $M^0_{\ 0} \approx 2/3 l_{\beta}^2$. 
The first term is $3/l_H^2$, so when $l_H(\tau) \ll l_{\beta}$, the massive 
term can be neglected in comparison with the first term. 
Then, the first term is balanced by the right-hand-side of Eq.~(\ref{c00f'}),
i.e. by the term $1/ l_0^2 a^{3(q+1)}$. From the comparison of these two terms 
one finds that the inequality $l_0 \ll l_H(\tau)$ must hold.  Thus, in the 
interval of evolution such that $l_0 \ll l_H(\tau) \ll l_{\beta}$, the 
$a(\tau)$ and $\varepsilon(\tau)$ are well approximated by their GR expressions, 
Eq.~(\ref{atau}) and Eq.~(\ref{epsi}).  

\begin{figure}[tbh]
\vspace*{13pt}
\centerline{\psfig{file=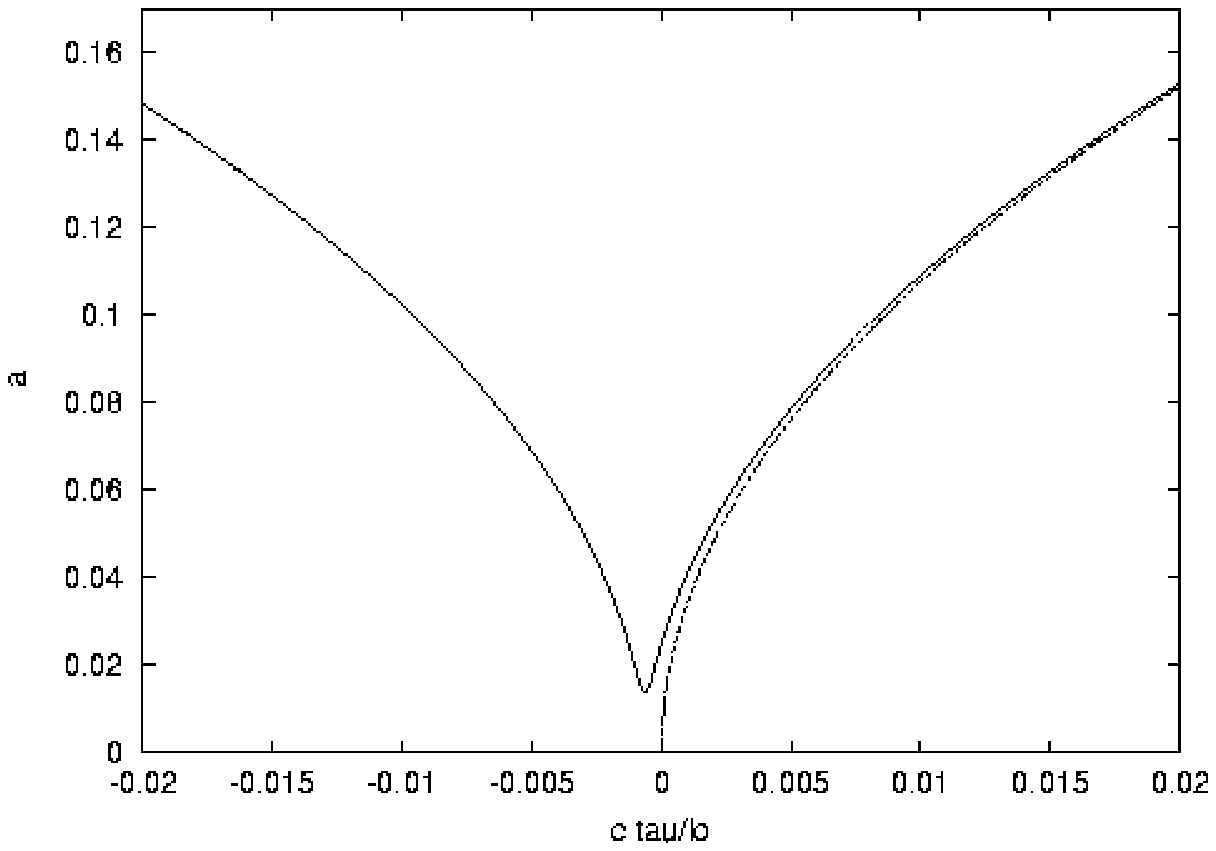}}
\vspace*{13pt}
\fcaption{The dashed line is a Friedmann solution with initial data $a=1$
at $c\tau/l_0 = \sqrt{3}/2$. The solid line is a numerical solution to 
Eq.~(\ref{c00f'}) for $\zeta=1/4, \alpha^2 l_0^2 =10^{-2}$
and the same initial data. Both solutions are found for $q=1/3$.}
\label{sing}
\end{figure}

We now turn to cosmological evolution at early times, when $a^2 \ll 1$. In GR, 
the early evolution begins with the singularity $a(\tau) =0$ at $\tau =0$. 
The scale factor cannot go through a regular minimum, 
where ${\dot a} =0$ and $a = a_{min}$. If it were possible that ${\dot a} =0$ 
at some moment of time, the l.h.s. 
of Eq.~(\ref{GRe}) would vanish at that moment of time, while the r.h.s. is 
strictly positive, so that Eq.~(\ref{GRe}) would not be satisfied. The situation 
changes in the finite-range gravity. It follows from Eq.~(\ref{c00f'}) that, 
in contrast to GR, $a(\tau)$ cannot be arbitrarily small. 
Indeed, if it were possible that $a^2 \rightarrow 0$, the r.h.s. of Eq.~(\ref{c00f'}), 
which is proportional to $1/a^{3(q+1)}$, would be negligibly small in comparison
with the massive term $M^0_{\ 0}$, which grows as $\sim 1/a^{6}$. But then, the two 
positive terms in Eq.~(\ref{c00f'}), the first and the second one, would not be
able to balance each other, and Eq.~(\ref{c00f'}) would not be satisfied. 
Instead, the scale factor of the finite-range cosmology goes through
a regular minimum, where ${\dot a}=0$. Near the minimum, the Hubble radius 
tends to infinity, so that $l_H \gg l_{\beta}$. The first term in Eq.~(\ref{c00f'})
can be neglected in comparison with the second (massive) term. From the comparison of 
the massive term with the r.h.s. of Eq.~(\ref{c00f'}), one can evaluate the minimum
value $a_{min}$ of the scale factor:
\[
a_{min} \approx \left(\frac{l_0}{l_{\beta}}\right)^{\frac{2}{3(1-q)}}.
\]
Since $l_0 \ll l_{\beta}$ and the exponent $2/3(1-q)$ is strictly positive,
we see that $a_{min} \ll 1$ as it should be. In the particular case of the early 
radiation-dominated era, i.e. for $q=1/3$, one finds that 
$a_{min} \approx l_0 /l_{\beta}$. The minimum of $a(\tau)$ is deeper for 
larger values of $l_{\beta}$ and, hence, for smaller values of the mass $m_0$ 
of the $spin-0$ graviton. (Certainly, the expected deep minimum of $a(\tau)$ 
does not invalidate the quantum-mechanical generation of cosmological
perturbations and their observational consequences \cite{BosG}.) The vicinity
of the minimum is shown in Fig.~\ref{sing} as a numerical solution to 
Eq.~(\ref{c00f'}) for $\zeta =1/4$, $\alpha^2 l_0^2 =10^{-2}$, and the
initial data $a=1$ at $c\tau/l_0 = \sqrt{3}/2$. The value of the parameter
$\alpha^2l_0^2$ is taken large, because otherwise the graph would be
superimposed on the Friedmann solution.

We shall now consider the late-time evolution in the finite-range cosmology. First,
we note that, in contrast to GR, the scale factor $a(\tau)$ cannot grow indefinitely.
Indeed, if it were possible that $a(\tau) \rightarrow \infty$, the r.h.s. of
Eq.~(\ref{c00f'}) would be negligibly small in comparison with the constant massive
term $M^0_{\ 0} \sim 1/ l_{\beta}^2$. But then, the two positive terms on the
l.h.s. of Eq.~(\ref{c00f'}) would not be able to balance each other. Instead,
the scale factor $a(\tau)$ goes through a regular maximum, where ${\dot a}=0$
and $a = a_{max}$. Near the maximum, the Hubble radius tends to infinity, 
and, similar to what takes place in the vicinity of the regular minimum, 
one has $l_H \gg l_{\beta}$.
The first term in Eq.~(\ref{c00f'}) can be neglected, and from the comparison
of the massive term with the r.h.s. of Eq.~(\ref{c00f'}) one can evaluate the
maximum value $a_{max}$ of the scale factor:
\[
a_{max} \approx \left(\frac{l_{\beta}}{l_0}\right)^{\frac{2}{3(1+q)}}.
\]
Since $l_0 \ll l_{\beta}$ and the exponent $2/3(1+q)$ is strictly positive,
one has $a_{max} \gg 1$, as it should be. In the particular case of the late 
matter-dominated era, i.e. for $q=0$, one finds that 
$a_{max} \approx (l_{\beta} /l_0)^{2/3}$. The maximum of $a(\tau)$ is higher for 
larger values of $l_{\beta}$ and, hence, for smaller values of the mass $m_0$ 
of the $spin-0$ graviton. The energy density at the maximum of expansion is 
given by the universal (valid for any $q$) formula 
$\kappa \varepsilon_{max} \approx 1/l_{\beta}^2$. 
The late-time behaviour of $a(\tau)$ admits an analytical treatment. 
At $a^2 \gg 1$, the approximate form of Eq.~(\ref{c00f'}) is
\bea
3\left( \frac{\dot a}{a}\right)^2+ \frac{3 \beta^2}{2(\zeta+2)} = 
\frac{1}{ l_0^2 a^3}~, \label{c00f'q0} 
\ena 
where we have taken $q=0$. This equation can be rearranged to read
\[
\frac{l_0 \sqrt{a} {\rm d}a}{\sqrt{1 - \frac{3 l_0^2 \beta^2}{2(\zeta+2)} 
a^3}} = \frac{c{\rm d}\tau }{\sqrt{3}}.
\]
The exact solution to Eq.~(\ref{c00f'q0}) takes the form

\be
\frac{1}{\sqrt{3}}c(\tau + \tau_{0}) =
\left\{
       \begin{array}{ll}
        \sqrt{\frac{2(\zeta+2)}{3\beta^2}} \arcsin\left[a^{3/2} 
        \sqrt{\frac{3 l_0^2 \beta^2}
        {2(\zeta+2)}} \right],  & \mbox{if $\beta^2 > 0$.} \\ 
        \sqrt{\frac{2(\zeta+2)}{-3\beta^2}} \ln\left[a^{3/2} 
        \sqrt{\frac{-3 l_0^2 \beta^2}
        {2(\zeta+2)}}+ \sqrt{\frac{-3 l_0^2 \beta^2}{2(\zeta+2)}a^3 +1}\right], 
        & \mbox{if $\beta^2 < 0$.}  
       \end{array}
\right. 
\label{deSit1}
\en

\begin{figure}[tbh]
\vspace*{13pt}
\centerline{\psfig{file=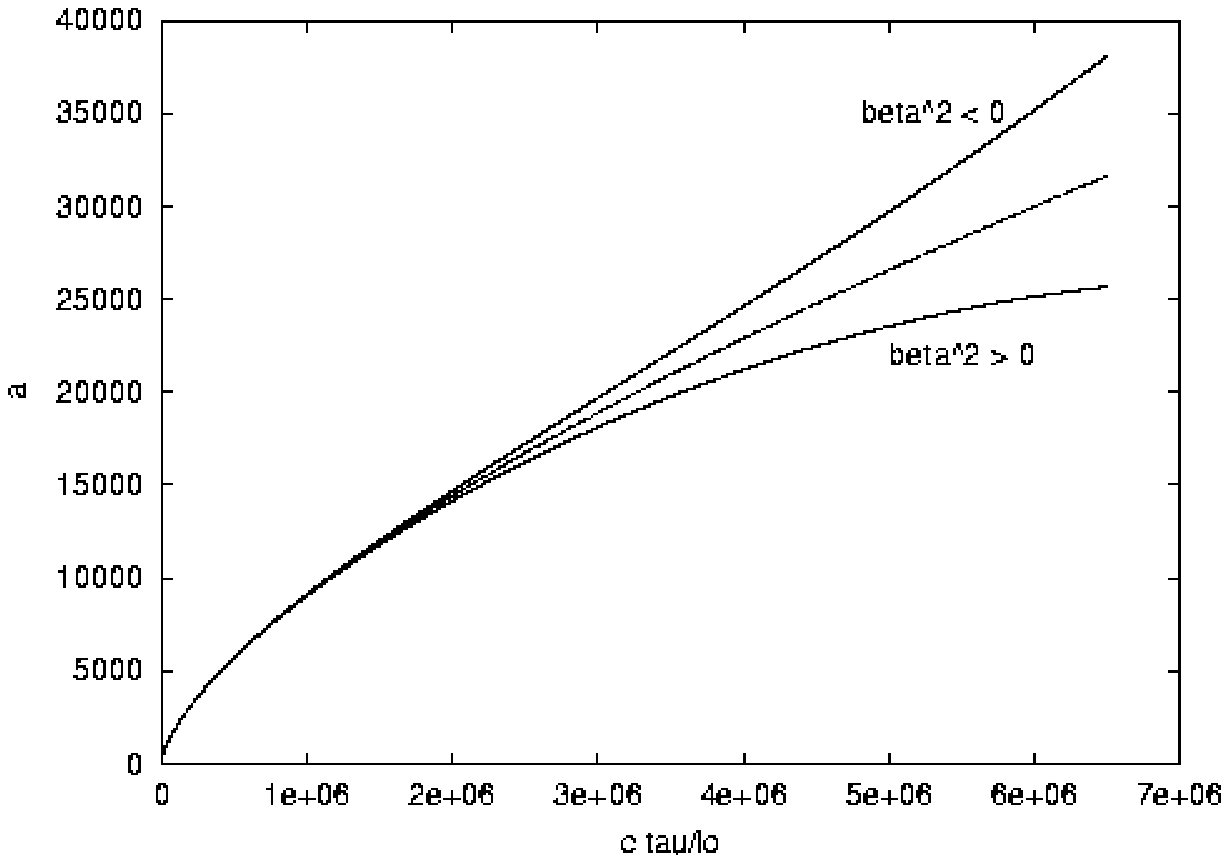}}
\vspace*{13pt}
\fcaption{The dashed line is a Friedmann solution of GR, the upper solid line 
is a solution with $\beta^2 < 0$ and the lower solid line is a solution
with $\beta^2 > 0$. The graphs are calculated for $ q=0,\; \zeta =1/4,\;
\alpha^2l_0^2 =10^{-12}$ and the initial data $a=1$ at $c\tau/l_0 = 2/\sqrt{3}$.}
\label{cosmol}
\end{figure}

Let us start the analysis of Eq.~(\ref{deSit1}) from the usual case $\beta^2 > 0$. 
The scale factor at late times can be written as 
\[
a(\tau) \approx \left(\frac{l_{\beta}}{l_0}\right)^{2/3} 
\left(\frac{2(\zeta +2)}{3}\right)^{1/3} \sin^{2/3}\left[\frac{c(\tau+\tau_0)}
{l_{\beta} \sqrt{2(\zeta +2)}}\right].
\]
It is clear from this formula that $a(\tau)$ goes through the regular maximum at
the moment of time when the argument of the $sine$ function reaches 
$\pi/2$. In agreement with the evaluations done above, the maximum value of 
the scale factor is given by $a_{max} \approx (l_{\beta}/l_0)^{2/3}$. On the 
other hand, as we have already shown, the scale factor experiences also a regular 
minimum. Thus, in a quite remarkable manner, the arbitrarily small mass-terms
(\ref{two}) make the
cosmological evolution oscillatory.
 \footnote{It is likely that, sooner or
later, this behaviour of $a(\tau)$ will be declared a prediction of ``inflation"; we are 
assured by previous inflationary literature that everything in life was predicted 
by ``inflation" or, at most, by alternatives to ``inflation".} The minima and maxima of
$a(\tau)$ are ``turning points'' of the effective potential in Eq.~(\ref{c00f'}),
which consists of $M^0_{\ 0}$ and the r.h.s. of that equation.

We now turn to the case $\beta^2 < 0$. As was explained in Introduction, the
interpretation of parameters $\alpha^2$ and $\beta^2$ in terms of masses requires 
them to be positive. However, the massive Lagrangian itself allows the
parameters $\alpha^2$ and $\beta^2$ to be negative. It is interesting 
to note that the case $\beta^2 <0$ makes the scale factor $a(\tau)$ 
exponentially growing with $\tau$ at very late times. Indeed, for 
$a^3 \gg 2(\zeta+2)/(-3 l_0^2 \beta^2)$, the second line of Eq.~(\ref{deSit1})
can be rearranged to read: 
\[
a(\tau) \approx \left( \frac{\zeta +2}{-6 l_0^2 \beta^2} \right)^{1/3}
e^{c(\tau+\tau_0)\sqrt{\frac{-2\beta^2}{9(\zeta+2)}}}.
\]
The part of evolution, where $a(\tau)$ experiences an exponential growth 
with $\tau$, mimics the contribution of the positive cosmological $\Lambda$-term.
This evolution is also similar to the ``accelerated expansion" driven 
(in framework of GR) by some speculative forms of matter known as ``quintessence" 
and ``dark energy". In contrast to these possibilities, the finite-range gravity
provides for the ``accelerated expansion" of the Universe at the expense of 
a specific modification of GR, without resorting to exotic forms of matter. 
The possible behaviour of the scale factor, under the assumption that
$\beta^2 <0$, is illustrated in Fig.~\ref{cosmol}. 
(Modifications to cosmological evolution caused by various alternative theories
of gravity have been discussed in references \cite{Freund}, \cite{Logunov},
\cite{GRS}, \cite{DKP}, \cite{DGS}.) 

For simplicity, we were considering here a conventional one-component matter. 
If, nevertheless, nature does allow for a cosmological $\Lambda$-term (of 
whatever origin, sign and value) as well as for various types of exotic matter, 
then, of course, the effects of massive gravity at early and late times can 
be partially or totally compensated by the $\Lambda$-term 
or exotic matter. In this case, the number of possibilities for cosmological 
evolution and its explanation increases greatly. For instance, the dynamical
effect of a huge positive $\Lambda$-term could be almost compensated by an
appropriate massive term with $\beta^2 >0$, with the net result of a modest  
``accelerated expansion" in the present era. Clearly, more definitive 
cosmological observations are badly needed.

\section{Conclusions}
\label{conc}

The internal logic of the field-theoretical formulation of the Einstein's
general relativity suggests certain modifications of GR, which take place
in the form of very specific mass-terms. These terms appear in addition 
to the field-theoretical analog of the usual Hilbert-Einstein Lagrangian. 
The arising finite-range gravity is a theory fully acceptable, both, 
from mathematical and physical points of view. Probably, the resulting 
theory can also be viewed as a phenomenological realisation of some 
macroscopic modifications of the 4-dimensional gravity, suggested by 
the M/string theory (see, for example, \cite{DDGV}, \cite{GRS}, \cite{KMP}, 
\cite{BKK}).

We have derived and studied the exact non-linear equations of the theory,
along with its linear approximation. The added terms have been interpreted
as masses of $spin-2$ and $spin-0$ gravitons. We have shown that the local
weak-field predictions of GR are fully recovered in the massless limit, that
is, when both masses are sent to zero. At the same time, the traditional
(Fierz-Pauli) way of including the mass-terms was shown to be very peculiar 
and unacceptable. It corresponds to sending the mass of the $spin-0$ graviton 
to infinity. As a result, the Fierz-Pauli theory contradicts the performed 
static-field experiments, as well as the indirect gravitational-wave 
observations. The contradiction stays even in the limit when the 
mass of the $spin-2$ graviton tends to zero. This fact rules out those 
variants of the candidate fundamental theories which suggest the macroscopic 
modifications of GR in the Fierz-Pauli form. At the same time, we have shown
that the non-Fierz-Pauli theories are free from ``negative energies",
``instabilities", etc. 

The most surprising deviations of the finite-range gravity from GR occur in 
strongly non-linear regime. We have considered static spherically-symmetric 
configurations and homogeneous isotropic cosmologies. We demonstrated that the 
mass-terms modify the Schwarzchild solution not only at very large 
distances (these are the expected Yukawa-type modifications that explain the 
name: finite-range gravity) but also in the vicinity of the Schwarzchild 
sphere $R= 2M$. The deviations near $R=2M$ are so radical that the 
event horizon does not form, and the massive solution smoothly continues 
up to $R=0$, where the curvature singularity develops. The result of this 
study is quite dramatic. In the astrophysical sense, the resulting massive 
configuration is still similar to the black hole configuration of GR. Namely, 
in the region of space just outside the $R=2M$, the gravitational field of the 
massive gravity solution is practically indistinguishable from the 
Schwarzschild solution. However, all conclusions of GR that rely specifically 
on the existence of the black hole event horizon, are likely to be abandoned. 
One can distinguish the two configurations observationally.
For instance, the gravitational waveforms emitted by a body inspiralling 
toward the center of configurations are expected to be different. In the 
finite-range gravity, in contrast to GR, the body continues to emit 
observable gravitational waves even from distances $R < 2M$.

Finally, we have considered cosmological solutions for homogeneous 
isotropic universes. We have shown that there is a long interval of 
evolution where cosmological solutions of the finite-range gravity 
are practically indistinguishable from those of GR. However, the 
arbitrarily small mass-terms lead to strong deviations from GR at very
early and very late times. We show in detail how the unlimited expansion 
is being replaced by a regular maximum of the scale factor, while the 
singularity is being replaced by a regular minimum of the scale factor. 
In other words, the arbitrary small mass-terms give rise to the oscillatory 
behaviour of the model universe. We show that when the gravitons are
traded for the ``tachyons", the cosmological scale factor exhibits an
interval of accelerated expansion instead of slowing down toward the 
maximum of expansion. This may explain the ``cosmic acceleration", if it
is observationally confirmed.

We believe that the solid theoretical motivations for the finite-range 
gravity, as well as its highly interesting conclusions derived so far, 
warrant further investigations in this area of research.


\appendix{\hspace{3mm} Equivalence of Eq.~(\ref{MBian}) and Eq.~(\ref{fbian})}.

First, we can rewrite Eq.~(\ref{MBian}) in a more convenient for our purposes form:

\bea
(\sg M^{\mu}_{\ \nu})_{|\mu}\equiv (\sg M^{\mu}_{\ \nu})_{,\mu}  
- \Gamma^{\sigma}_{\ \mu\nu}(\sg M^{\mu}_{\ \sigma}) = 0. \label{apMnum1}
\ena
Using $m_{\alpha\beta} \equiv 2(k_1 h_{\alpha\beta} + k_2 \gamma_{\alpha\beta}h )$
in Eq.~(\ref{1M}), we have
$$
M^{\mu}_{\nu} = (g^{\mu\alpha}\delta^{\beta}_{\nu} -\frac1{2}g^{\alpha\beta}
\delta^{\mu}_{\nu})m_{\alpha\beta}.
$$
Then, 
$$
\sg M^{\mu}_{\nu} = \sgm \left[ (\gamma^{\mu\alpha} + h^{\alpha\beta})
\delta^{\beta}_{\nu} -\frac1{2}(\gamma^{\alpha\beta} +h^{\alpha\beta})
\delta^{\mu}_{\nu}\right]m_{\alpha\beta}.
$$
The last expression is a tensor density with respect to the metric $\gamma_{\alpha\beta}$,
so we can rewrite Eq.(\ref{apMnum1}) as follows:
\bea
(\sg M^{\mu}_{\ \nu})_{|\mu} = (\sg M^{\mu}_{\ \nu})_{;\mu}  
- (\Gamma^{\sigma}_{\ \mu\nu} -C^{\sigma}_{\ \mu\nu})
(\sg M^{\mu}_{\ \sigma}) = 0.
\label{apMmaster}
\ena
Consider specifically the term $ \Gamma^{\sigma}_{\ \mu\nu} \sg M^{\mu}_{\ \sigma}$. 
Writing $\Gamma^{\sigma}_{\ \mu\nu}$ explicitly as a function of $g^{\mu\nu}$, we
obtain
\bea
 \Gamma^{\sigma}_{\ \mu\nu} \sg M^{\mu}_{\ \sigma} = 
\frac1{2}\sg \left( g^{\mu\alpha}\delta^{\beta}_{\nu} -
\frac1{2} g^{\alpha\beta}\delta^{\mu}_{\nu} \right) g^{\sigma\omega}
(g_{\omega\mu ,\nu} + g_{\omega\nu ,\mu} - g_{\mu\nu ,\omega}) m_{\alpha\beta}.
\label{apMnum2}
\ena
Using $\sg g^{\mu\alpha} g^{\beta\omega}
g_{\omega\mu ,\nu} = (\sg g^{\alpha\beta})_{,\nu} + \frac1{2}g^{\alpha\beta}
g_{\sigma\omega} (\sg g^{\sigma\omega})$ and $\sg g^{\sigma\omega}
g_{\sigma\omega ,\nu} = g_{\sigma\omega} (\sg g^{\sigma\omega})_{,\nu}$, 
Eq.(\ref{apMnum2}) can be transformed to
\bea
 \Gamma^{\sigma}_{\ \mu\nu} \sg M^{\mu}_{\ \sigma} =
-\frac1{2}(\sg g^{\alpha\beta})_{,\nu}m_{\alpha\beta}. \label{A5'}
\ena
The term $\sg g^{\alpha\beta}$ is again a tensor density as seen from 
Eq.~(\ref{g}). So we can trade the ordinary derivative of this term 
for the covariant one according to:
$$
(\sg g^{\alpha\beta})_{,\nu} = (\sg g^{\alpha\beta})_{;\nu} - 
\sg g^{\alpha\sigma} C^{\beta}_{\ \sigma\nu} - \sg g^{\beta\sigma} 
C^{\alpha}_{\ \sigma\nu} + \sg g^{\alpha\beta} C_{\nu}.
$$
Using this expression in (\ref{A5'}) and substituting (\ref{A5'}) into the 
last term of (\ref{apMmaster}), we get
\bea
\sg \left( g^{\mu\alpha}\delta^{\beta}_{\sigma} -
\frac1{2} g^{\alpha\beta}\delta^{\mu}_{\sigma} \right) 
 (\Gamma^{\sigma}_{\ \mu\nu} -C^{\sigma}_{\ \mu\nu})m_{\alpha\beta} &=& 
-\frac1{2}(\sg g^{\alpha\beta})_{;\nu}m_{\alpha\beta} \nonumber \\
&=&
\sgm h^{\alpha\beta}_{\ \ ;\nu}m_{\alpha\beta}.
\label{appMnum3}
\ena
The final step is to use the explicit form of $m_{\alpha\beta}$ and 
$\sgm h^{\alpha\beta}_{\ \ ;\nu}m_{\alpha\beta} = \sgm (k_1 h^{\alpha\beta}
h_{\alpha\beta} +k_2h^2)_{;\nu}$ in (\ref{appMnum3}) and (\ref{apMmaster}),
which leads us to the desired result:

\bea
\sg M^{\mu}_{\ \nu |\mu} &=& \sgm \left[ 2k_1h^{\mu}_{\ \nu} - 
\delta^{\mu}_{\nu} \left( k_1 + 2k_2\right)h + 2k_1h^{\mu\alpha}
h_{\nu\alpha}+ 2k_2h^{\mu}_{\ \nu}h - \right.\nonumber \\
& & \left. \frac1{2} \delta^{\mu}_{\nu}(k_1
h^{\alpha\beta}h_{\alpha\beta}+ k_2 h^2)\right]_{;\mu}=0  
\ena
One can see that the last equation proves the equivalence of Eq.~(\ref{MBian}) and
Eq.~(\ref{fbian}).

\appendix{\hspace{3mm} Emission of gravitational waves in the finite-range gravity}
\label{gwc}

In the presence of $T^{\mu\nu}$, the linearised field equations are
\bea
{h^{\mu\nu ,\alpha}}_{,\alpha} +
\eta^{\mu\nu}{h^{\alpha\beta}}_{,\alpha ,\beta} -
{h^{\nu\alpha ,\mu}}_{,\alpha}- {h^{\mu\alpha ,\nu}}_{,\alpha} + \nonumber \\
2\left[2k_1 h^{\mu\nu}-\eta^{\mu\nu}\left(k_1 + 2k_2 \right)h \right]=
2\kappa T^{\mu\nu}.
\label{Lpsgws}
\ena
Repeating the same steps that has led us to Eqs. (\ref{lsp0}), (\ref{lsp2}) and 
(\ref{dDL}), and taking into account the independent equations 
\be
\label{Tcons}
{T^{\mu\nu}}_{,\nu} = 0,
\en 
we again arrive at Eq.~(\ref{dDL}), but equations for $h$ and $H^{\mu\nu}$ 
become inhomogeneous:  
\bea
\Box h +\beta^2 h =\frac{2 \alpha^2 + \beta^2}{3 \alpha^2}2 \kappa T; \label{mlhws}   \\
\Box H^{\mu\nu} + \alpha^2 H^{\mu\nu} = 2\kappa \left( T^{\mu\nu}-
\frac1{3\alpha^2}T^{,\mu,\nu}+\frac1{6\alpha^2}\eta^{\mu\nu}\Box T -
\frac 1{6}\eta^{\mu\nu}T\right), \label{mlhttws}
\ena
where $T= g_{\mu\nu} T^{\mu\nu} \approx \eta_{\mu\nu}T^{\mu\nu}$.

The emitted field is determined by the retarded solutions to Eqs. (\ref{mlhws}),
(\ref{mlhttws}). When writing down these solutions, we closely follow the
recipes and conventions of the book \cite{Morse}, and we refer to this book for
further details. Let the distance between the point $\bf{r}_0$ 
within a compact source and the observation point $\bf{r}$ 
be $R=|\bf{r}-\bf{r}_0|$. The retarded
time $t_r$ is $t_r = t - R/c$. Let us start from Eq.~(\ref{mlhws}) and its
solution:  
\bea
h (t, {\bf r}) =\frac {2\kappa}{4\pi} \frac{2\alpha^2 +\beta^2}{3 \alpha^2}
\int_{0}^{t}dt_0 \int \left[ \frac{1}{R} \delta (t-t_0-R/c) - 
\right. \nonumber \\
 \left.\frac{\beta}{\sqrt{(t-t_0)^2-(R/c)^2}}J_1\left(\beta c
\sqrt{(t-t_0)^2-(R/c)^2}\right)u(t-t_0-R/c)\right] T\; d^3r_0, \label{EdPor}
\ena
where $J_1(x)$ is a Bessel function, $u(x)$ is a step function.
We assume that $r \gg r_0$, so that $R \approx R_0$, and we use the
small-argument approximation for the Bessel function, $J_1(x) \approx x/2$.
Then,
\bea
h (t, {\bf r}) =\frac {2\kappa}{4\pi} \frac{2\alpha^2 +\beta^2}{3 \alpha^2}
\left[ \frac1{R_0}\int T(t_r,r_0)d^3 r_0 -
\frac{\beta^2 c}{2}\int_0^{t_r}\int T(t_0,r_0)dt_0 d^3 r_0 +
O(\beta^4)\right]. \nonumber
\ena
In a similar manner, one finds the approximate solution to Eq.~(\ref{mlhttws}):
\bea
H^{\mu\nu}(t, {\bf r})= \frac{2\kappa}{4 \pi} \frac{1}{R_0} \left\{ 
\int T^{\mu\nu}(t_r,r_0)d^3 r_0- 
\frac 1{3\alpha^2}\partial^{\mu}\partial^{\nu}\int T(t_r,r_0)d^3 r_0 + \right. \nonumber \\
 \left. \frac 1{6\alpha^2}\eta^{\mu\nu}\Box\int T(t_r,r_0)d^3 r_0 - 
 \frac 1{6}\eta^{\mu\nu}\int T(t_r,r_0)d^3 r_0- \right. \nonumber \\
\frac{cR_0\alpha^2}{2}\left[\int^{t_r}_0\int T^{\mu\nu}(t_0,r_0)dt_0d^3r_0  
 -\frac 1{3\alpha^2}\partial^{\mu}\partial^{\nu}
\int^{t_r}_0\int T(t_0,r_0)dt_0d^3r_0 +  \right. \nonumber \\
\left.\left.\frac 1{6\alpha^2}\eta^{\mu\nu}\Box
\int^{t_r}_0\int T(t_0,r_0)dt_0d^3 r_0 -   \frac 1{6}\eta^{\mu\nu}\Box\int
^{t_r}_0\int T(t_0,r_0)dt_0d^3 r_0\right] +O(\alpha^4)\right\}.\nonumber
\ena
The terms involving integration over $t_0$ are ``tail effects" reflecting
the fact that the d'Alembert operator in Eqs. (\ref{mlhws}), (\ref{mlhttws})
is augmented by terms $\beta^2 h$ and $\alpha^2 H^{\mu\nu}$, respectively, 
causing dispersion of waves. Ironically, these cumbersome
and difficult to calculate ``tail effects" turn out to be unimportant for 
what follows below, but it was not easy to envisage this fact in advance.  

Having found $h$ and $H^{\mu\nu}$, we can write down the $h^{\mu\nu}$, 
using the relationship (\ref{dech}). We again assume that $h$ and 
$H^{\mu\nu}$ are Fourier decomposed in expansions similar to Eq.~(\ref{fexp}). 
Then, we obtain for the Fourier amplitudes:
\bea
\label{amn(k)}
a^{\mu\nu}(k^{\sigma})&=& \frac{1}{L} \left[ \hat{T}^{\mu\nu}(k^{\sigma})+
\frac{k^{\mu}k^{\nu}}{3\alpha^2}\hat{T}(k^{\sigma})-\frac1{3}\eta^{\mu\nu}
\hat{T}(k^{\sigma})-\frac{c R_0}{6}k^{\mu}k^{\nu}\tilde{T}(k^{\sigma})+
O(\alpha^2)\right],\nonumber \\
A(k^{\sigma})&=& \frac{1}{L} 
\left[-\frac1{6}\hat{T}(k^{\sigma}) +O(\beta^2)\right],\nonumber
\ena
where 
\bea
\frac{1}{L} = \frac{2\kappa}{4 \pi R_0}, ~~~~~
\hat{T}^{\mu\nu}(k^{\sigma})&=&\int\left( \int T^{\mu\nu}(t_r,r_0) d^3 r_0\right)
e^{-ik_{\sigma}x^{\sigma}} d^4x; \nonumber\\
\hat{T}(k^{\sigma})&=&\int\left( \int T(t_r,r_0) d^3 r_0\right)
e^{-ik_{\sigma}x^{\sigma}} d^4x; \nonumber \\
\tilde{T}(k^{\sigma})&=&\int\left( \int_{0}^{t_r}dt_0\int T(t_0,r_0) d^3r_0\right)
e^{-ik_{\sigma}x^{\sigma}} d^4x. \nonumber 
\ena
Clearly, since Eq.~(\ref{Tcons}) requires
\be
\label{Tcons1} 
\hat{T}^{\mu\nu} k_{\nu} =0,
\en
the derived matrix $a^{\mu\nu} (k^{\sigma})$ satisfies (in the leading order
by $\alpha^2$) the restrictions (\ref{5con}).    

We now consider a wave propagating in $z$-direction, and we neglect in $a^{\mu\nu}$
and $A$ the small contributions of order $\alpha^2$, $\beta^2$ and higher. Then,
\bea
A = \frac{1}{L} \left[-\frac1{6}\hat{T}\right];\;\;\; 
a^{11}= \frac{1}{L} \left[\hat{T}^{11}+
\frac1{3}\hat{T}\right]; \nonumber \\
a^{12}= \frac{1}{L} \hat{T}^{12}; \;\;\; 
a^{13}= \frac{1}{L} \hat{T}^{13}; \;\;\; 
a^{22}= \frac{1}{L} \left[\hat{T}^{22}+
\frac1{3}\hat{T}\right]; \nonumber \\
a^{23}= \frac{1}{L} \hat{T}^{23}; \;\;\; 
a^{33}= \frac{1}{L} \left[\hat{T}^{33}+
\frac{k^3k^3}{3\alpha^2}\hat{T} +\frac1{3}\hat{T}-
\frac{cR_0}{6}k^3k^3\tilde{T}\right]. \label{amnA} 
\ena
One can compare these amplitudes with those of GR, Eq.~(\ref{amplGR}). 
We see that the wave-field produced by one and the same distribution of matter is
different, depending on whether the emission is governed by equations
of GR or by equations of the finite-range gravity. The problem 
now is to quantify this difference in terms of observable effects, and to 
explore the massless limit $\alpha^2 \rightarrow 0$, $\beta^2 \rightarrow 0$ 
of the massive theory.

\appendix{\hspace{3mm} Gravitational energy-momentum tensor in GR and in finite-range gravity}
\label{emt}

The exact gravitational energy-momentum tensor $t^{\mu\nu}$ \cite{BG} in GR 
is quadratic in first derivatives of the field variables. In Lorentzian 
coordinates, the lowest non-vanishing approximation to the energy-momentum
tensor is given by the expression  
\bea
\label{gremtquadr}
\kappa t^{\mu\nu} = \frac{1}{4} \left[ h_{\alpha \beta ,}^{~~~\mu} 
h^{\alpha \beta , \nu} -\frac{1}{2} h_{,}^{~\mu}h_{,}^{~\nu} \right]+  
\nonumber \\ 
\frac{1}{2} \left[h^{\mu \nu}_{~~, \alpha} h^{\alpha \beta}_{~~, \beta} +
h^{\mu}_{~\alpha , \beta} h^{\nu \alpha , \beta}-
h^{\mu \alpha}_{~~, \alpha} h^{\nu \beta}_{~~, \beta} -
h^{\alpha \beta , \mu}h^{\nu}_{~\beta , \alpha}-  
h^{\alpha \beta , \nu}h^{\mu}_{~\beta , \alpha} \right]+  \nonumber \\ 
\frac{1}{4} \eta^{\mu\nu} \left[h_{\alpha \beta , \sigma} h^{\beta\sigma ,\alpha} - 
\frac{1}{2} h_{\alpha \beta , \sigma} h^{\alpha \beta, \sigma} +
\frac{1}{4} h^{, \alpha}h_{, \alpha} \right]. 
\ena
In the finite-range gravity, the gravitational energy-momentum tensor
consists of the GR part and the mass contribution (\ref{massemt}). The 
lowest (quadratic) approximation to the mass contribution is given by the 
expression
\bea 
\label{massemtq}
\kappa t^{\mu \nu}_{mass} = \frac{1}{2} \alpha^2 \left[ \frac{1-\zeta}
{2(2+\zeta)} h^{\mu\nu}h -h^{\mu}_{\alpha} h^{\nu \alpha}+ 
\frac{1}{4} \eta^{\mu \nu} \left( h^{\alpha\beta} h_{\alpha\beta}-
\frac{1+2\zeta}{2(2+\zeta)}h^2 \right) \right], 
\ena 
where we have used the relationships between $k_1, k_2$ and $\alpha^2, \beta^2$,
and $\zeta = \beta^2/\alpha^2$. These quadratic expressions (\ref{gremtquadr}),
(\ref{massemtq}) are sufficient for the calculation of energy-momentum
characteristics of weak gravitational waves.

In GR, we have to use the general solution (\ref{plw}) in
the expression (\ref{gremtquadr}). If, for the moment, we ignore in 
Eq. (\ref{plw}) the gauge terms with $c^{\alpha}$, then, because
of the conditions (\ref{ak}) and $k_{\alpha} k^{\alpha} =0$, 
the energy-momentum tensor (\ref{gremtquadr}) simplifies to 
\be
\label{gwtmn}
t_{\mu\nu} =\frac{1}{4 \kappa}\left[{h^{\alpha\beta}}_{, \mu}h_{\alpha\beta , \nu} -
\frac{1}{2} h_{, \mu}h_{, \nu} \right]. 
\en
Further calculation, using the relationships (\ref{a0mu}), (\ref{atild}),
(\ref{TTa}), leads to
\be
t^{\mu\nu}= \frac{1}{4 \kappa} k^\mu k^\nu \left[2 \tilde{a}^{ij} 
\tilde{a}^{*}_{ij}- \tilde{a}^{ij} \tilde{a}_{ij}e^{2ik_{\alpha}x^{\alpha}} -  
\tilde{a}^{*\ ij} \tilde{a}^{*}_{ij}e^{-2ik_{\alpha}x^{\alpha}}\right]. 
\en
Neglecting here the purely oscillatory terms (as we normally do in 
electrodynamics and other radiation theories) we derive formula (\ref{gwenm}).
In terms of the source characteristics, and for a wave traveling in
$z$-direction, we arrive at
\be
\label{emtz}
t^{\mu\nu}= \frac{1}{4 \kappa} k^\mu k^\nu \frac{1}{L^2} 
\left[ \left( \hat{T}^{11} - \hat{T}^{22} \right)^2 + 4 \left( \hat{T}^{12}
\right)^2 \right].
\ena
A remarkable fact is that even if we do not ignore the gauge terms 
with $c^{\alpha}$, and use the general solution (\ref{plw}) in the full 
expression (\ref{gremtquadr}), we still arrive at the same formula 
(\ref{gwenm}). The calculations show that the terms with $c^{\alpha}$ are 
only capable of producing purely oscillatory contributions, which we neglect 
anyway. Thus, only the TT-components contribute to the energy-momentum tensor.  
Of course, this result is fully consistent with the observational 
manifestations of g.w. in GR (see Sec. 5).

We now turn to the finite-range gravity. We have to use the general
solution (\ref{lad1}) and the energy-momentum tensor consisting of the
sum of Eq. (\ref{gremtquadr}) and Eq. (\ref{massemtq}). We neglect 
oscillatory contributions and retain only the lowest-order terms 
in mass parameters $\alpha^2, \beta^2$. In the massless limit, the
wave-vectors $k^{\alpha}$ and $q^{\alpha}$ coincide, but we retain
their symbols in order to keep track of terms of differing 
origins. Then, Eq. (\ref{gremtquadr}) gives 
\bea
\label{gremtq2}
\kappa t^{\mu\nu} = \frac{1}{4} k^{\mu} k^{\nu} 
\left[2 a^{\alpha \beta}a^{*}_{\alpha \beta} \right] -
(5+2 \zeta) |A|^2 q^{\mu} q^{\nu}.
\ena
Under the same conditions, Eq. (\ref{massemtq}) gives 
\bea 
\label{massemtq2}
\kappa t^{\mu \nu}_{mass} = 2(1+ \zeta) |A|^2 q^{\mu} q^{\nu}.
\ena
The terms with $\zeta$ in (\ref{gremtq2}) and (\ref{massemtq2}) originate
from the last two terms in (\ref{lad1}). We know (see Sec. 5) that those 
terms do not contribute to the g.w. observational effects. In a very
satisfactory way, those terms disappear also in the energy-momentum
tensor. Indeed, the sum of Eq. (\ref{gremtq2}) and Eq. (\ref{massemtq2}) gives 
the final result:
\bea
\label{emtqtot}
\kappa \left( t^{\mu\nu} +t^{\mu\nu}_{mass} \right) =  
\frac{1}{4} k^{\mu} k^{\nu} \left[2 a^{\alpha \beta}a^{*}_{\alpha \beta} \right] -
3|A|^2 q^{\mu} q^{\nu}.
\ena

This remarkable result requires a special discussion. At the first sight,
the Fierz-Pauli theory ($A=0$) is a satisfactory theory, whereas the non-Fierz-Pauli
theories ($A \neq 0$) are not. Indeed, in the first case, the energy density 
component $(\mu=0, \nu=0)$ is strictly positive and looks like that in GR, 
whereas in the second case there is a strictly negative contribution 
proportional to $|A|^2$, which seems to point out to ``negative energies" and 
associated ``instabilities". The reality, however, is diametrically opposite
to these naive expectations. The judgment on numerical values of the
energy-momentum tensor should be based on the relationship between the g.w. 
amplitudes and parameters of the source. In the Fierz-Pauli case, the 
amplitudes $a^{\alpha\beta}$ differ from those in GR, and therefore the 
numerical values of the gravitational energy-momentum tensor differ.
This is also reflected in differing deformation patterns of test particles.
On the other hand, in the non-Fierz-Pauli theories, the negative contribution
of $3|A|^2$ nicely cancels out with the similar term contained in
$a^{\alpha \beta}a^{*}_{\alpha \beta}$. Using Eqs. (\ref{amnA}), one can check
that expression (\ref{emtqtot}) reduces exactly to the GR expression 
(\ref{emtz}). This is in full agreement with the fact that observational
manifestations of GR and the finite-range gravity also coincide in the massless limit.

\appendix{\hspace{3mm} Static spherically-symmetric gravitational field in linear approximation}
\label{bhb}

In the beginning of this Appendix we perform calculations in Lorentzian 
coordinates (\ref{Mi}), but we later use also the spatially-spherical 
coordinates (\ref{sph}). Since the field is static and spherically-symmetric, 
all the components of the field are functions of $r=\sqrt{x^2 +y^2 +z^2}$.
The d'Alembert operator $\Box$ is replaced by the (negative) Laplace 
operator $\triangle \equiv \eta^{kl} \partial_{k} \partial_l$,
where $\partial_k$ is the ordinary partial derivative:
$
\partial_k \equiv \frac{\partial}{\partial x^k}.
$
Equations and solutions for linearised static fields 
are similar to equations and solutions for weak gravitational waves.
One will be able to see this similarity at every level of calculations.

First, we briefly summarise the situation in GR.
For the time-independent fields, the source-free equations (\ref{Lgw}) simplify, 
as they do not contain the time derivatives. The general static spherically-symmetric 
solution to these equations is given by
\be
\label{sle00}
h^{00} =  \frac{b}{r} + \triangle \psi, ~~~~~ 
h^{0k} = \left(\frac{a}{r}\right)^{,k}, ~~~~~ 
h^{kl} = -2 \partial^k \partial^l \psi +\eta^{kl}\triangle \psi, 
\en
where $b$ and $a$ are constants of integration, and $\psi$ is an arbitrary
function of $r$. The integration constants are determined by the source
of the field. Considering a static point-like source, 
that is, $T^{00}=M\delta^3(r), \; T^{0i}=0, \; T^{ij}=0$,
one identifies the integration constants: $b=4M,~ a=0$.  
One can now calculate the Riemann tensor (\ref{linriem}) and write down the
geodesic deviation equation (\ref{lgdein}). The result of this calculation 
is given by
\bea
\frac{d^2\xi^{i}}{dt^2}=-\left[\frac{M}{r}\right]^{,i}_{\ ,j} \xi^{j}.
\label{grgdssfin}
\ena

We now proceed to the source-free solutions of the massive gravity.
The full set of equations is the time-independent version of equations 
(\ref{dDL}), (\ref{lsp0}), (\ref{lsp2}), namely:
\bea
{h^{\mu\nu}}_{,\nu} -\frac{\alpha^2 -\beta^2}{2(2\alpha^2 +\beta^2)}h^{,\mu} = 0, 
\label{system2.0} \\
\triangle h  +  \beta^2 h =0,   \label{system2.1}\\ 
\triangle H^{\mu\nu} + \alpha^2 H^{\mu\nu} =0. \label{system2.2}
\ena
Let us start from the equation (\ref{system2.1}) for the trace $h$. 
The general solution to this equation is a linear combination of two 
Yukawa potentials:
\bea
h= b_1 Y(-\beta r) + b_2 Y(\beta r) \label{solh},
\ena
where
$$
 Y(-\beta r) \equiv \frac{e^{-\beta r}}{r}, \;\;\;\;\;\;\;\;
 Y(\beta r) \equiv \frac{e^{\beta r}}{r}.
$$
Here and below, we impose boundary conditions which require the solutions 
to vanish at infinity, i.e. for $r\rightarrow \infty$. In the case of
solution (\ref{solh}), this means that the constant $b_2$ must be put equal 
to zero, so that only the term with $Y(-\beta r)$ survives.  

We now turn to Eq.~(\ref{system2.2}). A tensor field is regarded
spherically-symmetric if it is built from a scalar function depending
only on $r$. The general expression for $H^{\mu\nu}$ is  
a combination of $Y(-\alpha r)$ and $Y(\alpha r)$, and their
derivatives. We retain only terms with $Y(-\alpha r)$ in order to satisfy
the boundary conditions. Then, the general and vanishing at infinity solution 
to Eq.~(\ref{system2.2}) has the form 
\bea
H^{\mu\nu}= A^{\mu\nu}Y(-\alpha r) + V^{\mu}\partial^{\nu}Y(-\alpha r)+
V^{\nu}\partial^{\mu}Y(-\alpha r) + B \partial^{\nu}\partial^{\mu}Y(-\alpha r),
\label{genH}
\ena
where $A^{\mu\nu},~ V^{\mu},~ B$ are constants. Since the $H^{\mu\nu}$ 
satisfies the conditions (\ref{TT}), not all constants are arbitrary; 
they are restricted by the relationships:
\bea
A^{00}= -2A,~~A^{0k}=0,~~ A^{kl}=A\eta^{kl},~~V^{\mu}=0,~~B=\frac{A}{\alpha^2},
\nonumber
\ena
where $A=\eta_{\mu\nu}A^{\mu\nu}$. With these constants, Eq. (\ref{genH})
takes on the form
\be
H^{00} =  -2 A Y(-\alpha r), ~~~ H^{0k} = 0, ~~~
H^{kl} = A\eta^{kl} Y(-\alpha r)  +
\frac{A}{\alpha^2} \partial^{k}\partial^{l} Y(-\alpha r). 
\label{solH11}
\en

Having found $h$ and $H^{\mu\nu}$, we can calculate $h^{\mu\nu}$. For this
purpose, we use formula (\ref{dech}) written as
$$
h^{\mu\nu}= H^{\mu\nu} + \frac{1}{2\alpha^2+\beta^2} h^{,\mu ,\nu} +
\frac{\alpha^2+\beta^2}{2(2\alpha^2 +\beta^2)} \eta^{\mu\nu} h.
$$
Substituting solutions for $h$ and $H^{\mu\nu}$ into this expression, and 
rearranging the terms, we arrive at
\bea
h^{00} &=& \left[-\frac{5}{2}AY(-\alpha r)+ DY(-\beta r)\right]
 + \triangle \left(\frac1{\alpha^2}\Psi\right),~~~~~h^{0k} =0, \label{ssol1}\\
h^{kl} &=& -\eta^{kl}\Psi -2 \partial^k\partial^l
\left(\frac1{\alpha^2}\Psi\right)  +\eta^{kl}
\triangle \left(\frac1{\alpha^2}\Psi\right), \label{ssol2}
\ena
where
\[
\Psi = -\frac1{2} A Y(-\alpha r)- D Y(-\beta r)~~~~{\rm and}~~~~
D = \frac{\alpha^2}{2(2\alpha^2+\beta^2)}b_1.
\]
With these $h^{\mu\nu}$, Eqs. (\ref{system2.0}) are satisfied automatically.
Thus, we are left with two arbitrary constants: $b_1$ and $A$.

The constants $b_1$ and $A$ are determined by the source of the field. Clearly,
the inhomogeneous equations for $h$ and $H^{\mu\nu}$ are the same as the 
previously discussed gravitational wave equations (\ref{mlhws}), (\ref{mlhttws}), 
but with the ``box" being replaced by the ``triangle":
\bea
\triangle h+ \beta^2 h &=&\frac{2 \alpha^2+ \beta^2}{3\alpha^2} 2 \kappa T, 
\label{mssswsh} \\
\triangle H^{\mu\nu} +\alpha^2 H^{\mu\nu} &=& 2\kappa\left( 
T^{\mu\nu}-\frac1{3\alpha^2}T^{,\mu,\nu}+\frac1{6\alpha^2}\eta^{\mu\nu}
\triangle T - \frac 1{6}\eta^{\mu\nu}T\right). \label{mssswstt}
\ena
We consider a static point-like source with $T^{\mu\nu}$ used before, 
namely, $T^{00}=M\delta^3(r),~T^{0i}=0,~T^{ij}=0$. Then, solution for $h$ 
satisfying the adopted boundary conditions at infinity is given by \cite{Morse}:
\bea
h=\frac{2\kappa}{4\pi} \frac{2 \alpha^2+\beta^2}{3\alpha^2}
\int \frac{e^{-\beta |{\bf r}-{\bf r}_0|}}
{|{\bf r} - {\bf r}_0|} T(r_0)d^3r_0 = 
 \frac{2(2 \alpha^2 +\beta^2)}{\alpha^2}\frac{2M}{3}Y(-\beta r). 
\nonumber
\ena
Comparing this solution with (\ref{solh}), we identify the constant $b_1$ 
and, hence, the constant $D$: $D=2M/3$.
In a similar manner, one finds solution to Eq.~(\ref{mssswstt}).
However, to identify the constant $A$, it is sufficient to write down 
only one component, $H^{00}$:
\bea
H^{00}=\frac{8M}{3}Y(-\alpha r).\nonumber
\ena
Comparing this solution with (\ref{solH11}), we identify $A$: $A= - 4M/3$.
As a result, function $\Psi$ takes the form
\be
\label{Psi}
\Psi=\frac{2M}{3}[Y(-\alpha r)-Y(-\beta r)].
\en

In Sec. 6 we will need solution (\ref{ssol1}), (\ref{ssol2})
written in spherical coordinates (\ref{sph}). Applying to tensor $h^{\mu\nu}$
a coordinate transformation from Cartesian to spherical coordinates,
one finds  
\bea
h^{00}&=& 4MY(-\alpha r) -\Psi + \triangle\left(\frac1{\alpha^2}\Psi\right), 
\label{sfh00}\\
h^{11}&=& \Psi -\frac2{\alpha^2}\ddot{\Psi} +
\triangle\left(\frac1{\alpha^2}\Psi\right), \label{sfh11}\\
h^{22}&=&\sin^2\theta~ h^{33}=\frac 1{r^2}\left(\Psi +
\frac1{\alpha^2}\ddot{\Psi}\right), \label{sfh22}
\ena
where the indeces $(1,2,3)$ correspond to the coordinates $(r, \theta, \phi)$,
and an over-dot denotes the derivative with respect to $r$.

We now turn to the geodesic deviation equation in the massive gravity.
We have to calculate the Riemann tensor participating in Eq.~(\ref{lgdein}).  
The last term of $h^{00}$ in (\ref{ssol1}), and the last two terms
of $h^{kl}$ in (\ref{ssol2}), do not contribute to the Riemann tensor.  
This fact simplifies the calculations. Having found the contribution of the 
remaining terms to the Riemann tensor, we can write:
\bea
\frac{d^2\xi^{i}}{dt^2}=\frac 1{2}\left[2AY(-\alpha r) +
DY(-\beta r)\right]^{,i}_{\ ,j}\xi^{j}.\label{mgdss}
\ena
Using $A= - 4M/3$,~$D =2M/3$, and expanding the Yukawa potentials
in powers of small parameters $\alpha r \ll 1$, $\beta r \ll 1$, we obtain
\bea
\frac{d^2\xi^{i}}{dt^2}=-\left[\frac{M}{r}+
O((\alpha r)^2,(\beta r)^2) \right]^{,i}_{\ ,j}\xi^{j}.\label{mgdssfin}
\ena
In other words, in the limit of vanishingly small parameters $\alpha^2$ and
$\beta^2$, we fully recover the GR result given by Eq.~(\ref{grgdssfin}).

The situation is dramatically different in the Fierz-Pauli case. 
As we have already discussed above, the trace $h$ is zero everywhere outside
the source. The solution (\ref{ssol1}), (\ref{ssol2}) retain its form,
but with $D =0$. At the same time, the constant $A$ is the same as in the
general case, i.e. $A= -4M/3$. Using $D=0$ and $A= -4M/3$ in the geodesic 
deviation equation (\ref{mgdss}), we arrive at 
\bea
\frac{d^2\xi^{i}}{dt^2}= - \frac{4}{3}\left[M Y(-\alpha r)
\right]^{,i}_{\ ,j} \xi^j = -\frac{4}{3}\left[\frac{M}{r} + 
O(\alpha^2 r^2)\right]^{,i}_{\ ,j} \xi^{j}. \nonumber
\ena
Comparing this expression with the GR result, Eq.~(\ref{grgdssfin}), we see that 
there exists a finite discrepancy, represented by the factor $4/3$, which does not 
vanish even in the limit of $\alpha^2 \rightarrow 0$. This discrepancy puts the 
Fierz-Pauli coupling in conflict with the available static field experiments. 
In particular, the massless limit of the Fierz-Pauli 
theory contradicts the observed value of  
the light deflection in the gravitational field of the Sun. This conclusion
has been previously reached by a number of authors \cite{VDV},~\cite{BD}.
 
Finally, we have to emphasize again that solution (\ref{ssol1}), (\ref{ssol2}) 
is valid at distances far away from the central source, i.e. at $r \gg M$. If, 
at the same time, $r$ is much smaller than $1/\alpha$ and $1/\beta$, then $r$ 
belongs to the region which we call the intermediate zone. In this zone, 
solution (\ref{ssol1}), 
(\ref{ssol2}) with $A= -4M/3$ and $D=2M/3$, is practically indistinguishable 
from the linearised GR solution (\ref{sle00}) with $b=4M$ and $a=0$. At very 
large distances, that is,
in the region where $\alpha r$ and $\beta r$ are comparable with 1 and much
greater than 1, solution (\ref{ssol1}), (\ref{ssol2}) of the massive gravity 
strongly deviates from that of GR, as the Yukawa potentials lead to the
exponential decrease of the field components $h^{\mu\nu}$ as functions of the 
increasing $r$. This is the expected behaviour, and it explains the name: 
the finite-range gravity. The region where $r$ is comparable with $M$ and 
smaller then $M$ is not covered by the linear approximation. To study the 
behaviour of solutions in this region, we need a non-linear treatment.

\end{document}